\documentclass[aps,prx,twocolumn,superscriptaddress,longbibliography]{revtex4-1}
\usepackage{amsfonts}
\usepackage{amsmath}
\usepackage{mathtools}
\usepackage{longtable}
\usepackage{amssymb}
\usepackage{graphicx}
\usepackage{bm}
\usepackage{color}
\usepackage{mathrsfs}
\usepackage[colorlinks,bookmarks=true,citecolor=blue,linkcolor=red,urlcolor=blue]{hyperref}
\usepackage{appendix}
\usepackage{float}
\usepackage{booktabs}
\usepackage[export]{adjustbox}
\usepackage{braket}
\newcommand{\RNum}[1]{\uppercase\expandafter{\romannumeral #1\relax}}

\begin{document}
\title{Superdiffusive transport protected by topology and symmetry in all dimensions}
\author{Shaofeng Huang}
\thanks{These authors contributed to this work equally.}
\affiliation{Beijing National Laboratory for Condensed Matter Physics and Institute of Physics, Chinese Academy of Sciences, Beijing 100190, China}
\affiliation{University of Chinese Academy of Sciences, Beijing 100049, China}
\author{Yu-Peng Wang}
\thanks{These authors contributed to this work equally.}
\affiliation{Beijing National Laboratory for Condensed Matter Physics and Institute of Physics, Chinese Academy of Sciences, Beijing 100190, China}
\affiliation{Institute of Science and Technology Austria (ISTA), Am Campus 1, 3400 Klosterneuburg, Austria}
\author{Jie Ren}
\affiliation{Beijing National Laboratory for Condensed Matter Physics and Institute of Physics, Chinese Academy of Sciences, Beijing 100190, China}
\affiliation{School of Physics and Astronomy, University of Leeds, Leeds LS2 9JT, United Kingdom}
\author{Chen Fang}
\email{cfang@iphy.ac.cn}
\affiliation{Beijing National Laboratory for Condensed Matter Physics and Institute of Physics, Chinese Academy of Sciences, Beijing 100190, China}
\affiliation{Kavli Institute for Theoretical Sciences, Chinese Academy of Sciences, Beijing 100190, China}

\begin{abstract}
    Superdiffusion is an anomalous transport behavior. Recently, a new mechanism, termed the ``nodal mechanism," has been proposed to induce superdiffusion in quantum models.
    However, existing realizations of the nodal mechanism have so far been proposed on fine-tuned, artificial Hamiltonians, posing a significant challenge for experimental observation. In this work, we propose a broad class of 
    models for generating superdiffusion potentially realizable in condensed matter systems across different spatial dimensions. A robust nodal structure emerges from the hybridization between the itinerant electrons and the local impurity orbitals, protected by the intrinsic symmetry and topology of the electronic band. We derive a universal scaling law for the conductance, $G \sim L^{-\gamma}$, revealing how the exponent is dictated by the dimensionality of the nodal structure ($D_{\text{node}}$) and its order $n$, and the dimensionality of the system $(D)$ at high temperatures or that of the Fermi surface ($D^F$) at low temperatures.
    Through numerical simulations, we validate these scaling relations at zero temperature for various models, including those based on graphene and multi-Weyl semimetals, finding excellent agreement between our theory and the computed exponents.
    Beyond the scaling of conductance, our framework predicts a suite of experimentally verifiable signatures, notably a new mechanism for linear-in-temperature resistivity ($\rho \sim T$) and a divergent low-frequency optical conductivity ($\sigma(\omega) \sim \omega^{\gamma -1}$), establishing a practical route to discovering and engineering anomalous transport in quantum materials.
\end{abstract}

\maketitle

\section{Introduction}
Transport behavior can be characterized by the scaling exponent $\gamma$ of the conductance $G$ with system size $L$ under a fixed bias, expressed as $G \sim L^{-\gamma}$\cite{boundarydrivenRMP,RMP23}. This exponent quantifies how the current decays as the system length increases along the direction of the bias. For $\gamma=0$, the transport is ballistic,  meaning the current remains independent of system size. This behavior is commonly observed in non-interacting and certain integrable systems, where non-decaying quasiparticles prevent current decay. $\gamma=1$ corresponds to diffusive transport, which follows Ohm’s law, the most common scenario and can be observed in general chaotic systems. When $\gamma>1$, the system exhibits subdiffusive transport, characterized by a current that decays faster than in the diffusive case. This phenomenon is usually found in systems with kinetic constraints\cite{sub1,sub2,sub3,sub4,sub5,sub6,sub7,sub8,sub9} or low-dimensional quantum systems with random disorder\cite{Anderson58,Mantica97,Thouless72,Hirota71,Ljubotina18,Reichman15,Demler15,Altshuler06,Karrasch17}.

A particularly intriguing regime occurs for $0<\gamma<1$, where the system exhibits superdiffusive transport. In classical systems, superdiffusion is known to occur in one-dimensional models that conserve total momentum\cite{Prosen2000,Narayan2002,Spohn,Prosen2003,Prosen2005}. For quantum systems, superdiffusion may arise in systems with long-range interactions\cite{longrange1,longrange2,longrange3,longrange4,longrange5,longrange6,zhicheng25}. However, for typical short-range lattice models, it typically requires specific structural properties: in non-interacting systems, it can occur in disordered systems with designed inhomogeneous chemical potential\cite{Fibonacci1,Fibonacci2,Fibonacci3,Phillips1990,Harmonic1971,Dhar2001,Cane2021,Titov16,gattenlohner2016levy}; in interacting systems, superdiffusion is present in integrable models with non-Abelian symmetries\cite{Marko11,Prosen17,Prosen19,Romain19,Brayden21,Claeys22,Lucas21,AnnualReviews24,RMP23,Norman22,Romain24,Moessner24,RVprx25}. A recent numerical study\cite{Papic23} indicates superdiffusive energy transport in PXP
and related models up to long time scales, though a theoretical understanding remains absent.

Recent studies have proposed a new mechanism, known as the ``nodal  mechanism," for inducing superdiffusive transport in various systems with disorder\cite{Chen24,Junaid24}, dephasing\cite{Jie24,Marko24}, and interaction\cite{Romain25}, respectively, at high temperatures. This mechanism involves two key ingredients. The first is a single-band free-fermion Hamiltonian, $\hat H =\sum_k\epsilon_k \hat c_k^\dagger \hat c_k  $ with dispersion $\epsilon_k$, which provides ballistic quasiparticles. The second is the dephasing, disorder, or interaction terms constructed from local ``nodal operators," in which each term commutes with particle number $\hat n_{k_0}$ at a specific momentum $k_0$. This nodal structure results in an asymptotically long quasiparticle lifetime for momenta near $k_0$, ultimately leading to superdiffusive transport. Moreover, a recent study\cite{Pal25} suggests the possibility that this nodal structure can also induce the L\'evy flights of quantum information.

Previous studies on this mechanism, however, have three primary limitations: (1) they mainly focus on one-dimensional models; (2) they primarily investigate transport at infinite temperatures; and crucially, (3)  all models are artificial and fine-tuned, making experimental realization challenging.

In this work, we introduce a class of physically grounded models where nodal structures emerge naturally. These models exhibit superdiffusive transport across various dimensions, and we investigate their transport behavior at both high and low temperatures.

Inspired by the Anderson model\cite{Anderson1961}, we consider the ``two-component" model consisting of: (1) itinerant conduction electrons ($c$-electrons) and (2) localized $f$-electrons subject to disorder or interaction, coupled via (3) local hybridization. Then we demonstrate that the hybridization process itself naturally creates a nodal structure in momentum space where the hybridization amplitude between $c$-electrons and the $f$-electron vanishes. Crucially, we establish that these nodes are robustly enforced by the fundamental ``selection rules" dictated by the crystal symmetry and band topology of the $c$-electron bands. This protected nodal structure can make the system superdiffusive.

The collection of these protected nodes forms a geometric object---a point, line, or surface---that we term the ``nodal manifold". Its dimension, $D_{\text{node}}$, and the power-law with which the scattering vanishes near it (the order of the node, $n$) are the key parameters. 
We can use this framework to identify the nodal structures in realistic materials. For instance, we predict: (i) in graphene, topology enforces order-1 ($n=1$) nodal arcs connecting the Dirac points; (ii) in 3D multi-Weyl semimetals, the interplay between topology and rotation symmetry stabilizes higher-order nodal arcs; and (iii) in cubic lattices, crystal symmetry can generate nodal lines along the Brillouin zone (BZ) boundaries.

Based on this framework, we derive the universal scaling laws for the conductance with the system size: $G \sim L^{-\gamma}$. We find that $\gamma$ is determined by the geometry of the nodal structure at both high and low temperatures. At high temperature, the exponent is given by:
\begin{equation}
    \gamma_{\text{high}} = \min\{ \frac{D-D_{\text{node}}}{2n}, 1\},
\end{equation}
where $D$ is the system dimension. At zero temperature, charge transport is dominated by the states on the Fermi surface, and the exponent becomes dependent on it geometry and the intersection with the nodal manifold:
\begin{equation}
    \gamma_{\text{low}} = \min\{\frac{D^F - D^F_{\text{node}}}{2n} ,1  \},
\end{equation}
where $D^F$ and $D^F_{\text{node}}$ are the dimensions of the Fermi surface and its intersection with the nodal manifold. Let's illustrate these formulas by an example. Consider a 2D system with a first order $(n=1)$ nodal arc ($D_{\text{node}}=1$), a scenario relevant to graphene. At high temperature, this directly gives $\gamma_{\text{high}} = 1/2$. At zero temperature, assuming a typical 1D Fermi surface that intersects the nodal arc at isolated points ($D^F = 1, D^F_{\text{node}} =0$), we find $\gamma_{\text{low}} = 1/2$ as well, indicating superdiffusion. When the $f$-electron states are described by the disordered non-interacting impurity levels, we validate this scaling relation through extensive numerical simulations (by the Kwant package\cite{KwantRef}) in diverse models, including the 2D graphene system, the 3D topological multi-Weyl semimetals (e.g., based on $\text{Hg}\text{Cr}_2\text{Se}_4$) and a single-band model on the simple square (cubic) lattice, where nodes are protected by the nontrivial band topology or point-group symmetry.

We then discuss the effect of ``non-nodal" perturbations, terms that do not preserve $\hat{n}_{\mathbf{k}_0}$ at the node. These terms may include the potential scattering and the interaction among the $c$-electrons.
While any such perturbation will ultimately drive the system to diffusion in the thermodynamic limit, the superdiffusion will persist over an intermediate length scale. This leads to predictable crossover from superdiffusion $(G \sim L^{-\gamma})$ to diffusion $(G\sim L^{-1})$, governed by a characteristic length scale $L_c$, that is inversely proportional to the non-nodal scattering rate. We formalize this crossover with a universal scaling law and verify it numerically with a data collapse.

Beyond the scaling of conductance, our framework predicts experimentally verifiable effects, offering pathways to directly probe the nodal structures and quantum superdiffusion.
Firstly, in disordered systems exhibiting nodal structure alongside residual electron-electron interactions (which typically yield a Fermi liquid resistivity $\rho(T) \propto T^2$), our theory predicts the resistivity to scale as $\rho(T) \propto T^{2(1-\gamma_{\text{low}})}$ under a crossover temperature $T^*$. Notably, for $\gamma_{\text{low}}=1/2$, this points to the emergence of linear-in-temperature resistivity.
Secondly, we predict a distinct signature in the low-frequency optical conductivity, $\sigma(\omega)$, at zero temperature. The presence of the nodal structure leads to a power-law scaling of the form $\sigma(\omega) \sim \omega^{\gamma-1}$. In the superdiffusive regime ($\gamma < 1$), this manifests as a power-law divergence of the conductivity as $\omega \to 0$, a hallmark of enhanced transport from long-lived quasiparticles near the nodes.
Thirdly, the nodal structure should leave an imprint in angle-resolved photoemission spectroscopy (ARPES) measurements. Specifically, the quasiparticle broadening is predicted to scale with momentum deviation $\delta k$ from the node as $|\delta k|^{2n}$, where $n$ is the order of the node. This offers a spectroscopic method to characterize the nodal structure.
These experimentally accessible predictions provide concrete ways for testing the proposed mechanism of nodal-hybridization-induced superdiffusion and for identifying its manifestation in candidate materials.


By establishing a link between fundamental symmetries, band topology, and transport exponents, this work opens a new avenue for the discovery and engineering of quantum materials with exotic transport properties. Our findings bridge the gap between abstract theoretical models and the experimentally accessible signatures of real materials, providing a predictive framework for identifying superdiffusion in a wide range of systems.

The remainder of this paper is organized as follows.
In Sec.~\ref{sec:nodal structure}, we review the previous works about the nodal mechanism and introduce our ``two-component" model.
Sec.~\ref{sec:transport analysis} is devoted to the analysis of conductance's scaling with system size in the presence of the nodal structure.
In Sec.~\ref{sec:symmetry_superdiff} and~\ref{sec:topology_superdiff}, we detail the symmetry and the topological selection rules that induce the nodal structure and numerically verify the scaling law for the conductance.
In Sec.~\ref{sec:non-nodal}, we analyze the effect of non-nodal perturbation.
Sec.~\ref{sec:experimenal} shifts focus to experimentally verifiable consequences stemming from the nodal structure.
Finally, Sec.~\ref{sec:conclusion and discussion} provides a conclusion and discussion of this work.

\section{The nodal structure and the two-component model}
\label{sec:nodal structure}
In this section, we first review the ``nodal mechanism" as established in \cite{Chen24,Jie24,Romain25}. We then a introduce physically grounded two-component hybridization model and demonstrate how nodal structures can emerge naturally within it, providing a robust platform for realizing anomalous transport.

\subsection{The nodal mechanism}
Recent studies have proposed a new mechanism, known as the nodal-structure mechanism or simply nodal mechanism, for inducing superdiffusive transport in various systems with disorder\cite{Chen24,Junaid24}, dephasing\cite{Jie24,Marko24}, and interaction\cite{Romain25}, respectively, at high temperatures. This mechanism involves two key ingredients. The first is a single-band free-fermion Hamiltonian, $\hat H_0 =\sum_k\epsilon_k \hat c_k^\dagger \hat c_k  $ with dispersion $\epsilon_k$, which provides ballistic quasiparticles. The second is the dephasing, disorder, or interaction terms constructed from local operators $\hat d_{R}$'s that have nodes.

Using the 1D case as examples, we define 
\begin{align}\label{eq:d_operactor}
    \hat d_R \equiv \sum_\delta \phi_\delta \hat c_{R+\delta}, 
\end{align}
which is a superposition of annihilation operators $\hat c$ centered around site $R$. We impose the condition that the amplitude $\phi_\delta$ be nonzero only when $|\delta|<r$ for some finite range $r$, ensuring locality. The commutation relation between the local $\hat{d}$ operators and the Bloch-band operators $[\hat d_R, \hat c_k^\dagger]= e^{ikR} \tilde \phi_k$, where $\tilde \phi_k = \sum_\delta e^{ik\delta} \phi_\delta$ is the Fourier transform of the coefficients $\phi_\delta$. Consequently, if $\tilde \phi_{k_0} = 0$ for some specific momentum $k_0$, we have $[\hat d_R, \hat n_{k_0}]=0$ for all sites $R$. 
Hence we call $\hat d_{R}$'s nodal operators with a node at $k_0$.
Then, we can use these nodal operators to construct the nodal dephasing through the Lindbladian operators $\hat L_R = \hat d_R^\dagger \hat d_R$\cite{Jie24}, the nodal disorder $\hat V = \sum_R \epsilon_R \hat d_R^\dagger \hat d_R $ with random coefficients $\epsilon_R$\cite{Chen24}, and nodal interaction $\hat U = W\sum_R \hat d_{R}^\dagger \hat d_{R} \hat d_{R+1}^\dagger \hat d_{R+1}$\cite{Romain25}. In each case, every term commutes with $\hat n_{k_0}$, preserving the nodal structure. 

We can intuitively understand how the superdiffusion results from nodal structure by considering an initial state as a Slater determinant of $N$ quasiparticles with distinct momenta: $|k_1, \cdots, k_N\rangle$. Each quasiparticle transports ballistically with finite momentum under the time evolution with the free fermion Hamiltonian $\hat H_0$. When dephasing, disorder, or interactions are turned on, these ballistic quasiparticles would be scattered. However, as the nodal structure ensures $[\hat n_{k_0}, \hat V_i]=0$, the lifetime of quasiparticles with momenta near $k_0$ diverges asymptotically. In the long-time limit, transport is dominated by the surviving  quasiparticles, resulting in superdiffusive transport.

This construction provides a framework for building superdiffusive models in various systems, and the scaling exponent $\gamma<1$ is determined by the order of the node, the band dispersion near the node and the nature of the perturbation being dephasing, disorder or interaction\cite{Chen24,Junaid24,Romain25,Marko24,Jie24}.

\subsection{The two-component model}
While the nodal structure is not limited to one dimension, previous work primarily focused on one-dimensional models. Moreover, the node in the $\hat d-$operators requires special forms of $\phi_\delta$ in Eq.~(\ref{eq:d_operactor}), resulting in models that are challenging to realize experimentally. 

\begin{figure}[t]
    \includegraphics[width=0.96\columnwidth]{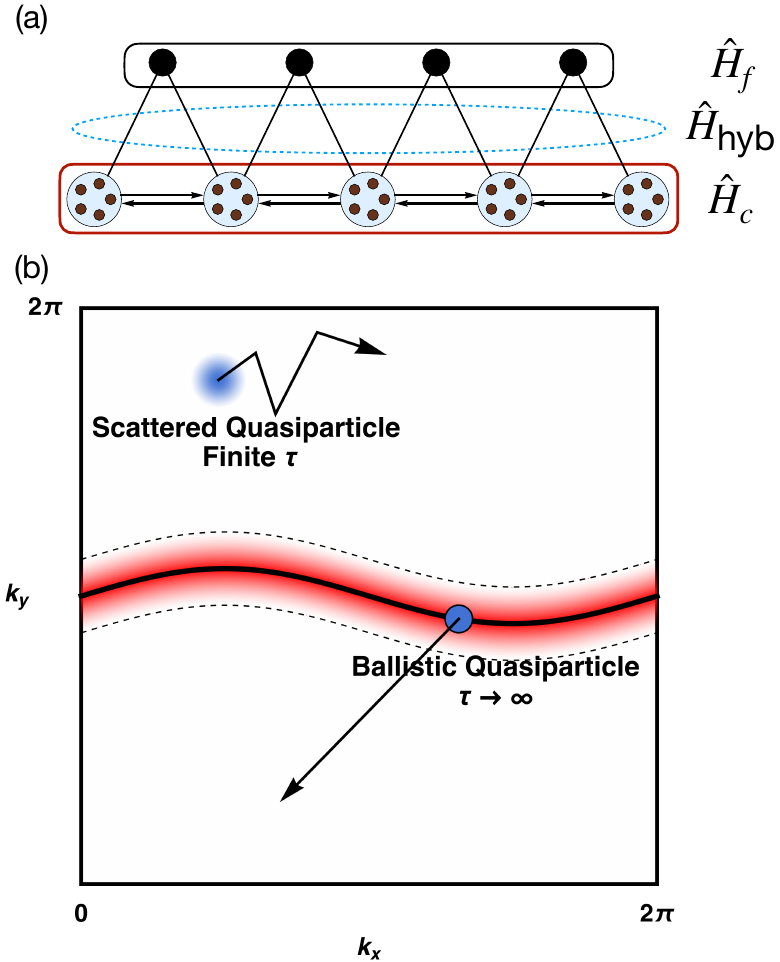}
    \caption{\textbf{The nodal hybridization mechanism for anomalous transport.} (a)A schematic of the two-component model. It consists of itinerant $c$-electrons (which may possess internal degrees of freedom like sublattice or orbital, represented by dots within circles) forming a conduction band ($\hat H_c$) and the $f$-electrons (black circles) as the scattering source ($\hat H_f$). The crucial element is the hybridization term ($\hat H_{\text{hyb}}$)  that couples the two systems. (b) The consequence of nodal structure in momentum space. The specific form of the hybridization creates ``nodal manifold" (thick black line) where the coupling vanishes. Quasiparticles on or near this manifold (red shaded area) have anomalously long lifetimes and dominate the transport. A state exactly on the node is ballistic ($\tau \to \infty$), while a state far from the node, represented by the blurred circle, is strongly scattered (finite $\tau$).}
    \label{fig:nodal_struc_ill}
\end{figure}

Here we propose a two-component model in which nodal operators emerge naturally in all dimensions.
Consider an electronic system conceptually divided into two subsystems: (1) itinerant free $c$-electrons and (2) localized $f$-electrons subject to disorder or interaction, as shown in Fig.~\ref{fig:nodal_struc_ill}(a). The total Hamiltonian of such a system is composed of three parts:
\begin{align}
    \hat H = \hat H_c + \hat H_f + \hat H_{\text{hyb}}. \label{eq:two_component}
\end{align}
$\hat H_c$, acting on the $c$-electrons, is a general multi-band non-interacting Hamiltonian. In the Wannier basis, the creation operator $\hat c^\dagger_{\mathbf{R},a}$ creates a $c$-electron in a state centered at the unit cell $\mathbf{R}$ with internal degrees of freedom (e.g., sublattice, orbital, spin) labeled by $a$. Its momentum-space representation is 
\begin{equation}
    \hat H_c = \sum_{\mathbf{k},a,b} \hat c^\dagger_{\mathbf{k},a} \mathcal{H}^{(c)}_{ab}(\mathbf{k}) \hat c_{\mathbf{k},b},
\end{equation}
where $\mathcal{H}^{(c)}(\mathbf{k})$ is the Bloch Hamiltonian matrix and $\hat c_{\mathbf{k},a}^\dagger = \frac{1}{\sqrt{N}}\sum_{\mathbf{R}} e^{i \mathbf{k}\cdot\mathbf{R}} \hat c_{\mathbf{R},a}^\dagger$. Upon diagonalizing this Hamiltonian, we obtain the band basis, where a Bloch state $\ket{\psi_{\mathbf{k},m}}$ with energy $E_m(\mathbf{k})$ is created by $\hat \psi_{\mathbf{k},m}^\dagger = \sum_a \psi^{(m)}_{\mathbf{k},a} \hat c_{\mathbf{k},a}^\dagger$, where $\psi^{(m)}_{\mathbf{k},a}$ are the components of the corresponding eigenvector of the Bloch Hamiltonian $\mathcal{H}^{(c)}(\mathbf{k})$. In the absence of other terms, the $c$-Bloch electron created by $\hat \psi^\dagger_{\mathbf{k},m}$ is a ballistic quasiparticle with velocity $\frac{1}{\hbar}\nabla_{\mathbf{k}}E_{\mathbf{k},m}$\cite{ashcroft2011solid}, and the system will exhibit ballistic transport supported by these Bloch electrons. 

The Hamiltonian for the $f$-electron, $\hat H_f$, which is the source of scattering, can be quite general. The only requirement is that $\hat H_f$ does not support ballistic quasiparticles; otherwise, the total system would exhibit ballistic transport.
For instance, $\hat H_f$ could consist of disordered impurities, the Hamiltonian of which is  represented by $\hat H_f = \sum_{\mathbf{R}\sigma} \epsilon_{\mathbf{R}} \hat f^\dagger_{\mathbf{R}\sigma} \hat f_{\mathbf{R}\sigma}$, where $\epsilon_{\mathbf{R}}$ are random on-site energies. Alternatively, the $f$-electron subsystem could be a flat band with Hubbard interaction, given by $\hat H_f= \sum_{\mathbf{R}\sigma} \epsilon_{f} \hat f^\dagger_{\mathbf{R}\sigma} \hat f_{\mathbf{R}\sigma}+U\sum_{\mathbf{R}} \hat n^{(f)}_{\mathbf{R}\uparrow}\hat n^{(f)}_{\mathbf{R}\downarrow}$, where $\epsilon_f$ is the flat band energy, $U$ is the interaction strength, $\hat n_{\mathbf{R}\sigma}^{(f)} = \hat f^\dagger_{\mathbf{R}\sigma} \hat f_{\mathbf{R}\sigma}$, and $\sigma = \uparrow/\downarrow$ denote the spin orientation. This resembles the famous periodic Anderson model\cite{hewson1997kondo}, where the $f$-electrons also form a periodic lattice. The operator $\hat f_{\mathbf{R}\sigma}^\dagger$ creates a localized $f$-electron at a specific Wyckoff position within the unit cell $\mathbf{R}$ with spin $\sigma$. For simplicity, we often suppress the spin index $\sigma$ in our notation, reinstating it when necessary.

The crucial coupling between the localized $f$-electrons and the itinerant $c$-electrons is described by the hybridization term $\hat H_{\text{hyb}}$. It arises from the quantum mechanical wavefunction mixing due to the fact that neither the localized $f$-electron orbital, nor the $c$-electron's wavefunction are exact eigenstates of the full single-particle Hartree-Fock Hamiltonian,  $h_{sp}(\mathbf{r},-i\boldsymbol{\nabla})$.
For an $f$-electron at the origin, $\varphi^f(\mathbf{r})$, and a $c$-electron of orbital character `$a$' at neighboring site $\boldsymbol{\delta}$, $\varphi^c_a(\mathbf{r}-\boldsymbol{\delta})$, the hybridization amplitude $V_a(\boldsymbol{\delta})$ is given by\cite{hewson1997kondo}:
\begin{equation}
     V_a(\boldsymbol{\delta}) = \int d \mathbf{r} \varphi^{f*}(\mathbf{r}) h_{sp}(\mathbf{r},-i\boldsymbol{\nabla}) \varphi^c_a(\mathbf{r} - \boldsymbol{\delta}).
\end{equation}
And the Hamiltonian for the hybridization part $\hat H_{\text{hyb}}$ is given by 
\begin{equation}
    \hat H_{\text{hyb}} = \sum_{\mathbf{R}}\sum_{\boldsymbol{\delta},a} (V_{a}(\boldsymbol{\delta})\hat f_{\mathbf{R}}^\dagger \hat c_{\mathbf{R}+\boldsymbol \delta , a} + h.c. ).
\end{equation}
We define the hybridization strength $V  = \sqrt{\sum_{\boldsymbol\delta,a} |V_{a}(\boldsymbol \delta)|^2}$ and normalized coefficients $\phi_{a}(\boldsymbol \delta) = {V_{a}(\boldsymbol{\delta})} / {V}$. This allows us to write the hybridization in a compact form, $\hat H_{\text{hyb}} = V\sum_{\mathbf{R}} \hat f_{\mathbf{R}}^\dagger \hat d_{\mathbf{R}} + h.c.$, where  $ \hat d_{\mathbf{R}} = 
\sum_{\boldsymbol \delta,a} \phi_{a}(\boldsymbol \delta) \hat c_{\mathbf{R}+\boldsymbol{\delta},a}$ defines a composite $c$-electron operator. 

To analyze transport, it is more convenient to rewrite $\hat H_{\text{hyb}}$ in the $c$-electron band basis. In this basis, $\hat H_{\text{hyb}}$ takes the form:
\begin{equation}
     \hat H_{\text{hyb}} = \sum_{\mathbf{R}} \sum_{\mathbf{k},n}  
e^{i \mathbf{k}\cdot \mathbf{R}} \mathcal{V}_{m}(\mathbf{k})    \hat f^\dagger_{\mathbf{R}}  \hat \psi_{\mathbf{k},m} + h.c. ,
\end{equation}
where the effective hybridization amplitude $\mathcal{V}_m(\mathbf{k})$ for band $m$ is given by the overlap between the Bloch state eigenvector $\boldsymbol{\psi}^{(m)}(\mathbf{k})$ and the hybridization form factor $\tilde{\boldsymbol{\phi}}(\mathbf{k})$:
\begin{equation}
    \mathcal{V}_{m}(\mathbf{k}) = V \sum_a \left(\sum_{\boldsymbol{\delta}} \phi_{a}(\boldsymbol{\delta}) e^{i\mathbf{k}\cdot\boldsymbol{\delta}}\right) \psi^{(m)}_{\mathbf{k},a} \equiv V \tilde{\boldsymbol{\phi}}(\mathbf{k}) \cdot \boldsymbol{\psi}^{(m)}(\mathbf{k}).
\end{equation}
Physically, this amplitude is equivalent to the wavefunction overlap $\braket{d|\psi_{\mathbf{k},m}}$, where $\ket{d} = \hat d_{\mathbf{R}=0}^\dagger \ket{0}$, and below we will refer to the state $\ket{d}$ as the ``hybridization orbital". 

In this model, the number operator of ballistic quasiparticle is $\hat n_{\mathbf{k},m} = \hat \psi^\dagger_{\mathbf{k},m} \hat \psi_{\mathbf{k},m}$. Since $c$-electron operators commute with $f$-electron operators, the 
non-conservation of the ballistic quasiparticle number, governed by $[\hat n _{\mathbf{k},m}, \hat H]$, 
arises solely from the hybridization, irrespective of the specific form of $\hat H_f$. 
Consequently, at ``nodes" $\mathbf{k}_0$ where the hybridization amplitude vanishes ($\mathcal{V}_m(\mathbf{k}) = 0$), the ballistic quasiparticle number is conserved: $[\hat n_{\mathbf{k}_0,m}, \hat H]=0$. These possible nodes constitute an emergent nodal structure illustrated in Fig.~\ref{fig:nodal_struc_ill}(b), formally equivalent to the one constructed artificially in prior work, but are protected by, to be shown in Sec.~\ref{sec:symmetry_superdiff} and Sec.~\ref{sec:topology_superdiff}, the intrinsic symmetries and/or band topology of the $c$-electrons. 

\section{Scaling of conductance}
\label{sec:transport analysis}
\subsection{Derivation of the scaling exponent}
Before illustrating how the nodal structures in the two-component model can be protected by symmetry and topology, we first investigate how they influence the 
exponent $\gamma$ in the scaling relation between the direct-current (dc) conductance and system size,  $G \sim L^{-\gamma}$. Our derivation proceeds
in three steps: first, we express the conductivity using the Kubo formula in terms of the $c$-electron's Green's function. 
Second, we give the self-energy correction to the Green's function arising from the $f$-electron scattering bath. 
Finally, we analyze the resulting conductivity integral to reveal its scaling with system size.

The dc longitudinal conductivity $\sigma_{\mu\mu}^{(\omega=0)}$ can be calculated using the Kubo formula\cite{altland2010condensed,akkermans2007mesoscopic}, where $\mu$ denotes the spatial direction. At finite temperature, $\sigma_{\mu\mu}^{(\omega=0)}$ is given by
\begin{equation}
    \sigma_{\mu\mu}^{(\omega=0)} = \int d \epsilon \sigma_{\mu\mu}(\epsilon)(-\frac{\partial n_F(\epsilon,T)}{\partial \epsilon}),
\end{equation}
where $n_F(\epsilon,T)$ is the Fermi-Dirac distribution function, and
\begin{equation}
    \sigma_{\mu\mu}(\epsilon) = \frac{\hbar}{\pi} \int \frac{d\mathbf{k}}{(2\pi)^D}\operatorname{Tr} \left[{j}_\mu\operatorname{Im} {G}^R_c(\mathbf{k},\epsilon)\right]^2,
    \label{eq:kubo_sigma_epsilon}
\end{equation}
is the energy-resolved conductivity in the Drude approximation.
Here ${G}^{R/A}_c(\mathbf{k},\epsilon)$ are the full retarded/advanced Green's function operators, $\operatorname{Im} G^{R}_c = \frac{1}{2i}(G^R_c - G^A_c) $ denotes the anti-Hermitian part of the full retarded Green's function\cite{PhysRevB.64.165303}, and ${j}_\mu = \frac{e}{\hbar}\frac{\partial {H}^{(c)}}{\partial k_\mu}$ is the current operator\cite{bernevig2013topological}. The Green's function in the momentum-basis is a matrix $G^R_c(\mathbf{k},\epsilon) = [\epsilon - \mathcal{H}^{(c)}(\mathbf{k}) - \Sigma^R_c(\mathbf{k},\epsilon)]^{-1}$, which is dressed by the self-energy $\Sigma^R_c$ due to the hybridization term. In the weak scattering limit, the nonsingular $G^R G^R$ and $G^A G^A$ pairs will drop out, leaving the standard $G^RG ^A$ pair used in our calculation\cite{akkermans2007mesoscopic}.

The central object is the $c$-electron self-energy, $\Sigma^R_c$, which arises from hybridizing with the $f$-electrons. For a wide range of physical scenarios, including $f$-electrons as disorder or as an interacting flat band at low temperatures (i.e., in the Fermi liquid regime), the self-energy can be shown to take a factorized form (see Appendix~\ref{app:self_energy}):
\begin{equation}
    [\Sigma_c^R(\mathbf{k},\epsilon)]_{mm'} = \mathcal{V}^*_m(\mathbf{k}) g_{f}^R(\epsilon) \mathcal{V}_{m'}(\mathbf{k}). \label{eq:self_energy_general}
\end{equation}    
Here, $\mathcal{V}_m(\mathbf{k}) = V \braket{d|\psi_{\mathbf{k},m}}$ is the hybridization amplitude, and $g^R_f(\epsilon)$ is the $f$-electron's Green's function,
whose detailed form depends on the nature of the localized $f$-electrons. 

The structure of Eq.~(\ref{eq:self_energy_general}) reveals the following observation: the momentum dependence of the scattering is entirely encoded in the hybridization
vertices $\mathcal{V}_m(\mathbf{k})$, while the nature of the scattering bath is encapsulated in the scalar propagator $g_f^R(\epsilon)$. The existence, location, and order of the hybridization nodes are universal properties of the $c$-electrons and hybridization, completely independent of the specifics of $\hat H_f$. 

Now we are ready to evaluate the conductivity by Eq.~(\ref{eq:kubo_sigma_epsilon}). The conductivity integral is determined by the poles of the $c$-electron Green's function, which correspond to the complex eigenvalues of an effective non-Hermitian Hamiltonian:
\begin{equation}
    \mathcal{H}^{\text{eff}} = \mathcal{H}^{(c)}(\mathbf{k}) + \Sigma_c^R(\mathbf{k},\epsilon).
\end{equation}
To elucidate the essential physics of the nodal structure, we project this effective Hamiltonian onto a single, isolated band $m$ that crosses the energy level $\epsilon$. Now the $c$-electron Green's function is a scalar: $G^R_{c,m}(\mathbf{k},\epsilon) = [\epsilon - E_m(\mathbf{k}) -\Sigma^R_{c,m}(\mathbf{k},\epsilon)]^{-1}$. Due to Eq.~(\ref{eq:self_energy_general}), $\Sigma^R_{c,m}(\mathbf{k},\epsilon)$ is nothing but $[\Sigma^R_c(\mathbf{k},\epsilon)]_{mm}$ Then we expand the Green's function around the quasiparticle pole\cite{bruus2004many} $\tilde E({\mathbf{k}}) - i \Gamma({\mathbf{k}})$ to first order:
\begin{equation}
    G^R_{c,m}(\mathbf{k},\epsilon) \approx \dfrac{Z_{\mathbf{k}}}{\epsilon - \tilde E({\mathbf{k}}) + i 
     \Gamma(\mathbf{k})} , \label{eq:expand_around_pole}
\end{equation}
where the renormalized band energy, $\tilde E({\mathbf{k}})$, is determined by $\tilde E({\mathbf{k}}) - \text{Re}[\Sigma^R_{c,m}(\mathbf{k},\tilde E({\mathbf{k}}))] =0$, the quasiparticle weight, $Z_{\mathbf{k}} = (1- \partial_\epsilon \Sigma^R_{c,m}(\mathbf{k},\epsilon)|_{\epsilon = \tilde E(\mathbf{k})})^{-1}$ and the inverse quasiparticle lifetime (scattering rate) $ \Gamma({\mathbf{k}}) = -Z_{\mathbf{k} } \text{Im}[\Sigma^R_{c,m}(\mathbf{k},\tilde E_\mathbf{k})]$. The crucial part comes from the scattering rate, which is given by $\Gamma({\mathbf{k}}) \propto |\mathcal{V}_m(\mathbf{k})|^2$. 
If $\mathcal{V}_m(\mathbf{k}_0 ) = 0$ for some $\mathbf{k}_0$, the scattering rate for this Bloch state is zero, signifying infinite lifetime.
Therefore, we define the nodal manifold as ${S}^{(m)}_{\text{node}}=\{\mathbf{k}_0 \in \text{BZ} \mid  \left|\mathcal{V}_m(\mathbf{k}_0)\right|=0\}$ for any $m$.

When nodes appear, the integral in Eq.~(\ref{eq:kubo_sigma_epsilon}) is dominated by regions in momentum space where the quasiparticle lifetime diverges. This occurs at the intersection, $\mathbf{k}_0$, of the equal energy surface $(\epsilon - \tilde E(\mathbf{k}_0) =0) $ and the nodal manifold. To analyze the integral behavior near $\mathbf{k}_0$, we expand the Green's function denominator
\begin{equation}
    \epsilon - \tilde E(\mathbf{k}_0 + \delta \mathbf{k}) + i \Gamma({\mathbf{k}_0 + \delta \mathbf{k}} )\approx - v(\mathbf{k}_0) \delta k_\parallel + i\Gamma |\delta\mathbf{k}_\perp|^{2n}. \label{eq:denominator_expansion}
\end{equation}
Here, $\delta k_\parallel$ is the momentum deviation along the velocity $\mathbf{v}(\mathbf{k}_0)$, $\delta \mathbf{k}_\perp$ is the momentum deviation away from the nodal manifold and $n$ is the order of the node. The singular region is anisotropic: the condition \( |v_\parallel(\mathbf{k}_0)\,\delta k_\parallel| \sim \Gamma\, |\delta \mathbf{k}_\perp|^{2n} \to 0\) controls the integrand.

In the thermodynamic limit, the node-induced contribution to the integral diverges. For a finite system with length $L_\parallel$ and transverse sizes $L_\perp$, infrared cutoffs ($|\delta k_i| \gtrsim 1/L_i$) regularize the integral. 
The anisotropic singularity mandates anisotropic regularization: we first take the transverse directions to the thermodynamic limit (\(L_\perp \to \infty\)) to form a continuum near the node, and then use \(L_\parallel\) to cut off the remaining divergence along the velocity direction, \(|\delta k_\parallel| \gtrsim 1/L_\parallel\)~\cite{gattenlohner2016levy}. The resulting scaling reads
\begin{equation}
    \sigma_{\parallel}(\epsilon,L_\parallel) \sim L_\parallel^{1-\gamma(\epsilon)},
    \qquad 
    \gamma(\epsilon) = \min\!\left\{\frac{D^{S(\epsilon)} - D^{S(\epsilon)}_{\text{node}}}{2n},\,1\right\},
    \label{eq:exponent_general}
\end{equation}
Here, $D^{S(\epsilon)}$ is the dimension of the equal energy surface of $\epsilon$, and $D^{S(\epsilon)}_{\text{node}}$ is its dimension of its intersection with the nodal manifold ${S}^{(m)}_{\text{node}}$. Notice that the final result is independent of the transverse dimensions $L_\perp$, consistent with the anisotropic nature of the singularity.

From Eq.~(\ref{eq:exponent_general}), we can determine the scaling of the conductance. At zero temperature, the charge transport is dominated by the states at the Fermi surface ($S(\epsilon_F) =F$). The conductance $G_\parallel$ with fixed transverse cross section $S_\perp$ is related to conductivity by $G_\parallel = \sigma_\parallel S_\perp /L_\parallel$. This gives the scaling relation:
\begin{equation}
    G_\parallel \sim L_\parallel^{-\gamma_{\text{low}}},
    \qquad
    \gamma_{\text{low}} = \min\!\left\{\frac{D^F-D^F_{\text{node}}}{2n},\,1\right\},
    \label{eq:exponent_low_temperature}
\end{equation}
which holds when the Fermi surface and the nodal manifold has non-zero intersection.
At high temperatures, the thermal smearing $(-\frac{\partial n_F(\epsilon)}{\partial\epsilon})$ makes states throughout the Brillouin zone relevant. Now the exponent is determined by the full dimensions of the Brillouin zone $(D)$ and the nodal manifold $(D_{\text{node}})$:
\begin{equation}
    G_\parallel \sim L_\parallel^{-\gamma_{\text{high}}},
    \qquad
    \gamma_{\text{high}} =  \min\!\left\{\frac{D-D_{\text{node}}}{2n},\,1\right\}.
    \label{eq:exponent_high_temperature}
\end{equation}
We see the exponent $\gamma$ is always determined by the codimension of the nodal manifold with respect to the manifold of the states contributing to the transport.

If $\gamma_{\text{high/low}}$ in Eq.~(\ref{eq:exponent_low_temperature},\ref{eq:exponent_high_temperature}) is less than 1, our theory predicts superdiffusion contributed by the $c$-electrons on or near the $S_{\text{node}}^{(m)}$. For example, in a 3D system, if the nodal manifold is 1D (nodal line) and the Fermi surface 2D, we have $D^F = 2, D^F_{\text{node}}=0$, if the nodal line crosses the Fermi surface. In this case, we have $\gamma_{\text{high}} = \gamma_{\text{low}} = 1$ (diffusion) for $n=1$; $\gamma_{\text{high}} = \gamma_{\text{low}}  =1/2$ for $n=2$ (superdiffusion).

\subsection{Anisotropic scaling and boundary conditions}
\begin{figure}[t]
    \includegraphics[width=0.8\linewidth]{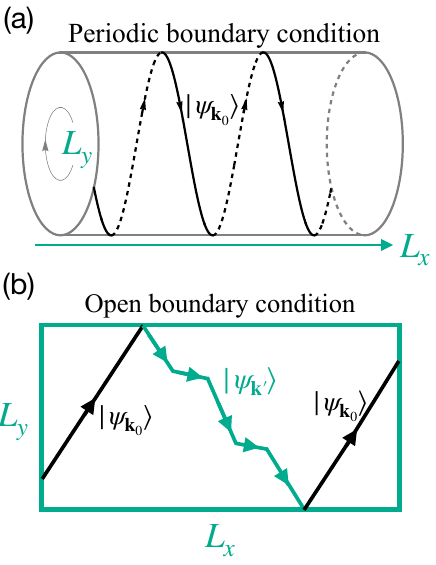}
    \caption{\textbf{Motion of nodal quasiparticle under different transverse boundary condtions}
    (a) \textbf{Periodic boundary condition (PBC) along $y$}: transverse momentum is conserved, and a quasiparticle in a protected nodal state \(\ket{\psi_{\mathbf{k}_0}}\) propagates ballistically, wrapping around the cylinder without changing its transverse momentum, allowing superdiffusive behavior to manifest.
    (b) \textbf{Open boundary condition (OBC) along $y$}: reflection at the boundaries does not conse
    rve transverse momentum, causing boundary-induced channel mixing, scattering a protected nodal quasiparticle $\ket{\psi_{\mathbf{k}_0}}$ to an unprotected state $\ket{\psi_{\mathbf{k}'}}$ with finite mean free path. Superdiffusive transport is therefore more robust in systems with $L_y \gg L_x$ where bulk properties dominate.
    }
    \label{fig:boundary_condition}
\end{figure}
The anisotropic singularity in Eq.~\eqref{eq:denominator_expansion} mandates a careful treatment of geometry. Unlike isotropic scaling where all dimensions are scaled equally\cite{Abrahams1979Scaling}, here transport is governed by the anisotropic scaling in Eq.~\eqref{eq:exponent_general}. To extract the scaling exponent \(\gamma\), we take the transverse dimensions \(L_\perp\) to the thermodynamic limit first (\(L_\perp\!\to\!\infty\)) to form a continuum near the nodal manifold and vary the longitudinal size \(L_\parallel\).

For simulations of finite lattices, boundary conditions in the transverse directions are also crucial (see Fig.~\ref{fig:boundary_condition}). The bulk Kubo calculation implicitly assumes periodic boundary condition (PBC), which preserves transverse momentum as a good quantum number; under PBC, a protected nodal channel is free from boundary scattering and remains protected. In contrast, open boundary condition (OBC)---as in realistic devices---can induce mixing between momentum channels near and far from the nodal structure due to reflection at physical edges. Due to this boundary-induced channel mixing, superdiffusion can be destroyed even in the presence of nodes. 

Therefore, to provide the verification of the bulk theory, all numerical simulations in this work employ PBC in the transverse directions. We further predict that superdiffusive signatures in experiments with OBC will be most robust for large aspect ratios \(L_\perp \gg L_\parallel\), where bulk properties dominate over boundary effects.

\subsection{Weak localization correction}
An important question is whether our prediction of superdiffusive transport—derived from a semiclassical Drude picture—remains robust against quantum interference. In disordered systems, such effects have been known to produce weak-localization (WL) corrections to the conductivity\cite{bruus2004many,RevModPhys.57.287}.
To assess their relevance, we adapt the general approach in \cite{gattenlohner2016levy,Wölfle1984interference} to calculate the leading-order WL correction from the maximally crossed diagrams within our model.
We find that the correction, which is also anisotropic in space, is independent of the transverse sizes \(L_\perp\) and scales with longitudinal size \(L_\parallel\) as (see Appendix~\ref{app:WL correction} for details)
\begin{equation}
    \delta  \sigma^{\text{WL}}_{\parallel} \propto -L_\parallel^{1-\frac{(\gamma+1)(D-1)}{2}}.
    \label{eq:weak_localization}
\end{equation}
The anisotropy of \(\delta \sigma^{\text{WL}}_\parallel\) follows from the anisotropy of the semiclassical transport. Comparing with the Drude conductivity \(\sigma_\parallel \sim L_\parallel^{1-\gamma}\), the ratio behaves as
\begin{equation}
    \left|\frac{\delta\sigma^{\text{WL}}_\parallel}{\sigma_\parallel}\right| \propto L_\parallel^{-\frac{(\gamma+1)(D-1)}{2} + \gamma} \;\;\xrightarrow[L_\parallel\to\infty]{D=2,3;0< \gamma<1}\; 0,
\end{equation}
provided the exponent is negative. For physical dimensions \(D=2,3\), this condition is satisfied throughout the superdiffusive regime \(0<\gamma<1\). We conclude that WL corrections are subleading; the anomalous scaling of the Drude conductivity is the defining feature of transport.

\section{Symmetry protected superdiffusion}
\label{sec:symmetry_superdiff}
In this section we show how the symmetry protects or enforces the nodal structures in the two-component model and lead to robust superdiffusion.
We begin by considering a system whose Hamiltonian is invariant under a space group $\mathcal{G}$. Each symmetry element $g \in \mathcal{G}$ is represented by a unitary operator that acts on the Hilbert space, such that $[\hat g,\hat H]=0$. Under a symmetry $\hat g$, the $c$-electron creation operators transform according to a unitary representation $D^{(c)}(g)$\cite{chiu2016classification}: 
\begin{equation}
    \hat g \hat c_{\mathbf{k},a}^\dagger \hat g^{-1} = \sum_{b} \hat c_{g\mathbf{k},b}^\dagger [D^{(c)}_{g\mathbf{k}}(g)]_{b,a}.
\end{equation}
Here, $g\mathbf{k}$ denotes the action of the point-group component of the symmetry element $g$ (a rotation or reflection matrix) on the momentum $\mathbf{k}$.

The $f$-electrons are assumed to occupy a specific Wyckoff position $\mathbf{r}_f$, and their on-site orbital is non-degenerate. This non-degeneracy requires that the $f$-electron state transforms as a one-dimensional irreducible representations (1D irrep) of its site-symmetry group, $\mathcal{G}(\mathbf{r}_f) \subset \mathcal{G}$\cite{po2020symmetry}. For any operation $h \in \mathcal{G}(\mathbf{r}_f)$ that leaves the site invariant, the transformation is:
\begin{equation}
    \hat h \hat f_{\mathbf{R}}^\dagger \hat h^{-1} = \chi_f(h) \hat f_{\mathbf{R}}^\dagger,
\end{equation}
where $\chi_f(h)$ is the character of the 1D irrep.

The invariance of the hybridization term, $[\hat h, \hat H_{\text{hyb}}] =0$, imposes a strict constraint on the transformation properties of the hybridization orbital $\ket{d_{\mathbf{R}}} $. Specifically, for $h \in \mathcal{G}(\mathbf{r}_f)$, $\ket{d_{\mathbf{R}}}$ must transform according to the same 1D irrep as $\ket{f_{\mathbf{R}}}$ (see Appendix~\ref{sec:irrep} for proof). This leads to the symmetry selection rule: hybridization between $c$- and $f$-electrons, given by the overlap $\braket{d|\psi_{\mathbf{k},m}}$, vanishes if the $c$-electron Bloch state $\ket{\psi_{\mathbf{k},m}}$ and the hybridization orbital $\ket{d}$ transform under different irreps of the site symmetry group. 

This can be seen by considering a high-symmetry momentum point $\mathbf{k}_h$ that is invariant under an operation $h \in \mathcal{G}(\mathbf{r}_f)$. If the c-Bloch state at this point is also non-degenerate, it transforms as $\hat h \ket{\psi_{\mathbf{k}_h,m}} = \chi^m_{c}(h) \ket{\psi_{\mathbf{k}_h,m}}$. A non-zero overlap requires $\chi_f(h) = \chi_c^{m}(h)$. If there is a symmetry mismatch, i.e., $\chi_f(h) \neq \chi^m_c(h)$, the overlap must vanish by orthogonality, forming nodes in the hybridization. For degenerate Bloch states, this argument generalizes via Schur's lemma\cite{dresselhaus2007group}: if the 1D irrep of the f-orbital is not contained within the representation of the $c$-electrons at $\mathbf{k}_h$, the hybridization is forbidden by symmetry mismatch and must be zero. Nodes from symmetry mismatch usually appear at high-symmetry points, lines, or planes in the Brillouin zone.

\begin{figure*}[t]
    \includegraphics[width=1.8\columnwidth]{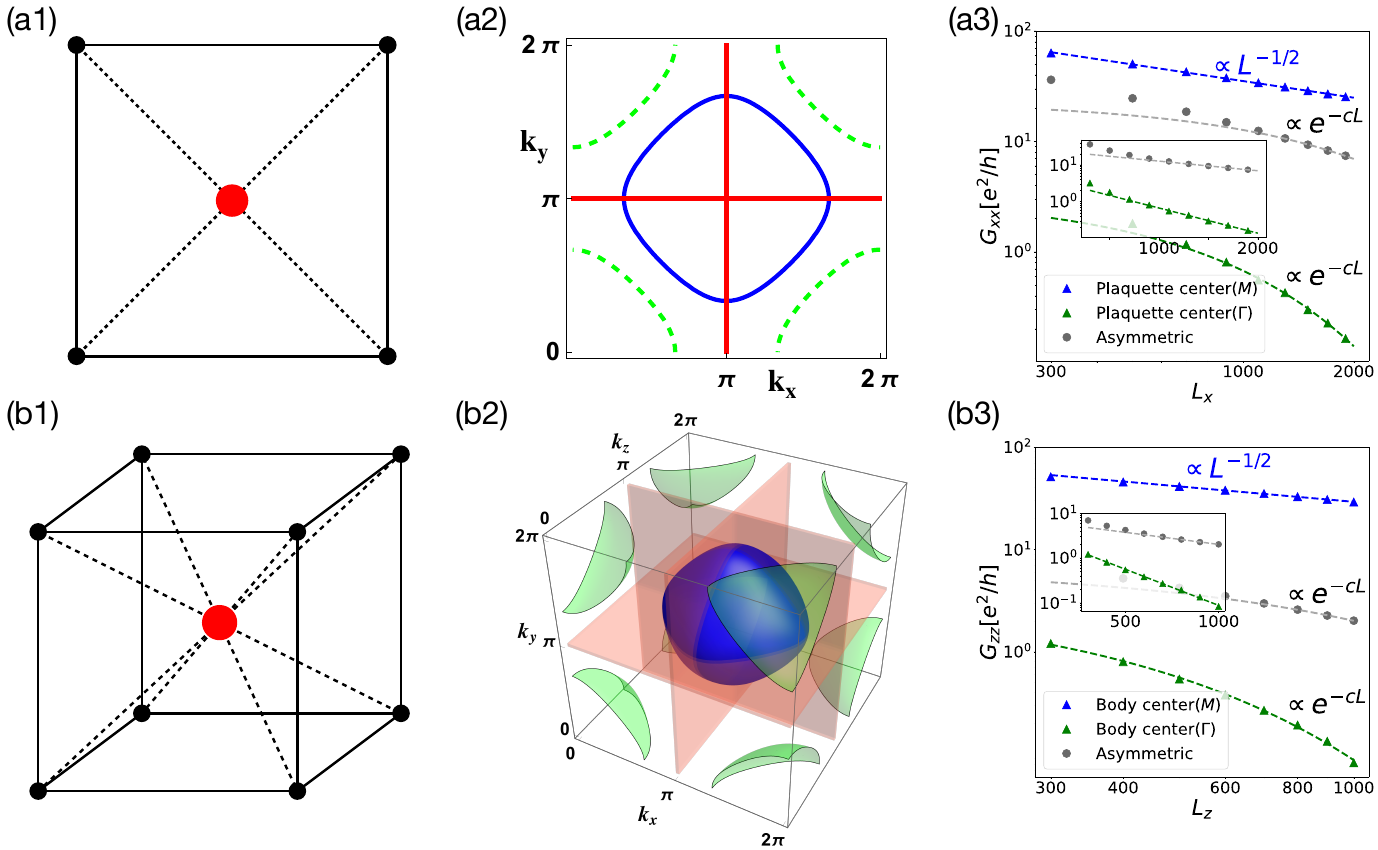}
    \caption{\textbf{Symmetry protected nodal structures and their transport signatures in 2D and 3D models.} This figure provides an overview of the symmetry selection rule, from the real-space geometry to the momentum-space nodal structure and its direct consequences for zero-temperature transport.
    \textbf{(a) Two-Dimensional Square Lattice} 
    (a1) The real-space geometry: an $f$-electron (red) at the plaquette center, hybridizing with its four neighboring $c$-electron (black). 
    (a2) The resulting momentum-space structure. The hybridization vanishes on a nodal manifold (red lines) at the Brillouin zone boundaries. The panel shows two representative Fermi surfaces. The $M$-pocket Fermi surface (blue solid line) with energy $E_F = 1$ centered at the $(\pi, \pi)$ point and intersects with nodal manifold and the $\Gamma$-pocket Fermi surface (green dash line) with energy $E_F = -1$ centered at $(0,0)$ does not.
    (a3) Numerical simulation of the transport. 
    The conductance $G_{xx}$ is calculated for the two Fermi energies shown in (a2). For energy $E_F = 1$ corresponding to the $M$-pocket Fermi surface intersecting with nodes (blue triangles), the transport is superdiffusive with $G \sim L_x^{-1/2} (\gamma = 1/2)$. For energy $E_F = -1$ corresponding to the non-intersecting $\Gamma$-pocket (green triangles), the conductance shows exponential decay and indicates localization. The case with hybridization orbital $\ket{d}$ breaking the mirror symmetry with Fermi energy $E_F = 1$ (gray circles) is also localization. The inset highlights the exponential decay on a semi-log plot.
    \textbf{(b) Three-Dimensional Simple Cubic Lattice} 
    (b1) The real-space model: an $f$-electron at the body center hybridizing with its eight neighboring $c$-electrons.
    (b2) The corresponding nodal structure, consisting of nodal planes (semi-transparent red) at Brillouin zone boundaries. Two possible Fermi surfaces are shown: the $M$-pocket Fermi surface centered at $(\pi,\pi,\pi)$ with $E_F = 3$ (blue) and the $\Gamma$-pocket Fermi surface centered at the origin with $E_F = -3 $ (green).
    (b3) Numerical simulation on the 3D model. The conductance $G_{zz}$ is computed at the Fermi energies corresponding to the surfaces in (b2). Superdiffusion with $\gamma = 1/2$ (blue triangles) is only observed with the Fermi surface and the nodal planes intersected. The cases where the intersection does not appear (green triangles) or when the $f$-orbitals are not mirror symmetric show exponential decay.
    The numerical result perfectly match the low-temperature prediction $\gamma_{\text{low}} = (D^F-D^F_{\text{node}})/(2n) = 1/2$ for the 2D and the 3D system.
    The transverse size of the system is fixed when doing the numerical simulation and $L_y =202$ for the 2D square lattice and $L_x = L_y = 21$ for the simple cubic lattice system. 
    Results are averaged over at least 700 disorder configurations for the 2D model and at least 300 disorder configurations for the 3D model. The $f$-electron energies $\epsilon_{\mathbf{R}}$ are drawn form a uniform distribution in the range $[-W,W]$, where $W_{2D} = 0.8$ and $W_{3D} = 2.0$ (for the $\Gamma$-pocket Fermi surface, $W$ is taken as $0.5$ to enhance the readibility of the data). The hybridization strength is set to be $V_{2D} =1$ and $V_{3D} = \sqrt{2}$.
    }
    \label{fig:symmetry}
\end{figure*}

To demonstrate the power of this principle, we first analyze a 2D square lattice model. The $c$-electrons are described by a single band tight-binding Hamiltonian with nearest neighbor hopping, $\hat H_c = -t \sum_{\mathbf{R},\mathbf{R}'} \hat c^\dagger_{\mathbf{R}} \hat c_{\mathbf{R}'} + h.c.$. The $f$-electrons, assumed to be of $s$-orbital character, reside on the plaquette centers, as shown in Fig.~\ref{fig:symmetry}(a).  
Let us focus on the hybridization amplitude $\braket{d|\psi_{\mathbf{k}}}$. Its site-symmetry group, $\mathcal{G}(\mathbf{r}_f)$, includes mirror reflections, $\mathcal{M}_{x/y}$, whose reflection planes pass through $\mathbf{r}_f$: the plaquette center in the unit cell at the origin. Without loss of generality, we take the lattice spacing as identity. The $s$-orbital $f$-state transforms trivially, with character $\chi_f(\mathcal{M}_{x/y}) = +1$. In contrast, a $c$-electron Bloch state $\ket{\psi_{\mathbf{k}}}$ transforms as $\hat {\mathcal{M}}_x \ket{\psi_{\mathbf{k}}} = e^{i k_y} \ket{\psi_{\mathcal{M}_x\mathbf{k}}}$ and $\hat {\mathcal{M}}_y \ket{\psi_{\mathbf{k}}} = e^{i k_x} \ket{\psi_{\mathcal{M}_y\mathbf{k}}}$. At the mirror invariant lines, $k_x = \pm \pi$ and $k_y=\pm\pi$, the c-Bloch state is an eigenstate of $\hat {\mathcal{M}}_{x/y}$ with eigenvalue $\chi_c(\mathcal{M}_{x/y}) = e^{i\pi} = -1$. This clear symmetry mismatch ($\chi_f \neq \chi_c$) dictates that the hybridization $\braket{d|\psi_{\mathbf{k}}}$ must vanish along the entire Brillouin zone boundary, forming a 1D nodal manifold which is illustrated by Fig.~\ref{fig:symmetry}(a2).

The symmetry-based conclusion can be directly verified by computing the hybridization form factor, $\tilde \phi(\mathbf{k})$. The hybridization couples each $f$-electron to the four $c$-electrons at the corner of it plaquette, defining the hybridization orbital $\ket{d_{\mathbf{R}}} = \frac{1}{2}(\ket{c_\mathbf{R}} + \ket{c_{\mathbf{R}+\mathbf{e}_x}}+\ket{c_{\mathbf{R}+\mathbf{e}_y}}+\ket{c_{\mathbf{R}+\mathbf{e}_x+\mathbf{e}_y}})$. The form factor is the Fourier transformation of these coefficients:
\begin{equation}
    \tilde \phi(\mathbf{k}) \propto (1+e^{ik_x})(1+e^{i k_y}).
\end{equation}
This expression directly shows the nodes lie on the Brillouin zone boundaries ($k_{x/y} = \pi$). Since near a nodal line (e.g. $k_x = \pi$), the amplitude vanishes linearly with the momentum deviation $\delta k_x = \pi - k_x$, establishing the order of the nodes as $n=1$.

According to our theory, this nodal structure can lead to superdiffusive transport. At high temperatures, the scaling exponent is predicted to be $\gamma_{\text{high}} = (D-D_{\text{node}})/(2n) = (2-1)/(2) =1/2$. At zero temperature, assuming a 1D Fermi surface that intersects this nodal manifold at discrete points $D^F_{\text{node}} =0$, the exponent is given by $\gamma_{\text{low}} = (D^F-D^F_{\text{node}})/(2n) = (1-0)/(2) = 1/2$. Both regimes indicate superdiffusion, and the low temperature transport is validated by the numerical simulation presented in Fig.~\ref{fig:symmetry}(a2) .

This mechanism readily generalizes to 3D. Consider a simple cubic lattice for the $c$-electrons, with $f$-electrons placed at the body center of each unit cell. The site symmetry group now contains mirror planes like $\mathcal{M}_{xy}/\mathcal{M}_{xz}/\mathcal{M}_{yz}$. A similar analysis shows that a symmetry mismatch occurs on the boundaries of the Brillouin zone (e.g. at $k_{x/y/z} = \pi$), forming a 2D nodal manifold, illustrated in Fig.~\ref{fig:symmetry}(b1). This again predicts $\gamma_{\text{high}/\text{low}} = 1/2$, a result that aligns well with the numerical calculations in Fig.~\ref{fig:symmetry}(b2). 

This symmetry-protected nodal structure in our 3D model maps naturally onto real materials. 
Consider a crystal with the B2 (CsCl) structure, composed of two interpenetrating simple-cubic sublattices with atoms on the cube corners (A sublattice) and at the body center (B sublattice). 
Introducing substitutional disorder on the B sublattice—e.g., replacing a fraction of B atoms with a different element~\cite{anderson1999site}—creates a random array of \(f\)-electron–like scattering centers at the body-centered sites. 
NiAl-based B2 alloys provide an experimentally accessible platform to search for the predicted superdiffusive transport \((\gamma=1/2)\) associated with symmetry-protected nodal planes.

When the (i) $f$-electron is placed at the generic, low-symmetric Wyckoff position whose site symmetry group is trivial, so the nodal manifold is absent, or (ii) the $c$-electron Fermi surface does not intersect with the nodal manifold, the scattering rate from hybridization is non-zero on the Fermi surface. For the 2D system (see Fig.\ref{fig:symmetry}) (a3), this will lead to the Anderson localization\cite{Anderson58}. The 3D system in quasi-1D geometry is expected to show localized transport for lengths $L$ exceeding the localization length\cite{muller2010disorder} (see Fig.~\ref{fig:symmetry}(b3)).

\section{Topology protected superdiffusion}
\label{sec:topology_superdiff}
In addition to crystal symmetries, the nontrivial band topology of the $c$-electron Hamiltonian, $\hat H_c$, provides another mechanism for generating hybridization nodes. The principle relies on the topological obstruction: one cannot deform symmetric Wannier functions to any band that has nontrivial topological invariants\cite{bradlyn2017topological,po2017symmetry}.

We consider $c$-electron bands that feature protected topological structures, such as topological band point touching (Dirac/Weyl points). The defining characteristic of these structures is that the Bloch states $\ket{\psi_{\mathbf{k},m}}$ carry a nontrivial topological invariant when evaluated over a closed manifold $\mathcal{M}$ in momentum space that encloses the band touching (e.g., a loop for a 2D Dirac point or a sphere for a 3D Weyl point). The nontrivial invariant (the Berry phase, the winding number, or the Chern number) signifies a ``twist" in the phase of the Bloch wavefunction\cite{chiu2016classification}. For example, in a system with time-reversal symmetry like graphene, this manifests as an obstruction to defining a globally smooth and single-valued and real-valued gauge for $\ket{u_{\mathbf{k},m}}$, which is the cell-periodic part of $\ket{\psi_{\mathbf{k},m}}$, over $\mathcal{M}$\cite{bernevig2013topological}. In a system without time-reversal symmetry like the Weyl system, it is an obstruction to fixing a globally smooth phase for $\ket{\psi_{\mathbf{k},m}}$\cite{bernevig2013topological}.

In contrast, the hybridization orbital $\ket{d}$ is topologically trivial, defined as a linear combination of a finite number of local Wannier orbitals ($\hat{d}_{\mathbf{R}} = \sum_{\boldsymbol{\delta},a} \phi_{a}(\boldsymbol{\delta}) \hat c_{\mathbf{R}+\boldsymbol{\delta},a}$). Its momentum representation corresponds to the hybridization form factor $\tilde{ \boldsymbol{\phi}}(\mathbf{k})$. As such, $\tilde{\boldsymbol{\phi}}(\mathbf{k})$ is guaranteed to be an analytic and single-valued function across the entire Brillouin zone and carries a trivial topological charge on any closed manifold.

This inherent mismatch in topological character leads to the topological selection rule: The hybridization amplitude $\mathcal{V}_m(\mathbf{k}) = \braket{d|\psi_{\mathbf{k},m}}$ must vanish at one or more points $\mathbf{k}_0$ on any closed manifold $\mathcal{M}$ where $\ket{\psi_{\mathbf{k},m}}$ carries a nontrivial topological charge. The proof follows from contradiction: If the overlap $\mathcal{V}_m(\mathbf{k})$ were non-zero everywhere on the closed manifold $\mathcal{M}$, one could use its phase to ``untwist" the Bloch state by defining a new state\cite{thonhauser2006insulator,soluyanov2011wannier} 
\begin{equation}
    \ket{\chi_{\mathbf{k}}}= \frac{\mathcal{V}_m^*(\mathbf{k})}{|\mathcal{V}_m(\mathbf{k})|}  =\frac{\hat{P}_{\mathbf{k},m} \ket{d}}{\sqrt{\braket{d|\hat{P}_{\mathbf{k},m}|d}}},
\end{equation}
where we use the definition of hybridization amplitude $\mathcal{V}_m(\mathbf{k}) = V \braket{d|\psi_{\mathbf{k},m}}$ and $\hat{P}_{\mathbf{k},m} = \ket{\psi_{\mathbf{k},m}} \bra{\psi_{\mathbf{k},m}}$ is the corresponding projector. 
Since the old and new state are related by the gauge transformation which is smooth on the whole Brillouin zone, they must possess the same topological invariant.
However, the projector $\hat{P}_{\mathbf{k},m}$ and the hybridization orbital $\ket{d}$ are globally smooth and well-defined and so by construction, this state $\ket{\chi_{\mathbf{k}}}$ is smooth and single-valued over the manifold $\mathcal{M}$ (or can be made so, satisfying the reality condition for time-reversal symmetric system). A globally smooth and well-defined state must, by definition, have a trivial topological invariant. Therefore, the premise is false, and the overlap must vanish somewhere. The topology of the $c$-bands is thus directly imprinted onto the nodal structure of the hybridization. The following examples of graphene and multi-Weyl semimetals are concrete illustrations of this principle.

\subsection{Graphene}
\begin{figure*}[t]
    \includegraphics[width=2\columnwidth]{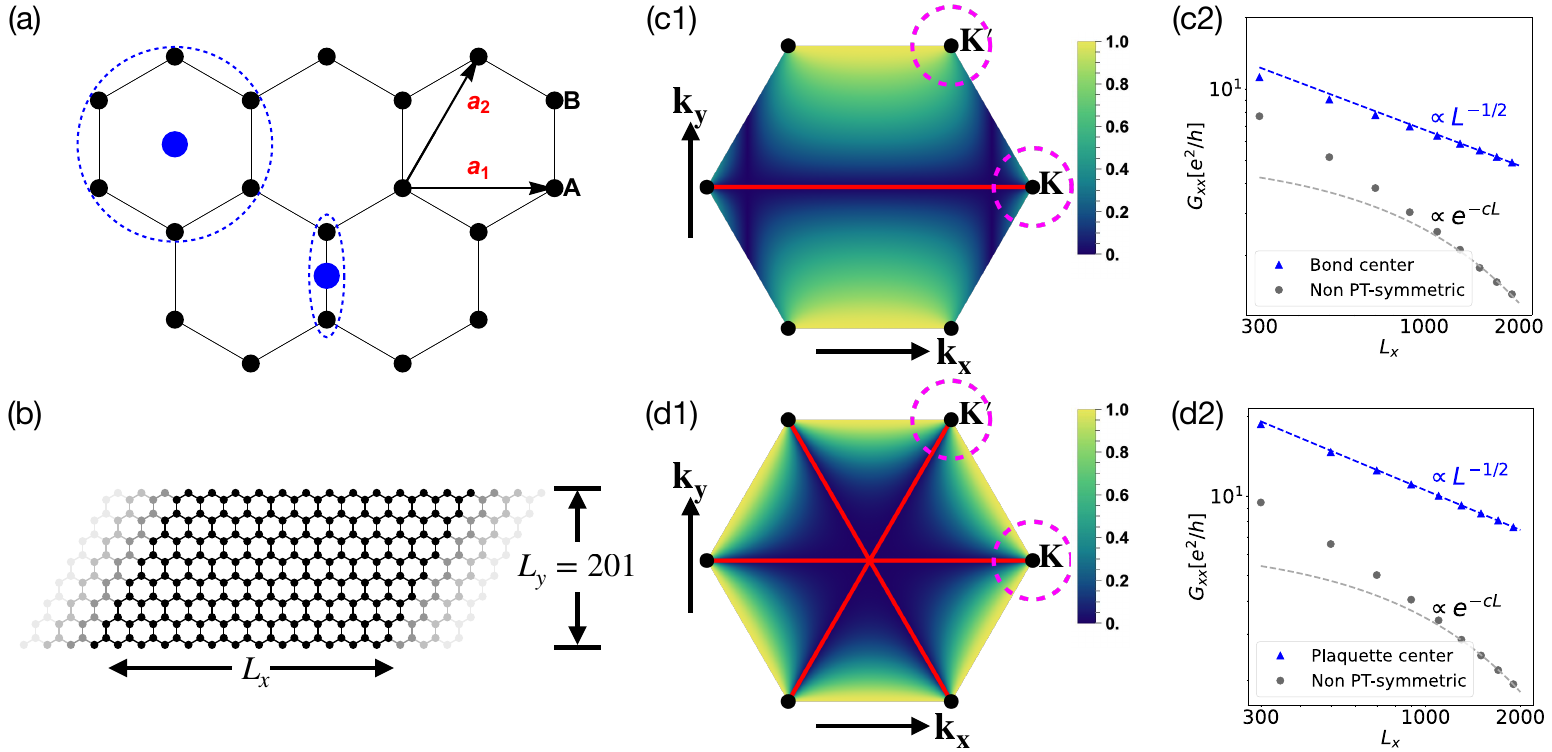}
    \caption{\textbf{Topologically protected nodal arcs and superdiffusive transport in graphene.} This figure provides an overview of the topological selection rule in graphene, from real-space geometries to momentum space nodal structures and their validation via zero-temperature transport simulations
    \textbf{(a)} Real-space illustration of two distinct, $PT-$symmetric $f$-electron placements on the honeycomb lattice: at the plaquette center and the bond center (indicated by blue circles). And the dashed loops enclose the $c$-electron orbitals that hybridize within them.
    \text{(b)} The quasi-one-dimensional stripe geometry used for two-terminal conductance calculation in Kwant. The centeral scattering region (black) has length $L_x$ and fixed width $L_y = 201$ (in the unit of lattice constant). The gray shaded area stands for the two semi-infinite clean leads connected to the scattering region.
    \text{(c1)} The nodal structrue when $f$-electrons are at the bond center. The color map shows the squared hybridiztion amplitude, which vanishes along nodal arcs (red lines) connecting Dirac points ($\mathbf{K},\mathbf{K}'$). The dashed magenta circles represented the Fermi surface corresponding to $E_F = 0.5$, which intersects the nodal arcs at discrete points.
    \textbf{(c2)} Conductance $G_{xx}$ versus length $L_x$ for the bond-center model. The $PT-$symmetric case (blue triangles) with Fermi energy $E_F = 0.5$ shows a power-law decay $G_{xx} \sim L_x^{-1/2}$, confirming the superdiffusive transport exponent $\gamma= 1/2$.
    \textbf{(d1)} Nodal structure when $f$-electrons are at the plaquette center. While the real-space geometry is different, topology again dictates the formation of nodal arcs connecting the Dirac points.
    \textbf{(d2)} The conductance scaling for the plaquette-center model is identical, with $\gamma = 1/2$. This confirms that the superdiffusive exponent is related to the topology, independent of specific hybridization geometry. In both (c2) and (d2), if $f$-electron breaks the $PT-$symmetry, the nodal structure will be destroyed and lead to conventional localization behavior (gray circles).
    The numerical results match the low-temperature prediction $\gamma_{\text{low}} = (D^F -D^F_{\text{node}})/(2n) =1/2$, where the 1D Fermi surface intersects the 1D nodal manifold at discrete points. Results are averaged over at 400 disorder configurations. The $f$-electron energies are drawn uniformly in $[-W,W]$ with disorder strength $W  =1.65$ for the bond-center model and $W = 1.0$ for the plaquette-center model. The hybridization strength is set as $V=\sqrt{2}$ for both models.
    }
    \label{fig:graphene}
\end{figure*}

The graphene system is modeled by a tight-binding Hamiltonian on the honeycomb lattice with nearest neighbor hopping. The Hamiltonian in the momentum-space is given by\cite{bernevig2013topological}
\begin{equation}
    \hat H_c =-t \sum_{\mathbf k} (\hat c^\dagger_{\mathbf k, A} \quad \hat c^\dagger_{\mathbf k, B}) \begin{pmatrix}
        0 && f(\mathbf k) \\ 
        f^*(\mathbf k) && 0
    \end{pmatrix} \begin{pmatrix}
        \hat c_{\mathbf k, A} \\ \hat c_{\mathbf k, B}
    \end{pmatrix},
\end{equation}
where $t$ is the hopping amplitude (below we set $t=1$), $A/B$ denotes the two sublattices in the unit cell, and $f(\mathbf{k}) = 1+ e^{-i k_2} + e^{i(k_1-k_2)}$. The momentum components $k_{1/2}$ are defined as $k_{1/2} = \mathbf{k}\cdot \mathbf{a}_{1/2}$, where $\mathbf{k}$ is the vector in the first BZ and $\mathbf{a}_{1/2}$ are the primitive lattice vectors of the honeycomb lattice, shown in Fig.~\ref{fig:graphene}(a). 
For clarity, we set the inversion center to be the bond center of the unit cell at the origin.
Graphene has the inversion symmetry, which interchanges $A$ and $B$ sublattice, $P = \tau_x$, and reverses the mometum, $k_{1/2} \to - k_{1/2}$; it also has the time-reveral symmetry $T = \mathcal{K}$ that also reverses the momentum. Therefore, graphene has the composite symmetry $PT$ that preserves the momentum
\begin{equation}
    PT = \tau_{x} \mathcal{K},
\end{equation}
where $\tau_{x,y,z}$ are the Pauli matrices acting in the sublattice space and $\mathcal{K}$ is the complex conjugate. 

The band structure features topological band touchings known as Dirac points in the Brillouin zone and are characterized by a nontrivial Berry phase protected by the $PT$-symmetry. For any closed loop $\mathcal{C}$ that enclosing a single Dirac point, the Berry phase is 
\begin{equation}
    \oint_{\mathcal{C}} d\mathbf{k} \cdot \braket{u_{\mathbf{k},m}|i\nabla| u_{\mathbf{k},m}} = \pi \pmod{2\pi},
\end{equation}
where $n=1,2$ corresponds to the lower and upper bands.
The nontrivial Berry phase implies a topological obstruction: it is impossible to define a globally smooth $PT$-symmetric gauge for $\ket{u_{\mathbf{k},m}}$, and therefore $\ket{\psi_{\mathbf{k},m}}$, over the entire loop $\mathcal{C}$. 

Now suppose the hybridization orbital $\ket{d}$ is also $PT$-symmetric. This condition can be met by placing the $f$-electrons at the bond centers or plaquette centers of the honeycomb lattice, see Fig.~\ref{fig:graphene}(a). If we assume, for the sake of contradiction, the wavefunction overlap $\braket{ \psi_{\mathbf{k},m}| d}$ is non-zero everywhere on the loop $\mathcal{C}$, we can use it to fix the globally $PT$-symmetric gauge by defining $\ket{\chi_{\mathbf{k}}} = \frac{\braket{\psi_{\mathbf{k},m}| d}}{|\braket{ \psi_{\mathbf{k},m}| d}|} \ket{\psi_{m,\mathbf{k}}}$. The Bloch state $\ket{\chi_\mathbf{k}}$ would be globally smooth and $PT$-symmetric, which implies a trivial Berry phase. This contradicts with the $\pi$ Berry phase of the Dirac points. So the overlap must vanish at least one point $\mathbf{k}_0$ on the loop.

The formation of the node $\mathbf{k}_0$ can be understood geometrically\cite{jin2022chern}. The two-band Hamiltonian allows us to associate pseudo-spin vectors with each Bloch state $\ket{\psi_{\mathbf{k},m}}$ and the hybridization orbital $\ket{d}$:  
\begin{equation}
    \mathbf{S}_\mathbf{k} = \sum_{a,b}{\psi^{(m)*}_{\mathbf{k},a}\boldsymbol \tau_{ab} \psi^{(m)}_{\mathbf{k},b}}; \quad    \mathbf{d}_\mathbf{k} = \sum_{a,b}\phi_{a}(\mathbf{k}) \boldsymbol {\tau}_{ab} \phi_{b}^* (\mathbf{k}).
\end{equation}
The $PT-$symmetry of the system constrains both $\mathbf{S}(\mathbf{k})$ and $\mathbf{d}(\mathbf{k})$ to lie on the equatorial plane of the Bloch sphere. And we can show the wavefunction inner product can be expressed by :
\begin{equation}
    |\braket{d|\psi_{\mathbf{k},m}}|^2 =\frac{|\mathbf{d}_\mathbf{k}|}{2} (1+\cos\theta_\mathbf{k}), \label{eq:graphene:geometric}
\end{equation}
where $\theta_{\mathbf{k}}$ is the relative angle between the two pseudo-spins and the node occurs when $\theta_{\mathbf{k}_0} = \pi$ ($\mathbf{S}_{\mathbf{k}_0}$ and $\mathbf{d}_{\mathbf{k}_0}$ are anti-parallel). 

The existence of such a point is ensured by the topological mismatch between $\ket{\psi_{\mathbf{k},n}}$ and $\ket{d}$. On a loop $\mathcal{C}$ enclosing a single Dirac point, the nontrivial Berry phase ensures the pseudo-spin has an odd winding number around the origin. In contrast, because the local state $\ket{d}$ is constructed from local orbitals with a finite cluster, it is topologically trivial. This means the corresponding form factor $\tilde{ \boldsymbol{\phi}}_{j}(\mathbf{k})$ has a trivial Berry phase on the loop $\mathcal{C}$ and the associated pseudo-spin vector $\mathbf{d}_\mathbf{k}$ must have an even winding number. The total change of the angle between $\mathbf{S}_{\mathbf{k}}$ and $\mathbf{d}_{\mathbf{k}}$ upon trasversing the loop is $\Delta \theta = 2\pi \times(\text{odd integer})$. The net winding guarantees a point where these two pseudo-spins are anti-parallel. This prove that a node, where the hybridization amplitude vanishes $\braket{d|\psi_{\mathbf{k}_0,m}}=0$, must exist on the loop $\mathcal{C}$.

To determine the order of the node, we expanding the cosine for small deviation from the node, $\theta(\mathbf{k}) = \pi+\delta \theta$ shows
\begin{equation}
|\braket{d|\psi_{\mathbf{k},m}}|^2 \propto 1 + \cos(\pi + \delta\theta) \approx \frac{(\delta\theta)^2}{2}.
\end{equation}
Near the Dirac point, the angle deviation is linear in the momentum deviation from the node, $\delta k$ (see Appendix~\ref{sec:proof_linear_momentum} for proof). Consequently, the squared hybridization amplitude scales quadratically with momentum deviation:
\begin{equation}
|\mathcal{V}_m(\mathbf{k}_0+\delta\mathbf{k})|^2 \propto (\delta k)^2.
\end{equation}
From which we identify the node is first order, $n=1$.

If we enlarge the radius of the loop enclosing the Dirac point, there is always at least one node until the contour encloses another Dirac point so the Berry phase on the contour now is trivial. 
This implies that the topological charge of the Dirac points acts as sources and sinks for the nodal lines. The nodes cannot terminate in the middle of the Brillouin zone; they are forced to form `nodal arcs' connecting Dirac points with opposite topological charges.
Fig.~\ref{fig:graphene}(c1)/(d1) shows the nodal arcs when the $f$-electrons are placed at the bond center and the plaquette center of the honeycomb lattice and the corresponding hybridization orbital is $PT-$symmetric. 

This ``nodal arc" structure will lead to superdiffusive transport. In this model with finite doping, $0<E_F<1$, we have $D=2,D_{\text{node}}=1,D^F=1$ and $D^F_{\text{node}} = 0$. At high temperature, the scaling exponent to predicted to be $\gamma_{\text{high}} = (D-D_{\text{node}})/(2n)=1/2$. At zero temperature, the scaling exponent is $\gamma_{\text{low}} = (D^F-D^F_{\text{node}})/(2n) = 1/2$, indicating superdiffusion. Our numerical simulations (see Fig.~\ref{fig:graphene} (c2)/(d2) of the zero temperature transport aligns well with the theoretical predictions above.

\subsection{Multi-Weyl semimetal}
We now demonstrate how the interplay of symmetry and topology can give rise to robust, higher-order nodes ($n>1$). Our example is a 3D multi-Weyl semimetal\cite{fang2012multi}, whose low-energy physics is described by a two-band effective Hamiltonian protected by a crystalline rotation symmetry about the $z-$axis, $C_m$:
\begin{equation}
    H^{(c)}_{\text{eff}}(\mathbf{k}) = a k_+^{n} \sigma_{+} + a^*k_-^n \sigma_- - v_z k_z\sigma_z.
\end{equation}
Here, $a\neq0$ is a arbitrary complex number, $\mathbf{k}$ is the momentum deviation from the multi-Weyl point, $n \in \{1,2,3   \}$ is an integer dictated by the rotation symmetry representations of the two bands, and $\sigma_{\pm} = \sigma_x\pm i\sigma_y$. This Weyl point is a monopole of Berry curvature carry the Chern number $|C| = n$. Our goal is to show that $\mathcal{V}_m(\mathbf{k}) \propto \braket{d|\psi_{\mathbf{k},m}}$ develops a higher-order node if the $f$-orbital also has the rotation symmetry. 

\begin{figure*}[t]
    \includegraphics[width=1.7\columnwidth]{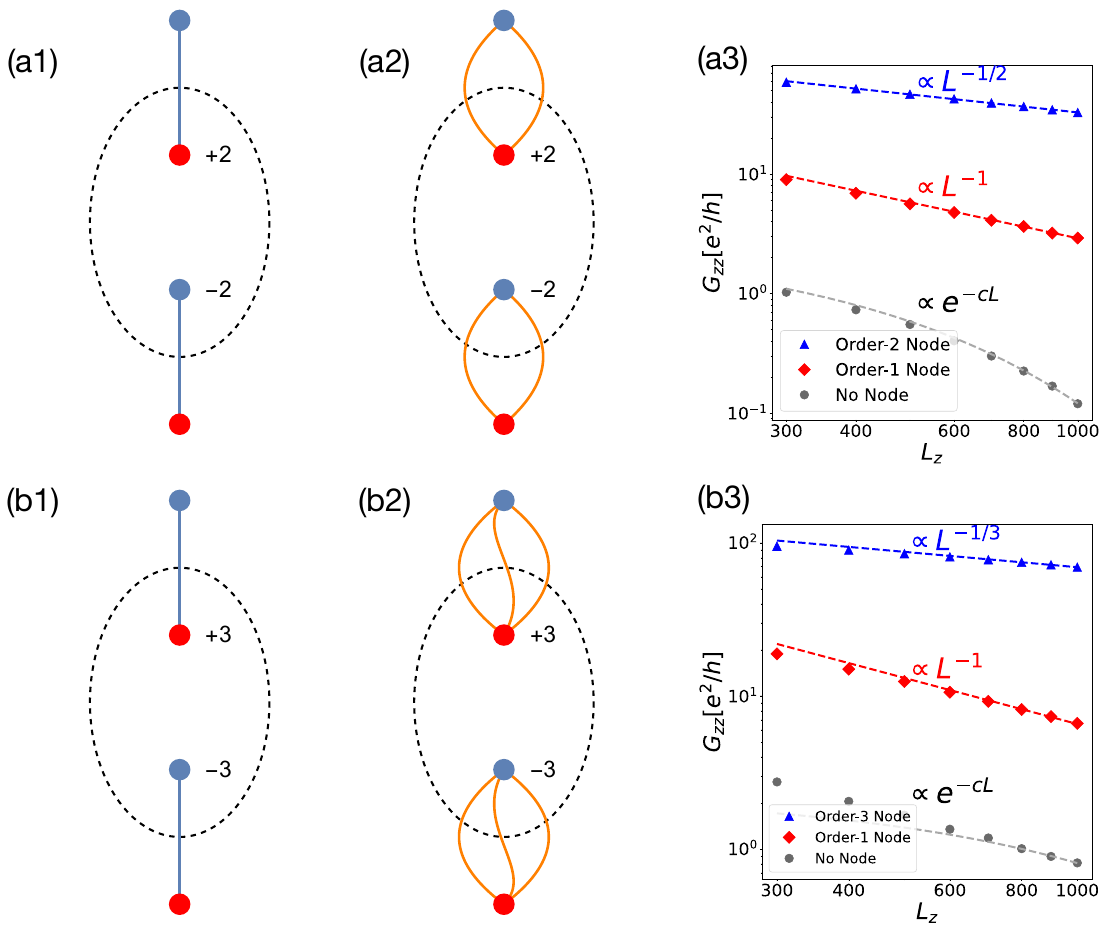}
    \caption{\textbf{Higher-order nodal arcs and symmetry-switched transport regimes in multi-Weyl semimetals} A demonstration of the interplay between topology and symmetry and the direct numerical verification of the resulting transport exponents. All simulations are performed at zero temperature.
    \textbf{(a) Double-Weyl Semimetal($|C|=2$).} (a1) Conceptual schematic of a stable, higher-order ($n=2$) nodal arc (blue line) pinned at the rotation axis connecting double-Weyl points of charge $\pm 2$, protected by the $C_4$ rotation symmetry. The dashed circle stands for the Fermi surface.
    (a2) When the hybridization orbital breaks the rotation symmetry, the higher-order arc splits into two first order ($n=1$) arcs (orange lines). 
    (a3) The resulting transport regimes. The second-order node leads to the superdiffusion with $G_{zz} \sim L_z^{-1/2}$ (blue triangles), matching the predicted exponent at low temperature $\gamma = (D^F - D^F_{\text{node}})/(2n) = 1/n = 1/2$. The split first-order nodes lead diffusion with $G_{zz} \sim L_z^{-1}$ (red diamonds). Nodes are absent for a scalar potential without any internal structures, for example, $ \hat  V_{\mathbf{R}} \propto \hat I$ and $\hat I$ is the identity operator. This will lead to localization behavior (gray circles).
    \textbf{(b) Triple-Weyl Semimetal($|C|=3$).}
    (b1-b2) The same principle of symmetry-protected fusion and splitting applies to a triple-Weyl point.
    (b3) The resulting transport regimes. The third-order node leads to the superdiffusion with $G_{zz} \sim L_z^{-1/3}$ (blue triangles), matching the predicted exponent at low temperature $\gamma  = 1/n = 1/3$. For split first order nodes and absence of nodes, we again have diffusion with $G_{zz} \sim L_z^{-1}$ (red diamonds) and localization (gray circles).
    We adapt a two-band effective lattice model to simulate the conductance and we leave the detailed setup in the Appendix~\ref{app:numerics}.
    The Fermi energy is set to be $E_F = 2$ corresponding to the Fermi surface (dashed circle) in (a1-a2) and (b1-b2). The system size along transverse directions are fixed: $L_x = L_y = 20$. The $f$-electron energies are drawn from a unifrom distribution $[-W,W]$, where $W=3.5$ for the double-Weyl case and $W=2.5$ for the triple Weyl case. Results are averaged over at least $100$ disorder configurations.
    }
    \label{fig:Weyl}
\end{figure*}

The existence of a hybridization node is guaranteed by the topology. On any sphere $S^2$ enclosing the Weyl node, the non-zero Chern number $|C|\neq 0$ implies a topological obstruction to defining a globally smooth gauge for the Bloch state $\ket{\psi_{\mathbf{k},m}}$ for $\mathbf{k} \in S^2$. As argued for the graphene case, this obstruction forces the overlap with any smooth, topologically trivial reference state (which is the hybridization orbital $\ket{d}$ in our case) to vanish at some point on the sphere.

Furthermore, this connection is qualitative. The Chern number of the $c$-electron band on the sphere is precisely equal to the winding numbers of the wavefunction overlap $\braket{\psi_{\mathbf{k},m}|d}$ around its zeros (the nodes) on that sphere. This is given by the ``topological conservation law"(see Appendix~\ref{app:topo_conservation}):
\begin{equation}
    C = \sum_i \oint_{c_i} \frac{d \mathbf{k}}{2 \pi i} \cdot \nabla_{\mathbf{k}} \log \braket{\psi_{\mathbf{k},m}|d} = \sum_i w_i, \label{eq:topo_conservation_law}
\end{equation}
where $c_i$ is an infinitesimal loop around the $i-$th node, and $w_i$ is its integer winding number. The order of the $i-$th node is given by $n_i = |w_i|$.  On the sphere enclosing the Weyl point, nodes with opposite windings can annihilate but the total winding must equal the Chern number. This identity dictates that the topological charge of the band structure must be mirrored in the nodal structure of the hybridization.

The rotation symmetry further constrains the locations of these nodes. Let the hybridization orbital $\ket{d}$ be the eigenstate of the same $C_m$ rotation, a condition met if $C_m \subset \mathcal{G}(\mathbf{r}_f)$: the rotation group is a subgroup of the site-symmetry group. At the poles of the sphere enclosing the multi-Weyl point (i.e., along the $k_z$ axis where $k_{x/y} =0$), the off-diagonal terms of $H_{\text{eff}}^{(c)}$ vanishes, and the eigenstates are simply the basis states, which we denote as $\ket{\uparrow},\ket{\downarrow}$. Let us choose the hybridization orbital $\ket{d}$ to be compatible with $ \ket{\uparrow}$ (i.e., $\braket{d|\uparrow} \neq 0, \braket{d|\downarrow}= 0$. Consequently, at the north pole $\mathbf{k}_n$, where the eigenstate of the upper band is purely $\ket{\downarrow}$: $\ket{\psi_{\mathbf{k},+}} = \ket{\downarrow}$, the overlap must be identically zero. These nodes are pinned to the $k_z-$rotation axis by symmetry. As in the Dirac case, these pinned nodes form an arc connecting multi-Weyl points of opposite charges. In Fig.~\ref{fig:Weyl}(a1/2, b1-2), we demonstrate the nodal arcs in multi-Weyl system for different types of hybridization orbitals.

Now we analyze the overlap near the node $\mathbf{k}_n$. 
For small transverse momentum $k_\perp = \sqrt{k_x^2+k_y^2}$, the overlap $\braket{d|\psi_{\mathbf{k},+}}$ is given by
\begin{equation}
|\braket{d|\psi_{(\mathbf{k}_s+\mathbf{k}_\perp),+}}| \propto |\braket{\downarrow|\psi_{(\mathbf{k}_s+\mathbf{k}_\perp),+}}| \sim k_\perp^n.
\end{equation}
This demonstrate our central result: for multi-Weyl semimetal systems, the interplay of topology and rotation symmetry generates a higher-order nodal line on the rotation axis, whose order $n$ is precisely the topological charge of the multi-Weyl point.

This result exemplifies a connection between the topology and symmetry. While the topological selection rule dictates that $\sum_i w_i = C$, point-group symmetries effectively ``fuse" the possible lower order nodes into a single, stable higher order node. This provides a robust, non-fine-tuned mechanism for generating higher-order nodal structures in condensed matter systems. 

For a 3D system with such a nodal arc $D_{\text{node}} = 1$, our theory predict superdiffusion at high temperatures with the exponent $\gamma_{\text{high}} = \frac{3-1}{2\times n} = \frac{1}{n}$. At zero temperature, 
for a Fermi level away from the Weyl point, the system has a 2D Fermi surface $D^F_{\text{node}} = 2$.
Supposing the Fermi surface intersects the 1D nodal line at discrete points, the exponent is given by $\gamma_{\text{low}} = \frac{2-0}{2\times n} = \frac{1}{n}$. For the double-Weyl semimetal, this yields $\gamma_{\text{low}/\text{high}} = \frac{1}{2}$. For the triple-Weyl semimetal, we predict a more exotic superdiffusion with $\gamma_{\text{low}/\text{high}} = \frac{1}{3}$. The numerical simulations of the zero-temperature transport align well with the theoretical predictions, as shown in Fig.~\ref{fig:Weyl}(a3,b3).
Possible material realizations contain the double-Weyl semimetal $\text{HgCr}_2\text{Se}_4$\cite{fang2012multi,xu2011chern} and triple-Weyl semimetal $\text{Rb}(\text{MoTe}_3)$\cite{liu2017predicted}, with $f$-orbitals (the impurity level wavefunctions) having rotation symmetries.

When the \(f\)-orbital lacks rotational symmetry, a higher-order nodal arc splits into several lower-order arcs (typically \(n=1\)), as shown in Fig.~\ref{fig:Weyl}(a1/2, b1/2). 
Consistent with the analysis above, the high-temperature transport is diffusive, \(\gamma_{\text{high}}=1\). 
At \(T=0\), for a 2D Fermi surface that encloses the Weyl point and intersects the nodal arcs at isolated points, the transport also remains diffusive with \(\gamma_{\text{low}}=1\). 
Numerical simulations align with these predictions; see Fig.~\ref{fig:Weyl}(a3,b3).

When the nodal arcs are absent, the scattering rate is non-zero in BZ and the conductance will show exponential decay (see Fig.~\ref{fig:Weyl}(a3,b3)). This occurs if the effective scattering becomes isotropic in the internal space of the $c$-electrons. A possible physical scenario is the presence of multiple, distinct scattering channels whose individual nodes do not coincide, thereby providing a finite scattering rate for all states on the Fermi surface. 

\section{The Effect of Non-Nodal perturbation}
\label{sec:non-nodal}
Our analysis so far has focused on the ideal case where nodal hybridization is the sole scattering mechanism. In any realistic material, however, there will be a background of conventional, non-nodal perturbation. For instance, it could be the scalar potential disorder: $\hat H_{\text{dis}} = \sum_{\mathbf{R}} W_\mathbf{R} \hat c_{\mathbf{R}}^\dagger \hat c_{\mathbf{R}}$, where $W_\mathbf{R}$ is a random potential with zero mean and variance $\langle  W_{\mathbf{R}}W_{\mathbf{R}'}\rangle = W^2 \delta_{\mathbf{R},\mathbf{R}'}$. The corresponding scattering rate is proportional to the disorder strength: $\Gamma_{\text{dis}} \propto W^2$\cite{bruus2004many}. In this section, we show that its presence leads to a predictable crossover from superdiffusive to conventional diffusive transport, governed by a universal scaling law.

The presence of non-nodal perturbation changes the nature of the quasiparticles. Even at the hybridization node, the quasiparticle now acquires a finite lifetime, $\tau_{\text{dis}}$, and a corresponding finite mean free path, $l_{\text{dis}}$. In the thermodynamic limit, the finite mean free path will always lead to conventional diffusive transport\cite{ashcroft2011solid}.

However, anomalous superdiffusive transport can still be observed in a physical regime defined by the competition between the mean free path, $l_{\text{dis}}$, and the system length, $L$. 
\begin{itemize}
    \item \textbf{Effective superdiffusive Regime}($L \ll l_{\text{dis}}$): When the system size is much smaller than the mean free path set by the non-nodal disorder, a quasiparticle is more likely to traverse the entire system than it is to scatter off a potential impurity. In this regime, the finite system size $L$ acts as the dominant limiting factor for transport. The physics is effectively that of the ideal nodal system, and the conductance is well-described by the superdiffusive scaling law, $G \propto L^{-\gamma}$.
    \item \textbf{Diffusive Regime}($L \gg l_{\text{dis}}$): For large system sizes, quasiparticles are scattered many times from the potential disorder transversing the system. The transport is now limited by $l_{\text{dis}}$, and the system obeys Ohm's law, leading to conventional diffusive scaling, $G \propto L^{-1}$.
\end{itemize}

This intuitive picture can be formalized by analyzing the conductivity integral Eq.~(\ref{eq:kubo_sigma_epsilon}). The presence of a constant scattering rate $\Gamma_{\text{dis}}$ acts as an additional infrared cutoff. A scaling analysis (detailed in Appendix~\ref{app:non-nodal}) shows that the conductance should obey a universal scaling form:
\begin{equation}
    G(L, W^2)  \propto W^{2\gamma} F(L \cdot W^2),
    \label{eq:scaling_form}
\end{equation}
where $\gamma$ is the superdiffusive exponent and 
$F(x) = \frac{Ax^{-1/2}}{(x^2+B^2)^{1/4}}$
is the universal scaling function with $A,B$ being two positive, real parameters. And we have the asymptotic behaviors: $F(x) \sim x^{-\gamma}$ for $x \ll 1$ and $F(x) \to x^{-1}$ for $x \gg 1$. This scaling function predicts a crossover from superdiffusion to diffusion when $L \sim B/W^2$.

\begin{figure}[t]
    \includegraphics[width=0.95\columnwidth]{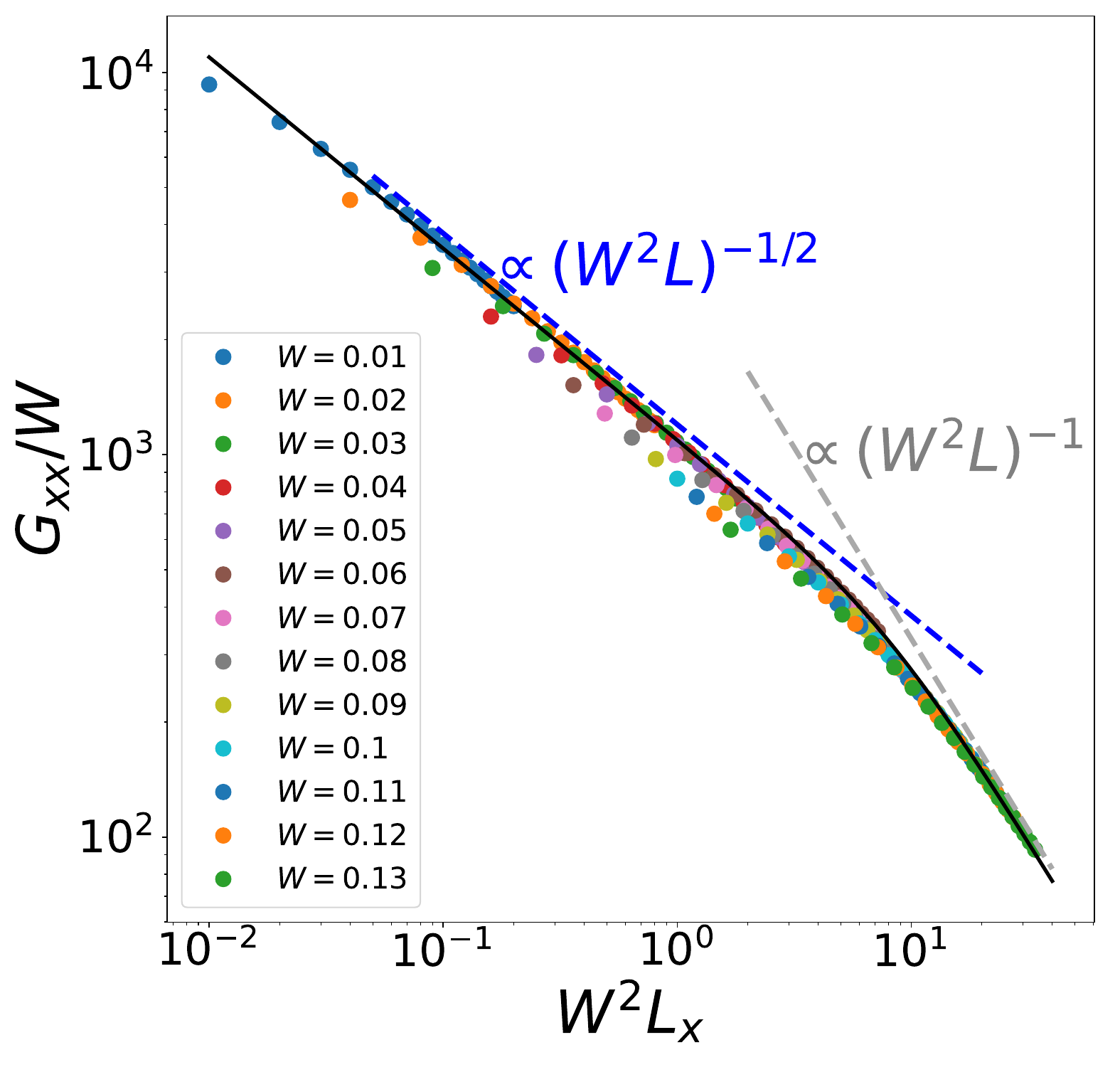} 
    \caption{\textbf{Universal scaling collapse for the superdiffusive-to-diffusive crossover.}
    The figure demonstrates the universal scaling behavior of the conductance in the 2D square lattice model with nodal lines ($\gamma=1/2$) in the presence of additional non-nodal (potential) disorder. The strength of this disorder is parameterized by $W^2$, which sets the background scattering rate, $\Gamma_{\text{dis}} \propto W^2$.
    The rescaled conductance, $G_{xx}/W$, is plotted against the rescaled system length, $x = W^2 L_x \propto \Gamma_{\text{dis}} L_x$. Data for thirteen different disorder strengths (from $W=0.01$ to $W=0.13$, as shown in the legend) collapse onto a single universal curve. 
    The solid black line is a fit to the universal scaling function, $F(x) = \frac{A x^{-1/2}}{(x^2+B^2)^{1/4}}$, which correctly captures the behavior across all regimes. The dashed lines show the theoretically expected asymptotic power laws:
    (i) In the \textbf{superdiffusive limit} ($x \ll 1$), the function follows $F(x) \propto x^{-1/2}$ (blue dashed line).
    (ii) In the \textbf{diffusive limit} ($x \gg 1$), the function follows $F(x) \propto x^{-1}$ (gray dashed line).
    }
    \label{fig:data_collapse}
\end{figure}

We verify this prediction numerically using the 2D square lattice model with nodal lines along the BZ boundaries with $\gamma=1/2$ (see Fig.~\ref{fig:symmetry}). In addition to the nodal hybridization, we introduce a weak scalar potential disorder characterized by the rate $\Gamma_{\text{dis}}$. Fig.~\ref{fig:data_collapse} shows the result of this scaling analysis. The collapse of all data onto a single universal curve provides a confirmation of our scaling theory for the superdiffusive-to-diffusive crossover.

\section{Experimental observables}
\label{sec:experimenal}
Beyond the scaling of the dc conductance, our theory of the nodal structure in the two-component model leads to a variety of experimentally verifiable consequences. These signature provide pathways to identify and characterize nodal structures in topological materials using standard experimental techniques. In this section, we outline three key predictions. While we present the derivations in the case where $\hat H_f$ is the Hamiltonian of onsite disorder, we emphasize that the predicted power-law signatures for spectroscopy and optical conductivity are universal consequences of the nodal geometry and are expected to hold when the $\hat H_f$ involves interaction.

\subsection{Spectroscopic Signature: Direct Visualization of Nodes with ARPES}
The most direct evidence for a nodal structure would be its visualization in momentum space. Angle-resolved photoemission spectroscopy (ARPES), which measures the single-particle spectral function $A(\mathbf{k},\omega)$\cite{sobota2021angle}, is the ideal tool for this purpose. The spectral function for a given band $n$ is a Lorentzian centered at the quasiparticle energy, with a width determined by the imaginary part of the self-energy (the scattering rate). By expanding the Green's function around its quasiparticle pole Eq.~(\ref{eq:expand_around_pole})
\begin{equation}
    A_m(\mathbf{k},\omega) \propto \frac{ Z_{\mathbf{k}} \Gamma( \mathbf{k} )}{(\omega-\tilde E_m(\mathbf{k}))^2 + \Gamma(\mathbf{k})^2}.
\end{equation}
Our result is this scattering rate $\Gamma(\mathbf{k})$ is directly proportional to the squared hybridization amplitude, $\Gamma(\mathbf{k}) \propto |\mathcal{V}_m(\mathbf{k})|^{2}$. Near a node of order $n$, this vanishes as $|\delta k|^{2n}$, where $\delta k$ is the momentum deviation from the nodal manifold.

This leads to a clear prediction for ARPES: the width of the measured quasiparticle peaks should exhibit a strong momentum dependence: these peaks will be infinitely sharp (resolution-limited) for momenta on the nodal manifold and will broaden as the momentum moves away, following the power-law $|\delta k|^{2n}$.

\subsection{Dynamic Signature: Divergent Low-Frequency Optical Conductivity}
The long-lived quasiparticles near the nodal manifold also leaves a distinct signature in the low-frequency optical responses of the material. At zero temperature, a finite frequency $\omega$ acts as an effective infrared cutoff, regularizing the dc conductivity integral. A standard calculation using the Kubo formula for optical conductivity (see Appendix~\ref{app:conductivity_integral}) yields a universal power-law scaling at low frequencies:
\begin{equation}
    \text{Re}[\sigma(\omega)] \sim |\omega|^{\gamma_{\text{low}}-1}.
\end{equation}
In the normal diffusive case $(\gamma = 1)$, this recovers the familiar finite Drude peak. However, in the superdiffusive regime ($0<\gamma_{\text{low}}<1$), this formula predicts a power-law divergence of the optical conductivity as $\omega \to 0$. For the common case of $\gamma = 1/2$, this means $\sigma(\omega) \sim 1/\sqrt{\omega}$.

\subsection{Thermal Signature: Anomalous Temperature-Dependent Resistivity}
\label{sec:exp_resistivity}

Finally, the nodal structure profoundly alters the temperature dependence of the resistivity, $\rho(T)$, when residual electron-electron (e-e) interactions are present alongside the primary nodal disorder scattering. At low temperatures, the transport is dominated by the physics near the nodal manifold, where the weak e-e scattering rate, $\Gamma_{ee} \propto (k_B T)^2$\cite{luttingerpr}, acts as the primary infrared cutoff for the conductivity integral.

A scaling analysis of the conductivity integral with this $T^2$ cutoff (detailed in Appendix~\ref{app:finite temperature cond}) predicts an anomalous power-law dependence for the resistivity at low temperatures:
\begin{equation}
    \rho(T) \propto T^{2(1-\gamma_{\text{low}})}.
    \label{eq:rho_T_scaling}
\end{equation}
This result is remarkable. Most notably, for the superdiffusive case of $\gamma_{\text{low}}=1/2$, our theory predicts the emergence of linear-in-temperature resistivity, $\rho(T) \propto T$. This provides a new mechanism for linear-in-temperature resistivity, originating from the nodal structure.

This anomalous scaling is expected to hold up to a crossover temperature, $T^*$. Above this temperature, the uniform thermal scattering from electron-electron interaction will dominate over the hybridization, leading to the more conventional metallic behavior. 
As shown in the Appendix~\ref{app:conductivity_integral}, this crossover scale is set by the strength of the nodal disorder scattering $\Gamma_{\text{node}}$ and the Fermi energy $E_F$ and
the crossover temperature $T^*$ is predicted to scale as:
\begin{equation}
    k_B T^* \propto \sqrt{\Gamma_{\text{node}} E_F}.
    \label{eq:T_star}
\end{equation}
Therefore, a key experimental signature would be the observation of the anomalous power law, Eq.~\eqref{eq:rho_T_scaling}, at low temperatures, followed by a crossover to a different behavior above $T^*$.

\section{Conclusion and Discussion}
\label{sec:conclusion and discussion}

In this work, we have introduced and analyzed the nodal hybridization mechanism in the two-component model, a general and physically grounded framework for generating anomalous superdiffusive transport in condensed matter systems. We have demonstrated that the hybridization between itinerant $c$-electrons and localized $f$-electrons can naturally give rise to a robust nodal structure in the effective $c$-electron scattering rate. Our central result is a set of universal scaling laws that directly link the superdiffusive transport exponent $\gamma$ to the geometric properties---dimension $D_{\text{node}}$ and order $n$---of this emergent nodal manifold at both low and high temperatures. 

Our work demonstrates that the fragile, fine-tuned ``nodal structure" in previous studies\cite{Jie24,Chen24,Romain25,Marko24} can be protected by symmetry and topology in the two-component model, making its realization in condensed matter experiments potentially feasible. In addition, we propose several models potentially realizable in corresponding candidate materials, including graphene, double Weyl semimetal $\text{HgCr}_2\text{Se}_4$\cite{fang2012multi}, triple Weyl semimetal $\text{Rb}(\text{MoTe})_3$\cite{liu2017predicted} and B2-structure crystals\cite{anderson1999site}. Besides the scaling of the conductance, we propose other experimentally verifiable signatures, including anomalous temperature-dependence of resistivity, the characteristic power laws in optical conductivity and momentum dependent quasiparticle weight.

Finally, we address the generality of the two-component model. Although the numerical results in this paper focus on non-interacting systems, the nodal structure itself doesn't depend on the specific details of the $f$-electron Hamiltonian. Crucially, our analytical calculation of the scaling exponent $\gamma$ remains valid even when $f$-electrons are interacting. This suggests that the nodal hybridization mechanism could play an important role in the transport properties of strongly-correlated materials such as heavy-fermion compounds or Kondo systems\cite{hewson1997kondo,coleman2015introduction}. For instance, in heavy-fermion materials like CeCoIn$_2$, the distinct orbital symmetries of conduction band electrons and local moments prevent hybridization along high-symmetry lines\cite{hybridization_nodes}.


\appendix
\section{Physical origin of the two-component model}
\label{app:origin}

To ground the two-component model used in the main text, we begin with a general first-quantized Hamiltonian describing electrons in a host crystal with impurities\cite{hewson1997kondo}:
\begin{equation}
    H = \sum_{i}\left(-\frac{\hbar^2\nabla_i^2}{2m} + U(\mathbf{r}_i) + V_{\text{imp}}(\mathbf{r}_i)\right) + \frac{1}{2} \sum_{i \neq j} \frac{e^2}{4\pi\epsilon_0|\mathbf{r}_i - \mathbf{r}_j|}.
\end{equation}
Here, $U(\mathbf{r}_i)$ is the periodic potential of the host lattice, $V_{\text{imp}}(\mathbf{r}_i)$ is the additional potential from the impurities, and the final term is the electron-electron Coulomb interaction.

To proceed to a second-quantized description, we first define an effective single-particle Hamiltonian, $h_{sp}(\mathbf{r})$, by treating the electron-electron interactions at a mean-field level (e.g., via a Hartree-Fock potential $U_{HF}(\mathbf{r})$):
\begin{equation}
    h_{sp}(\mathbf{r}) = -\frac{\hbar^2\nabla^2}{2m} + U(\mathbf{r}) + U_{\text{HF}}(\mathbf{r}) + V_{\text{imp}}(\mathbf{r}).
\end{equation}
We then partition the single-particle Hilbert space into two basis sets: the itinerant Bloch wavefunctions of the host crystal, $\psi^c_{\mathbf{k},m}(\mathbf{r}) = \braket{\mathbf{r}|\psi_{\mathbf{k},m}}$, which we associate with $c$-electrons, and localized atomic-like orbitals centered on the impurity sites, $\phi^f_{\mathbf{R}}(\mathbf{r}) = \braket{\mathbf{r}|f_{\mathbf{R}}}$, which we associate with $f$-electrons. Note that neither of these basis states are exact eigenstates of the full single-particle Hamiltonian $h_{sp}$.

The second-quantized Hamiltonian, expressed in this basis, naturally takes the form $\hat{H} = \hat{H}_c + \hat{H}_f + \hat{H}_{\text{hyb}}$. The diagonal parts are given by:
\begin{equation}
    \hat{H}_c = \sum_{m,\mathbf{k}}E_m(\mathbf{k}) \hat{\psi}^\dagger_{\mathbf{k},m} \hat{\psi}_{\mathbf{k},m} \quad \text{and} \quad \hat{H}_f = \sum_{\mathbf{R}} \epsilon_f(\mathbf{R}) \hat{f}^\dagger_{\mathbf{R}} \hat{f}_{\mathbf{R}}.
\end{equation}
Here, $E_m(\mathbf{k})$ are the Bloch bands of the host, and $\epsilon_f(\mathbf{R})$ is the on-site energy of the f-orbital at site $\mathbf{R}$. 

In $\hat H_c$, there is an additional direct potential scattering due to impurity potential $V_{\text{imp}}$:
\begin{equation}
    \hat H^{\text{pot}}_c = \sum_{\mathbf{R}} \sum_{ m\mathbf{k},m'\mathbf{k}'} u e^{i(\mathbf{k}-\mathbf{k}')\cdot \mathbf{R}} \hat{\psi}^\dagger_{\mathbf{k},m} \hat{\psi}_{\mathbf{k}',m'},
\end{equation}
where we approximate the matrix elements $\braket{\psi_{\mathbf{k},m}|V_{\text{imp},\mathbf{R}}|\psi_{\mathbf{k}',m}} \approx u$

The crucial coupling between the two subsystems arises from the off-diagonal matrix elements of $h_{sp}$, which form the hybridization Hamiltonian:
\begin{equation}
    \hat{H}_{\text{hyb}} = \sum_{\mathbf{k},n}\sum_{\mathbf{R}} \left( \mathcal{V}_m(\mathbf{k}) e^{i \mathbf{k}\cdot \mathbf{R}} \hat{f}^\dagger_{\mathbf{R}} \hat{\psi}_{\mathbf{k},m} + \text{h.c.} \right),
\end{equation}
where the hybridization amplitude $\mathcal{V}_m(\mathbf{k})$ is given by the integral:
\begin{equation}
    \mathcal{V}_n(\mathbf{k}) = \int d\mathbf{r} \, \phi^{f*}_{\mathbf{R}=0}(\mathbf{r}) h_{sp}(\mathbf{r}) \psi^c_{\mathbf{k},n}(\mathbf{r}) \equiv \braket{f_{\mathbf{R=0}}|h_{sp}|\psi_{\mathbf{k},n}}.
\end{equation}
In the main text, we also express this amplitude via the hybridization orbital, $\ket{d}$, which is a superposition of $c$-electron Wannier states, $\ket{c_{\boldsymbol{\delta},a}}$. These two pictures are equivalent. The real-space hybridization amplitudes are $V_a(\boldsymbol{\delta}) = \braket{f_{\mathbf{R=0}}|h_{sp}|c_{\boldsymbol{\delta},a}}$. Using the definition of the Hybridization Orbital from the main text, $\mathcal{V}_m(\mathbf{k}) = V\braket{d|\psi_{\mathbf{k},m}}$, and expanding in the Wannier basis, we confirm the identity:
\begin{equation}
\begin{split}
    V\braket{d|\psi_{\mathbf{k},m}} &= \sum_{\boldsymbol{\delta},a} V_a(\boldsymbol{\delta}) \braket{c_{\boldsymbol{\delta},a}|\psi_{\mathbf{k},m}} \\
    &= \sum_{\boldsymbol{\delta},a} \braket{f_{\mathbf{R}=0}|h_{sp}| c_{\boldsymbol{\delta},a}}\braket{c_{\boldsymbol{\delta},a}|\psi_{\mathbf{k},m}} \\
    &= \braket{f_{\mathbf{R}=0}|h_{sp}|\psi_{\mathbf{k},m}}.
\end{split}
\end{equation}
This confirms that the two-component model, with its characteristic hybridization orbital, is a well-founded effective description emerging from a more fundamental quantum mechanical picture.

\section{Derivation of the \(c\)-electron Self-Energy}
\label{app:self_energy}

In this appendix, we provide a detailed derivation of the $c$-electron self-energy, $\Sigma_c$, arising from hybridization with the $f$-electron subsystem. We consider both the case of $f$-electrons as static disorder and as an interacting flat band.

\subsection{Disordered \(f\)-electron Impurities}

We begin with the case where $\hat H_f$ describes a random distribution of non-interacting impurities. It is instructive to first solve the problem of a \textit{single} \(f\)-impurity located at the origin ($\mathbf{R}=0$). The equations of motion for the Green's functions, in the $c$-electron band basis ($n, n'$), are:
\begin{equation}
\begin{aligned}
&(\epsilon - E_m(\mathbf{k}))G_{c}(m\mathbf{k},m'\mathbf{k}';\epsilon) = \delta_{m\mathbf{k},m'\mathbf{k}'} + \mathcal{V}_m^*(\mathbf{k}) G_{fc}(m'\mathbf{k}';\epsilon), \\
&(\epsilon - \epsilon_f)G_{fc}(m'\mathbf{k}';\epsilon) = \sum_{m'',\mathbf{k}''} \mathcal{V}_{m''}(\mathbf{k}'')G_{c}(m''\mathbf{k}'',m'\mathbf{k}';\epsilon).
\end{aligned}
\end{equation}
Solving this coupled system for $G_c$ yields the standard \(T\)-matrix form of the Dyson equation:
\begin{equation}
\begin{split}
&G_{c}(m\mathbf{k},m'\mathbf{k}';\epsilon) ={} \delta_{m,m'}\delta_{\mathbf{k},\mathbf{k}'}g_{c}(m\mathbf{k},\epsilon) \\
& + g_{c}(m\mathbf{k},\epsilon)  T(m\mathbf{k},m'\mathbf{k}';\epsilon) g_{c}(m'\mathbf{k}',\epsilon),
\end{split}
\end{equation}
where $g_{c}(m\mathbf{k},\epsilon) = [\epsilon - E_m(\mathbf{k}) + i0^+]^{-1}$ is the bare $c$-electron propagator, and the single-impurity \(T\)-matrix is given by:
\begin{equation}
    T(m\mathbf{k},m'\mathbf{k}';\epsilon) = \mathcal{V}^*_m(\mathbf{k}) G_{f}(\epsilon) \mathcal{V}_{m'}(\mathbf{k}').
\end{equation}
Here, $G_{f}(\epsilon) = [\epsilon - \epsilon_f - \Sigma_f(\epsilon)]^{-1}$ is the \textit{full} Green's function of the single $f$-impurity, dressed by its hybridization to the $c$-electron sea via its own self-energy, $\Sigma_f(\epsilon) = \sum_{\mathbf{k},m} |\mathcal{V}_m(\mathbf{k})|^2 g_{c}(m\mathbf{k},\epsilon)$. In the wide-band limit, $\Sigma_f(\epsilon) = -i\Delta(\epsilon)$, where $\Delta(\epsilon) = \pi V^2 N_{c}(\epsilon)$ and $N_c(\epsilon)$ is the density of states of $c$-electrons at the unit volume.

Now, we consider a finite concentration, $n_{\text{imp}} = N_{\text{imp}}/N$, of such impurities distributed randomly. Here, \(N\) is the number of unit cells in the system. In the dilute limit, we can neglect interference between scattering events from different impurities (the independent scattering approximation). The disorder-averaged $c$-electron self-energy is then given by $\Sigma_c = N_{\text{imp}} \langle T \rangle$. To ensure a well-defined thermodynamic limit, we must properly normalize the hybridization amplitude. Let $\mathcal{V}_m(\mathbf{k})$ be defined for a single impurity in a finite system of volume $N$. We then define a volume-independent amplitude $\tilde{\mathcal{V}}_m(\mathbf{k}) \equiv \sqrt{N}\mathcal{V}_m(\mathbf{k})$. This is equivalent to redefining the Bloch wave: $\ket{\psi_{\mathbf{k},m}} \to \ket{\tilde{\psi}_{\mathbf{k},m}} =\sqrt{N} \ket{\psi_{\mathbf{k},m}}$
The self-energy becomes:
\begin{equation}
    [\Sigma_c^R(\mathbf{k},\epsilon)]_{nn'} = n_{\text{imp}}\tilde{\mathcal{V}}^*_n(\mathbf{k}) G_f^R(\epsilon)  \tilde{\mathcal{V}}_{n'}(\mathbf{k}).
    \label{eq:app_self_energy_disorder}
\end{equation}
And everything in the self-energy is finite in the thermodynamic limit. 
This is the general form presented in the main text. The quasiparticle scattering rate is proportional to its imaginary part:
\begin{equation}
\begin{aligned}
     \Gamma_\mathbf{k}^{(n)} \propto -\mathrm{Im}[\Sigma_c^R(\mathbf{k},\epsilon)]_{nn} & = -n_{\text{imp}} |\tilde{\mathcal{V}}_n(\mathbf{k})|^2 \mathrm{Im}[G_f^R(\epsilon)]\\
     & =-n_{\text{imp}} |\tilde{\mathcal{V}}_n(\mathbf{k})|^2 \frac{\Delta(\epsilon)}{(\epsilon-\epsilon_f)^2+\Delta^2}.
\end{aligned}
\end{equation}

\subsection{Interacting $f$-electron System}

We now consider the case where $\hat H_f$ describes an interacting flat band, as in the periodic Anderson model. We can write the problem in a matrix form in the (f, c) basis:
\begin{equation}
    \begin{pmatrix}
    \epsilon - \epsilon_f - \Sigma_{f,\text{int}}^R(\epsilon) &  -\mathcal{V}(\mathbf{k})\\
     -\mathcal{V}^\dagger(\mathbf{k}) & \epsilon - \mathcal{H}^{(c)}(\mathbf{k})
    \end{pmatrix}
    \mathbf{G}^R(\mathbf{k},\epsilon) = \mathbf{I}.
\end{equation}
Here, $\Sigma_{f,\text{int}}^R$ is the proper self-energy of the $f$-electrons due to their local interactions (e.g., the Hubbard U), calculated \textit{for the isolated $f$-electron subsystem}. The hybridization matrix $\mathcal{V}(\mathbf{k})$ has components $[\mathcal{V}(\mathbf{k})]_{\sigma n} = \braket{d_\sigma|\psi_{\mathbf{k},n}}V$.

By inverting this matrix blockwise, we can solve for the $c$-electron Green's function, $G_c^R = [\epsilon - \mathcal{H}^{(c)} - \Sigma_c^R]^{-1}$. We immediately identify the $c$-electron self-energy as:
\begin{equation}
    \Sigma_c^R(\mathbf{k},\epsilon) = \mathcal{V}^\dagger(\mathbf{k}) \frac{1}{\epsilon - \epsilon_f - \Sigma_{f,\text{int}}^R(\epsilon)} \mathcal{V}(\mathbf{k}).
\end{equation}
Rewriting this in the $c$-electron band basis yields the same factorized form as Eq.~\eqref{eq:self_energy_general} in the main text:
\begin{equation}
    [\Sigma_c^R(\mathbf{k},\epsilon)]_{mm'} = \mathcal{V}_m^*(\mathbf{k}) g_{f,\text{int}}^R(\epsilon) \mathcal{V}_{m'}(\mathbf{k}),
    \label{eq:app_self_energy_interacting}
\end{equation}
where we have defined the local Green's function of the isolated, interacting f-system as $g_{f,\text{int}}^R(\epsilon) = [\epsilon - \epsilon_f - \Sigma_{f,\text{int}}^R(\epsilon)]^{-1}$.

At low temperatures, using the Luttinger's result\cite{luttingerpr}, the $f$-electron self-energy near the Fermi level behaves as $\mathrm{Im}[g_{f,\text{int}}^R(\epsilon)] \propto -{(\epsilon-\epsilon_F)^2}$. This implies that the $c$-electron scattering rate will also follow this quadratic energy dependence, but crucially, it will still be modulated by the momentum-dependent factor $|\mathcal{V}_m(\mathbf{k})|^2$. Thus, the nodal structure and the resulting superdiffusive scaling exponents remain robust.

\section{Evaluation of the conductivity integral} \label{app:conductivity_integral}
In this appendix, we provide the detailed derivations for the scaling of the dc, optical, and finite-temperature conductivities, which are presented as main results in the main text.

\subsection{DC conductivity}  
\label{app:dc_cond}
Our starting point is the expansion of the $c$-electron Green's function denominator around a point $\mathbf{k}_{\boldsymbol{\theta}}$ on the intersection of an equal-energy surface and the nodal manifold. As derived in the main text, this takes the form $v(\mathbf{k}_{\boldsymbol{\theta}}) \delta k_\parallel - i \Gamma_{\boldsymbol \theta} |\delta \mathbf{k}_\perp|^{2n}$, where $\delta k_\parallel$ is the momentum deviation parallel to the group velocity and $\delta \mathbf{k}_\perp$ is the deviation transverse to it.

The singular part of the dc conductivity integral from Eq.~\eqref{eq:kubo_sigma_epsilon} is therefore given by:
\begin{equation}
    \sigma_{xx}^{\text{sing}} \propto \int d\boldsymbol{\theta} \int d\delta k_\parallel \int d\delta \mathbf{k}_\perp  \frac{|v_x(\mathbf{k_{\boldsymbol{\theta}}})|^2 \theta(|\delta k_\parallel - 1/L_\parallel|)}{|v(\mathbf{k}_{\boldsymbol{\theta}}) \delta k_\parallel - i \Gamma_{\boldsymbol \theta} |\delta \mathbf{k}_\perp|^{2n}|^2},
\end{equation} \label{eq:detailed_integral}
where $\boldsymbol{\theta}$ parameterizes the $(D^S_{\text{node}})$-dimensional intersection manifold, and the dimension of the transverse momentum space is $D' = D^S - D^S_{\text{node}} $ and $\theta(x)$ is the Heaviside step function.

Integrating over $\delta \mathbf{k}_\perp$ gives 
\begin{equation}
    \begin{aligned}
         \int d\delta \mathbf{k}_\perp  \frac{|v_x(\mathbf{k_{\boldsymbol{\theta}}})|^2}{|- v(\mathbf{k}_{\boldsymbol{\theta}}) \delta k_\parallel + i \Gamma_{\boldsymbol \theta} |\delta \mathbf{k}_\perp|^{2n}|^2} = \frac{ C_{\boldsymbol \theta}}{|\delta k_\parallel|^{2 - \gamma_{\epsilon}}},
    \end{aligned}
\end{equation}
where the coefficient 
\begin{equation}
    C_{\boldsymbol{\theta}}(\epsilon) =  \frac{\pi^{\gamma_{S}+1} \left|v_x(\mathbf{k}_{\boldsymbol{\theta}})\right|^2}{2n \Gamma\left(\gamma_{\epsilon}\right) \sin\left(\frac{\pi}{2}  \gamma_\epsilon\right) \Gamma_{\boldsymbol \theta}^{\gamma_\epsilon} \left[v(\mathbf{k}_{\boldsymbol{\theta}})\right]^{2 - \gamma_\epsilon} }, 
\end{equation}
and the exponent $\gamma_\epsilon = \frac{D^S-D^S_{\text{node}}}{2n}$. This exponent is determined by the geometry of the intersection between the equal energy surface and the nodal manifold. 

The remaining integral over $\delta k_\parallel$ is regularized by the finite system size $L_\parallel$, which imposes an infrared cutoff $|\delta k_\parallel| \ge 1/L_\parallel$. This gives the scaling for the energy-resolved conductivity:
\begin{equation}
\begin{aligned}
    \sigma_{xx}(\epsilon) &\propto \int d\boldsymbol{\theta} |v_x|^2 C_{\boldsymbol{\theta}}\int d\delta k_\parallel \theta(|\delta k_\parallel - 1/L_\parallel|) |\delta k_\parallel|^{\gamma_\epsilon-2}\\
    &\propto L_\parallel^{1-\gamma_\epsilon}, \label{eq:app_sigma_epsilon}
\end{aligned}
\end{equation}
where we have defined the energy-resolved scaling exponent $\gamma_\epsilon = \frac{D^S-D^S_{\text{node}}}{2n}$, which is determined by the codimension of the nodal manifold's intersection with the equal-energy surface.

Notice that when doing the conductivity integral, we assume that the direction deviating from the nodal manifold lies in the plane perpendicular to $\mathbf{v}_{\boldsymbol{\theta}}$, which does not hold true for a tilted nodal manifold. This problem can be fixed by doing a integration variable substitution. Suppose the direction deviating from the nodal manifold can be decomposed as $\mathbf{e}_\perp' = a  \mathbf{e}_\parallel + b \mathbf{e}_\perp $, where $a,b$ are two real coefficients. Then we do variables transformation: $(\delta k_\parallel,\delta \mathbf{k}_\perp) \to (\delta k, \delta \mathbf{k}_\perp')$ which is a constant. The integral now is given by
\begin{equation}
\begin{aligned}
    \sigma_{xx}^{\text{sing}} &\propto \operatorname{Jacobian} \times \int d\boldsymbol{\theta} \int d\delta k_\parallel \int d\delta \mathbf{k}_\perp' \\ &\frac{|v_x(\mathbf{k_{\boldsymbol{\theta}}})|^2 \theta(|\delta k_\parallel - 1/L_\parallel|)}{|v(\mathbf{k}_{\boldsymbol{\theta}}) \delta k_\parallel - i \Gamma_{\boldsymbol \theta} |\delta \mathbf{k}_\perp'|^{2n}|^2}.
\end{aligned}
\end{equation}
Since $\delta k_\parallel$ and $\delta \mathbf{k}_\perp'$ are independent variables, this integral bares the same form with Eq.~(\ref{eq:detailed_integral}) and gives the same scaling with system size.

The final transport exponents are determined by how this result is integrated over energy.

\paragraph{Zero Temperature:}
At $T=0$, transport is confined to the Fermi surface ($S=F$). The scaling of the total conductance is therefore directly given by the exponent evaluated at the Fermi energy:
\begin{equation}
    \gamma_{\text{low}} = \gamma_{\epsilon_F} = \frac{D^F-D^F_{\text{node}}}{2n}.
\end{equation}

\paragraph{High Temperature:}
At high temperatures, thermal smearing ensures that states across the entire Brillouin zone contribute to transport. The total dc conductivity is obtained by integrating Eq.~\eqref{eq:app_sigma_epsilon} over all energies: $\sigma_{\text{tot}} \propto \int d\epsilon \, L_\parallel^{1-\gamma_\epsilon}$.

For nodal structures with dimension $D_{\text{node}} \ge 1$ (lines or surfaces), the intersection geometry is stable over a finite energy window. The dimension of the intersection of two generic manifolds is given by $D^S_{\text{node}} = D^S + D_{\text{node}} - D$. We can rewrite the exponent $\gamma_\epsilon$ in a remarkable way:
\begin{equation}
\begin{aligned}
     \gamma_\epsilon &= \frac{D^S - D^S_{\text{node}}}{2n} = \frac{D^S - (D^S + D_{\text{node}} - D)}{2n}\\
     &= \frac{D - D_{\text{node}}}{2n}.
\end{aligned}
\end{equation}
Crucially, this result is independent of the energy $\epsilon$. Therefore, the exponent for the total, energy-integrated conductivity is simply this constant value:
\begin{equation}
    \gamma_{\text{high}} = \frac{D-D_{\text{node}}}{2n}.
\end{equation}
This demonstrates the unifying geometric principle: the transport exponent is always determined by the codimension of the nodal manifold with respect to the manifold of states contributing to transport (the Fermi surface at low $T$, and the full Brillouin zone at high $T$). The derivation for isolated nodal points ($D_{\text{node}}=0$) is more subtle but yields a result consistent with this general formula.

\subsection{Weak localization correction} \label{app:WL correction}
The weak localization correction to the Drude conductivity at frequency $\omega$ is given by 
\begin{equation}
    \delta\sigma_{\alpha\alpha}^{\text{WL}} = -\frac{4e^2}{h} \sum_{\mathbf{q}} \frac{D_{\alpha \alpha}(\mathbf{q})}{-i \omega +\sum_\beta D_{\beta\beta} q_{\beta}^2},
\end{equation}
where $D_{\alpha\alpha}$ is the anisotropic Drude diffusion constant. In the two-component model, the $c$-electron will show anomalous diffusion, which reveals itself in the momentum-dependence of $D_{\parallel}$ such that $D_\parallel(q_\parallel) = D_\parallel^0 (q_\parallel l_0)^{\gamma -1}$ and $l_0$ is the short-distance cutoff for the diffusion process. The diffusion constant along other directions remains momentum-independent, which we denote as $D_\perp$. The correction for the dc conductivity $(\omega = 0)$ is given by the integral
\begin{equation}
    \delta\sigma^{\text{WL}}_\parallel = -\frac{4e^2}{h} \sum_{\mathbf{q}} \frac{D^0_\parallel(q_\parallel l_0)^{\gamma-1}}{D^0_\parallel(q_\parallel l_0)^{\gamma-1} q_\parallel^2 + D_\perp|\mathbf{q}_\perp|^2}.
\end{equation}
By first integrating over $\mathbf{q}_\perp$ and take a cutoff at small $q_\parallel$: $|q_\parallel-1/L_\parallel|>0$, we can get the scaling of $\delta \sigma^{\text{WL}}_{\parallel}$:
\begin{equation}
    \delta \sigma^{\text{WL}}_{\parallel} \sim - L_\parallel^{1-\frac{(\gamma+1)(D-1)}{2}}.
\end{equation}

\subsection{Optical conductivity} \label{app:optical}
The low-frequency optical conductivity at $T=0$ is given by the Kubo formula. Within the Drude approximation, the key part of the integrand involves the product of a retarded and an advanced Green's function, $G^R(\mathbf{k}, \epsilon_F) G^A(\mathbf{k}, \epsilon_F - \omega)$. Expanding around a node gives the singular contribution:
\begin{equation}
\begin{aligned}
    \sigma_{xx}(\omega) \propto &\mathrm{Re} \bigg[ \int d\boldsymbol{\theta} \, d\delta k_\parallel \, d\delta \mathbf{k}_\perp \times \\
    &\frac{|v_x|^2}{(-v \delta k_\parallel + i \Gamma_{\boldsymbol{\theta}} |\delta \mathbf{k}_\perp|^{2n})(-\omega - v \delta k_\parallel - i \Gamma_{\boldsymbol{\theta}} |\delta \mathbf{k}_\perp|^{2n})} \bigg].
\end{aligned}
\end{equation}
The finite frequency $\omega$ prevents the two poles from coinciding, regularizing the integral. 
Integrating over the momentum variable yields the power-law divergence for $\omega \to 0$:
\begin{equation}
    \mathrm{Re}[\sigma_{xx}({\omega})] \sim |\omega|^{\gamma_{\text{low}}-1}.
\end{equation}

\subsection{Finite temperature conductivity}
\label{app:finite temperature cond}

We now analyze the finite-temperature resistivity, specifically for the case where the $f$-electron bath consists of  disorder, and we introduce residual electron-electron interactions among the $c$-electrons. Assuming the scattering processes are independent (Matthiessen's rule)\cite{akkermans2007mesoscopic}, the total $c$-electron self-energy is $\Sigma_{\text{tot}}^R = \Sigma_{\text{hyb}}^R + \Sigma_{ee}^R$. 

The residual electron-electron interactions provide the dominant low-temperature dependence. For a Fermi liquid, this interaction introduces a scattering rate\cite{luttingerpr,coleman2015introduction} 
\begin{equation}
    \tau_{ee}^{-1} = -2\mathrm{Im}[\Sigma_{ee}^R]  = 2\pi (\frac{w^2}{16 E_F}) [\frac{(\epsilon-E_F)^2 + (\pi k_B T)^2}{2}],
\end{equation}
where $w$ is dimensionless parameter related to the matrix element of the two-body potential.

The denominator of the Green's function in the conductivity integral is now modified to $|v \delta k_\parallel - i \Gamma|\delta\mathbf{k}_\perp|^{2n} - i/(2\tau_{ee})|^2$. The temperature-dependent term $1/\tau_{ee}$ acts as the dominant infrared cutoff for the integral, regularizing the divergence that would otherwise occur near the nodal manifold.

Conceptually, the contribution to the conductivity comes from two parts: the region in momentum space near the node and away from the node (where $\Sigma^R_{\text{hyb}}$ can be treat as momentum-independent). 

The contribution from the region near the node can be calculated following the same scaling logic as the dc conductivity calculation (Appendix~\ref{app:dc_cond}), but with the cutoff now set by $T^2$ instead of $1/L$. Now let's analyze the integral. First integrating out the $\delta k_\parallel$ variable gives us:
\begin{equation}
\begin{aligned}
   \sigma_{\text{node}}= &\int d\boldsymbol{\theta} d\delta\mathbf{k}_\perp  \frac{\pi}{v_{\boldsymbol{\theta}}} \frac{1}{\Gamma_{\boldsymbol{\theta}} |\delta \mathbf{k}_\perp|^{2n}+1/\tau_{ee}} \\
    =&\int d\boldsymbol{\theta} \frac{\pi}{v_{\boldsymbol{\theta}}}\Gamma_{\boldsymbol{\theta}}^{-\gamma_{\text{low}}} (\frac{1}{\tau_{ee}})^{\gamma_{\text{low}}-1} \int d\mathbf{x}  \frac{1}{1+|\mathbf{x}|^{2n}} \\
    \propto& \overline{1/(v_{\boldsymbol{\theta} } \Gamma_{\boldsymbol{\theta}}^{\gamma_{\text{low}}})} (\frac{1}{\tau_{ee}})^{\gamma_{\text{low}}-1}.
\end{aligned}
\end{equation}
Here $\overline{f(\boldsymbol{\theta})}$ is the typical value of the function $f$ on the intersection of the nodal manifold and the Fermi surface.
To get the finite temperature conductivity, we need integrate over the energy $\epsilon$:
\begin{equation}
\begin{aligned}
    \sigma_{\text{node}}(T) &= \int d\epsilon (-\frac{\partial n_F}{\partial \epsilon}) \sigma_{\text{node}}(\epsilon,T)\\
    & \propto \frac{1}{(v\Gamma^{\gamma_{\text{low}}})} \left[\frac{(k_BT)^2}{E_F} \right]^{(\gamma_{\text{low}}-1)} \label{eq:sigma_node}
\end{aligned}
\end{equation}
As we see, at low temperatures $\sigma_{\text{node}}(T)$ diverges and dominate the conductivity.
Inverting this gives the resistivity, $\rho(T) = 1/\sigma(T)$, which exhibits the anomalous temperature dependence:
\begin{equation}
    \rho(T) \propto T^{2(1-\gamma_{\text{low}})}.
\end{equation}
For the superdiffusive case $\gamma_{\text{low}}=1/2$, this yields a linear---in---$T$ resistivity, $\rho(T) \propto T$.

When temperature rises, the large, uniform thermal scattering will swamp the momentum-dependent scattering from the hybridization, and the temperature dependence of the conductivity will deviate from Eq.~(\ref{eq:sigma_node}). Before we derive the finite temperature conductivity, let us rewrite the Kubo formula. Notice the relation
\begin{equation}
\begin{aligned}
      &G^R_{c,m}(\mathbf{k},\epsilon) G^A_{c,m}(\mathbf{k},\epsilon) = |G^R_{c,m}(\mathbf{k},\epsilon)|^2 \\
      &\equiv \left| \frac{1}{\epsilon-E_m(\mathbf{k})-\Sigma^R_{c,m}(\mathbf{k},\epsilon)}  \right|^2 \\
      &=\frac{1}{\text{Im}\Sigma^R_{c,m}} \text{Im}\frac{1}{\epsilon - E_{m}(\mathbf{k}) - \Sigma_{c,m}^R(\mathbf{k},\epsilon)}\\
      &\equiv \frac{1}{-2\text{Im}\Sigma^R_{c,m}(\mathbf{k},\epsilon)} A_m(\mathbf{k},\epsilon),
\end{aligned}
\end{equation}
where $A_m(\mathbf{k},\epsilon)$ is the spectral function. If we are at the low-temperature regime and consider the scattering is weak, the broadening of the Lorentzian-like spectral function is small compared the the Fermi energy, and it can be approximated by the delta function
\begin{equation}
    A_m(\mathbf{k},\epsilon) \approx \delta(\epsilon-E_m(\mathbf{k})).
\end{equation}
And 
\begin{equation}
    -2\text{Im}\Sigma^R_{c,m}(\mathbf{k},\epsilon) = \tau_{ee}^{-1} + \tau_{\text{hyb}}^{-1},
\end{equation}
where
\begin{equation}
\begin{aligned}
    \tau_{\text{hyb}}^{-1} &= -2 \text{Im}\Sigma^R_{\text{hyb}}  \\
    &= n_{\text{imp}} |\braket{d|\tilde{\psi}_{\mathbf{k},m}}|^2 \frac{V^2\Delta(\epsilon)}{(\epsilon-\epsilon_f)^2+\Delta^2} \\
    &=\Gamma(\epsilon,V) |\braket{d|\tilde{\psi}_{\mathbf{k},m}}|^2.
\end{aligned}
\end{equation}
The nodal impurity scattering strength is dependent on the energy $\epsilon$ and the hybridization strength $V$.

Now the energy-resolved conductivity is given by 
\begin{equation}
    \begin{aligned}
        \sigma(T) &= \int d \epsilon \sigma_{\mu\mu}(\epsilon)(-\frac{\partial n_F(\epsilon,T)}{\partial \epsilon}) \\
        & = \int d \epsilon (-\frac{\partial n_F(\epsilon,T)}{\partial \epsilon}) \\
        & \times\int_{BZ} d\mathbf{k} \frac{v^2(\mathbf{k})}{\tau_{ee}^{-1}(\epsilon) + \tau_{\text{hyb}}^{-1}(\mathbf{k},\epsilon)} \delta(\epsilon-E_m(\mathbf{k})) \\
        & = \int_{BZ} d \mathbf{k} \left(-\frac{\partial n_F(\epsilon,T)}{\partial \epsilon}\right)_{\epsilon = E_m(\mathbf{k})} \\
        &\times\frac{v^2(\mathbf{k})}{\tau_{ee}^{-1}(E_m(\mathbf{k})) + \tau_{\text{hyb}}^{-1}(\mathbf{k},E_m(\mathbf{k}))}.
    \end{aligned}
\end{equation}
At low temperatures, we approximate $ (-\frac{\partial n_F(\epsilon,T)}{\partial \epsilon})$ as a narrowly peaked function centered at the Fermi energy $E_F$ and this effectively restricts the $\mathbf{k}$-space integral to a thin shell with thickness $~k_B T$ around the Fermi surface. Within this approximation, we can do the following replacement
\begin{equation}
    E_m(\mathbf{k}) \to E_F,
\end{equation}
and we are left with an integral over the Fermi surface
\begin{equation}
    \sigma(T) \propto \int_{FS} d S_{\mathbf{k}} \frac{v^2(\mathbf{k})}{\tau_{ee}^{-1}(E_F) + \tau_{\text{hyb}}^{-1}(\mathbf{k},E_F)}.
\end{equation}
The denominator of the integrand is
\begin{equation}
\begin{aligned}
        \tau_{ee}^{-1}(E_F) &= 2\pi \frac{w^2}{16 E_F} \frac{(\pi k_B T)^2}{2} \\
        \tau_{\text{hyb}}^{-1}(\mathbf{k},E_F) & = \Gamma(E_F,V) |\braket{d|\tilde{\psi}_{m\mathbf{k}}}|^2  \\
\end{aligned}
\end{equation}
and
\begin{equation}
\begin{aligned}
       &\tau_{ee}^{-1}(E_F) + \tau_{\text{hyb}}^{-1}(\mathbf{k},E_F) \\
    =& {\Gamma}(E_F,V)(|\braket{d|\tilde{\psi}_{\mathbf{k},m}}|^2 + \pi (\frac{w^2}{16}) [\frac{\pi k_B T}{\sqrt{E_F \Gamma}}]^2)
\end{aligned}
\end{equation}
Here, we define $\Gamma(E_F,V)$, which has the dimension of energy, as the nodal disorder scattering strength $\Gamma_{\text{node}}$.
And the conductivity integral is given by
\begin{equation}
    \sigma(T) \propto \frac{1}{\Gamma_{\text{node}}} \int_{FS}d S_{\mathbf{k}} \frac{v^2(\mathbf{k})}{|\braket{d|\tilde{\psi}_{\mathbf{k},m}}|^2 + \pi(\frac{w^2}{16})x^2} \equiv \frac{1}{\Gamma_{\text{node}}} F(x),
\end{equation}
where $x = \frac{\pi k_B T}{\sqrt{E_F \Gamma_{\text{node}}}}$, and $F(x) = \int_{FS} d S_{\mathbf{k}} \frac{v^2(\mathbf{k})}{|\braket{d|\tilde{\psi}_{\mathbf{k},m}}|^2 + \pi (\frac{w^2}{16})x^2}$. 

\begin{figure}[t]
    \includegraphics[width=0.95\linewidth]{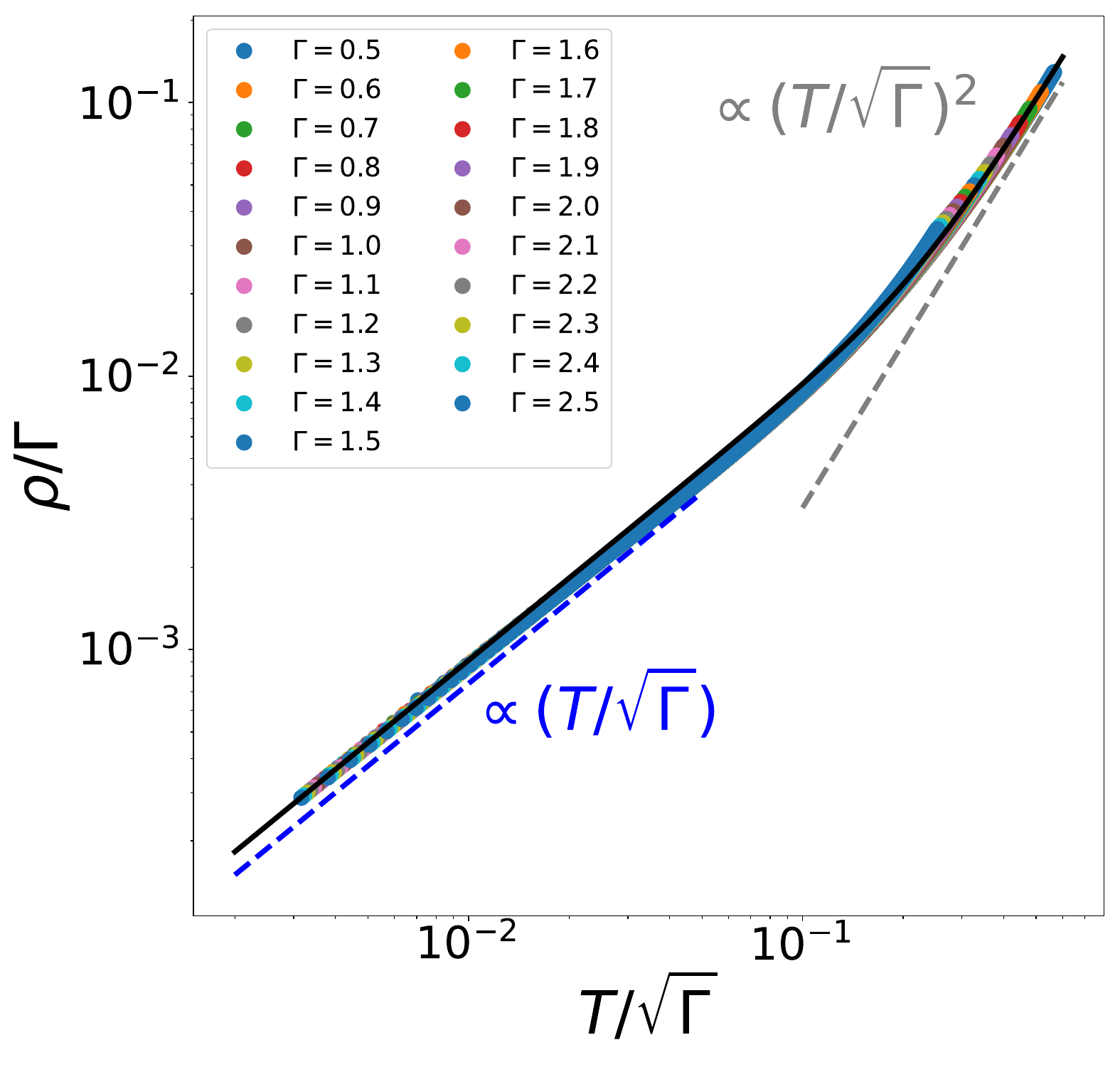}
    \caption{\textbf{Universal scaling collapse for the temperature dependence of resistivity}
    The figure demonstrates the universal scaling behavior of the resistivity in the 2D square lattice model with nodal lines ($\gamma_{\text{low}} = 1/2$) in the presence of electron-electron interaction. The strength of the hybridization is characterized by $\Gamma$. The rescaled resistivity, $\rho(T) / \Gamma$, is plotted against the rescaled temperature, $x = T/\sqrt{\Gamma}$. Data with different hybridization strength collapses onto a single universal curve. The solid back line is a fit to the universal scaling function, $F(x) = (C_1 x^3+C_2 x^6)^{1/3}$. The dashed lines show the theoretically expected asymptotic power
laws: (i) linear-in-temperature resistivity at low temperature (the blue dashed line); (ii) the $T^2$ law for usual Fermi liquid systems.     
    }
    \label{fig:temperature_transition}
\end{figure}

When $x$ is small, the integral is dominated by the regions near the node, and we have
\begin{equation}
    \begin{aligned}
        F(x) &\approx \int d \delta \mathbf{k}_\perp \frac{v_0^2}{ |\delta \mathbf{k}_\perp|^{2n} + A x^2} \\
        & \propto (Ax^2)^{\gamma_\text{low}-1}\quad x\to 0.
    \end{aligned}
\end{equation}
And it matches the $T^{2(\gamma_{\text{low}}-1)}$ law as we calculated using the Kubo formula. As the temperature rises, the uniform thermal scattering will dominate and we have
\begin{equation}
    F(x) \to x^{-2}, \quad x\to \infty,
\end{equation}
and this corresponds to $\sigma(T) \sim T^{-2}$. The asymptotic behavior is validated by the numeric, see Fig.~\ref{fig:temperature_transition}.

\section{Proof of the irrep}
\label{sec:irrep}
In this section, we prove that due to the symmetry constraint, $\hat{f}_{\mathbf{R}}$ and $\hat{d}_{\mathbf{R}}$ transform under the same irrep of the site-symmetry group.

By applying $h \in \mathcal{G}(\mathbf{r}_f)$ to the $f$-electron operators with a general lattice site, we have
\begin{equation}
    \hat{h} \hat{f}^\dagger_{\mathbf{R}} \hat{h}^{-1} = \chi_f(h) \hat{f}^\dagger_{\mathbf{R} + \mathbf{l}},
\end{equation}
where $\mathbf{l}$ is a Bravais lattice vector. 
Now apply $\hat{h}$ to $\hat{H}_{\text{hyb}}$ and assume that $[\hat{h},\hat{H}_{\text{hyb}}] = 0$, we have
\begin{align*}
\hat{h} H_{\mathrm{hyb}} \hat{h}^{-1}
&=V\sum_{\mathbf{R}}\!\left((\hat{h} \hat f_{\mathbf{R}}^{\dagger}\hat{h}^{-1})(\hat{h} \hat d_{\mathbf{R}}\hat{h}^{-1})+\text{h.c.}\right) \\
&=V\sum_{\mathbf{R}}\!\left(\chi_f(h)\, \hat f_{\mathbf{R}+\mathbf{l}}^{\dagger}\, (\hat{h} \hat d_{\mathbf{R}}\hat{h}^{-1})+\text{h.c.}\right) \\
&=V\sum_{\mathbf{R}}\!\left(\chi_f(h)\, \hat f_{\mathbf{R}}^{\dagger}\, (\hat{h} \hat d_{\mathbf{R}-\mathbf{l}}\hat{h}^{-1})+\text{h.c.}\right).
\end{align*}
Since $\hat{H}_{\text{hyb}}$ is invariant under $\hat{h}$, we have
\begin{equation}
    \sum_{\mathbf{R}} \hat{f}_{\mathbf{R}}^\dagger [\chi_f(h) (\hat{h}\hat{d}_{\mathbf{R}-\mathbf{l}} \hat{h}^{-1}) - \hat{d}_{\mathbf{R}}] + h.c. = 0.
\end{equation}
Since all the $f$-electron operators $\{\hat{f}^\dagger_{\mathbf{R}}\}$ are independent, we must have
\begin{equation}
    \chi_f(h) (\hat{h} \hat{d}_{\mathbf{R}-\mathbf{l}} \hat{h}^{-1} ) = \hat{d}_{\mathbf{R}},
\end{equation}
which is equivalent to 
\begin{equation}
    \hat{h} \hat{d}_{\mathbf{R}} \hat{h}^{-1} = \chi^*_f(h) \hat{d}_{\mathbf{R} + \mathbf{l}}.
\end{equation}
This completes the proof.

\section{Real gauge and node order in graphene} \label{sec:proof_linear_momentum}
\subsection{Real gauge}
We define a new set of basis by the unitary transformation:
\begin{equation}
\ket{\tilde{c}_{i}} = \sum_{j=A,B}U_{i,j}\ket{c_{j}}.
\end{equation}
The $PT$-operation is the $\ket{\tilde{c}_{i}}$ basis is given by
\begin{equation}
    \tilde{(PT)} = U^*\tau_xU^\dagger \mathcal{K} = U^*\tau_x (U^*)^T \mathcal{K}.
\end{equation}
Now our goal is find a matrix $U^*$, such that under the congruent transformation the Pauli matrix $\tau_x$ becomes an identity matrix. One can check the matrix $U^*$ is given by
\begin{equation}
    U^* = \frac{1}{\sqrt{2}}\begin{pmatrix}
        1 &1\\ i &-i
    \end{pmatrix}.
\end{equation}
And we have $\tilde{(PT)} =\mathcal{K}$. Since the graphene Hamiltonian has the $PT$-symmetry, in the new basis the Bloch Hamiltonian matrix $\tilde{H}(\mathbf{k})$ can be chosen to be real. 

\subsection{Proof of the order of node}

Near the Dirac point $\mathbf{K}=(\frac{4\pi}{3},\frac{2\pi}{3})$, we can expand the function $f(\mathbf{k})$:
\begin{equation}
    f(\delta \mathbf{k}+\mathbf{K}) = i \exp(i{2\pi}/{3})\delta k_1 + i \delta k_2.
\end{equation}
The corresponding pseudo-spin for the eigenvector of the upper band is given by 
\begin{equation}
    \mathbf{S}(\delta\mathbf{k}) = (\frac{\sqrt{3}}{2} \delta k_1,\delta k_2-\frac{1}{2}\delta k_1) / \mathcal{N},
\end{equation}
where $\mathcal{N} = (\delta k_1^2+\delta k_2^2-\delta k_1\delta k_2)^{1/2}$ is the normalization constant.
Without loss of generality, we consider the bond impurity, whose form factor is given by $\boldsymbol{\phi} = (1,1)/\sqrt{2}$. The corresponding pseudo-spin is given by
\begin{equation}
    \mathbf{d} = (1,0). 
\end{equation}
The relative angle between these two pseudo-spins is given by 
\begin{equation}
    \theta(\delta\mathbf{k}) = \arctan(\frac{\sqrt{3}/2}{\tan \phi-1/2}),
\end{equation}
where $\tan \phi = \delta k_2 /\delta k_1$. The overlap vanishes when these two pseudo-spins are anti-parallel, $\theta(\delta \mathbf{k}_0) = \pi$. Expanding $\theta(\delta \mathbf{k})$ around the node gives us 
\begin{equation}
    \theta(\delta\mathbf{k}) = \pi -\frac{2}{\sqrt{3}} \delta \phi + O(\delta \phi^2).
\end{equation}
According to the definition of $\phi$, the momentum deviation from the node $\tilde{\delta k}$ is given by
\begin{equation}
    \tilde{\delta k} = (\delta k_1^2 + \delta k_2^2)^{1/2} \times \delta \phi \propto \delta \phi.
\end{equation}

\section{Topological conservation law}\label{app:topo_conservation}
In this section, we prove the ``Topological conservation law" in Eq.~(\ref{eq:topo_conservation_law}). Let $\ket{\psi_{n,\mathbf{k}}}$ be the normalized Bloch state for a non-degenerate band over a closed 2D manifold, and let this band have a Chern number $C$. The proof relies on partitioning the manifold into two regions. Region I consists of unions of $N$ small, disjoint disks, $D_i$, each centered around a node $\mathbf{k}_i$. The boundary of each disk is the loop $c_i$. Within each disk, we can adapt a smooth gauge, and let us call this gauge-$I$, and the wavefunctions in this gauge are $\ket{\psi_{n,\mathbf{k}}^I}$. Region II is the rest of the manifold, $S^2 / (\cup_i D_i)$ and in this region, the inner product, $\braket{d|\psi_{n,\mathbf{k}}}$ is never zero. The boundary of Region II is the set of all loos $c_i$, but transversed in the opposite direction.

We observe that the gauge choice $I$ cannot be continued to Region II: because the inner product is non-zero and has non-zero winding number, it will be ill-defined at some point in this region. So we need a new gauge $II$ to untwist the phases in the wavefunction. This is achieved by defining
\begin{equation}
    \ket{\psi^{II}_{n,\mathbf{k}}} = \frac{\braket{\psi_{n,\mathbf{k}}|d}}{|\braket{\psi_{n,\mathbf{k}}|d}|} \ket{\psi_{n,\mathbf{k}}}.
\end{equation}
This is possible because $|\braket{\psi_{n,\mathbf{k}}|d}|$ is never zero in Region II. Since $\ket{d}$ is globally smooth, single-valued function, $\ket{\psi^{II}_{n,\mathbf{k}}}$ is also smooth and single-valued in Region II. 

The Chern number is defined as the integral of the Berry curvature over the entire sphere. We can split the integral into two parts\cite{bernevig2013topological}:
\begin{equation}
\begin{aligned}
        C &= \frac{1}{2\pi } \int_{S^2} d^2 k \mathcal{F} \\
        &= \frac{1}{2\pi } ( \sum_i \int_{D_i} d^2 k \mathcal{F} + \int_{\text{Region II}} d^2k \mathcal{F} )\\
        &=\frac{1}{2 \pi } (\sum_i \oint_{c_i}d\mathbf{k}\cdot (\mathbf{A}^I-\mathbf{A}^{II})).
\end{aligned}
\end{equation}
In the last equation, we used the Stokes' theorem and used the fact that boundary of Region II is the set of all loos $c_i$, but transversed in the opposite direction. By noticing that 
\begin{equation}
    \mathbf{A}^{II} = \mathbf{A}^I -\nabla \chi(\mathbf{k}),
\end{equation}
where $e^{i \chi (\mathbf{k})} \equiv\frac{\braket{\psi_{n,\mathbf{k}}|d}}{|\braket{\psi_{n,\mathbf{k}}|d}|} $. So the Chern number is given by $C = \frac{1}{2\pi} \sum_i\oint _{c_i}d\mathbf{k}\cdot \nabla \chi(\mathbf{k}) = \frac{1}{2\pi i}\oint_{c_i} d \log(\braket{\psi_{n,\mathbf{k}}|d})=\sum_i w_i$.

\section{Effect of non-nodal perturbation}\label{app:non-nodal}
Suppose the relevant non-nodal perturbation is from ordinary disorder given in Sec.~\ref{sec:non-nodal}. The relavant scattering rate $\Gamma_{\text{dis}} = 2\pi N_c(E_F) W^2$. Together with the finite size $L_\parallel$, now we have an effective cutoff given by $|v/L_\parallel + i \Gamma_{\text{dis}}|$ for the integral of $\delta k_\parallel$ in Eq. (\ref{eq:app_sigma_epsilon}). Evaluating Eq.~(\ref{eq:app_sigma_epsilon}) with the given cutoff gives us 
\begin{equation}
\begin{aligned}
     \sigma(L,\Gamma_{\text{dis}}) \propto &(v^2/L_\parallel^2+\Gamma_{\text{dis}}^2)^{(\gamma_{\text{low}}-1)/2} .
\end{aligned}
\end{equation}
And since $G_{\parallel} = \sigma_{\parallel} S_\perp/ L_\parallel$, we have 
\begin{equation}
\begin{aligned}
    G(L,\Gamma_{\text{dis}}) &\propto  \Gamma_{\text{dis}}^{\gamma_{\text{low}}} \frac{(\Gamma_{\text{dis}}L)^{-\gamma_{\text{low}}}}{\left[v^2 + (\Gamma_{\text{dis}}L)^{2}\right]^{1/2(1-\gamma_{\text{low}})}}\\
    &=   \Gamma_{\text{dis}}^{\gamma_{\text{low}}} F(\Gamma_{\text{dis}} L),
\end{aligned}
\end{equation}
where $F(x) = \frac{x^{-\gamma_{\text{low}}}}{[v^2 + x^2]^{1/2(1-\gamma_{\text{low}})}}$. We see there is a transition from the superdiffusive transport, $F(x) \sim x^{-\gamma_{\text{low}}} (x \ll 1)$, to diffusive transport, $F(x) \sim x^{-1}(x \gg 1)$ when $x \sim 1$. That corresponds to $L^* \sim 1/\Gamma_{\text{dis}}$.

\section{Numerical Simulation Details}
\label{app:numerics}

The numerical transport calculations presented in this work were performed using the open-source quantum transport package Kwant~\cite{KwantRef}. We simulated two-terminal conductance through quasi-one-dimensional systems, as illustrated in the main text figures. This appendix provides the specific models and parameters used.

\subsection{Disordered $f$-electron Models on Lattices}

For the models on the square, cubic, and honeycomb lattices, we implemented the c-f hybridization model directly. The procedure involves two steps:
\begin{enumerate}
    \item A tight-binding model for the $c$-electrons is built on the primary lattice. For all cases, we use a simple nearest-neighbor hopping with amplitude $t=1$.
    \item A second set of lattice sites for the $f$-electrons is created, with one f-site per unit cell at a specific Wyckoff position.
\end{enumerate}

The disorder is implemented by modulating both the on-site energy and the hybridization strength of the $f$-electrons with the same random variable, $\xi_\mathbf{R}$, drawn from a uniform distribution $[-W, W]$ with disorder strength as give in the main text. The Hamiltonian terms associated with the f-site at $\mathbf{R}$ are:
\begin{align}
    \hat{H}_{\text{f-site}, \mathbf{R}} &={} (\epsilon_F + \xi_\mathbf{R}) \hat{f}^\dagger_\mathbf{R} \hat{f}_\mathbf{R} \nonumber \\ 
    & + \sum_{\boldsymbol{\delta},a} (\xi_\mathbf{R} V_a(\boldsymbol{\delta}) \hat{f}^\dagger_\mathbf{R} \hat{c}_{\mathbf{R}+\boldsymbol{\delta},a} + \text{h.c.}).
\end{align}
This ensures that the f-sites only scatter $c$-electrons when the random potential is non-zero. The vectors of hybridization coefficients, $V_a(\boldsymbol{\delta})$, which define the Hybridization Orbital, were chosen as follows for each case.
\begin{itemize}
    \item \text{Square Lattice (2D):}
    \begin{itemize}
        \item \text{Symmetric (Plaquette Center):} $V = -(0.5, 0.5, 0.5, 0.5)$. The four components correspond to the four corner sites of the plaquette, ordered counter-clockwise.
        \item \text{Asymmetric (Generic Wyckoff position):} $V = (0.21, 0.35, 0.88, 0.25)$.
    \end{itemize}

    \item \text{Cubic Lattice (3D):}
    \begin{itemize}
        \item \text{Symmetric (Body Center):} $V$ is a vector of 8 identical components, $V_i = -0.5$.
        \item \text{Asymmetric (Generic Wyckoff position):} $V = (-0.67201,-0.31163,-0.35527, -0.41728\\,-0.37583,-0.33265,-0.29311,-0.4645)$ .
    \end{itemize}

    \item \text{Graphene (Honeycomb Lattice):}
    \begin{itemize}
        \item \text{Symmetric (Bond Center):} $V = -(1, 1)$ for the two sites sharing the bond.
        \item \text{Symmetric (Plaquette Center):} $V$ is a vector of 6 identical components for the sites around the hexagon, $V_i = -1/\sqrt{3}$.
        \item \text{Non PT-symmetric (Generic Wyckoff position):} For both cases, the symmetric vector was perturbed with small random numbers to break the protecting symmetry while keeping the norm approximately constant.
    \end{itemize}
\end{itemize}

\subsection{Multi-Weyl Semimetal (Continuum Discretization)}

For the multi-Weyl simulations, we used an effective continuum model which is then discretized onto a cubic lattice. The low-energy $\mathbf{k}\cdot\mathbf{p}$ Hamiltonian is given by:
\begin{equation}
    H^{\text{eff}}_{\text{k}\cdot \text{p}}(\mathbf{k}) = \begin{pmatrix}
        m-\beta k^2 & D k_z (k_x -ik_y)^n\\
        Dk_z(k_x + ik_y)^n & -(m - \beta k^2)
    \end{pmatrix} \label{eq:weyl_kp},
\end{equation}
where $n=2$ for the double-Weyl and $n=3$ for the triple-Weyl case. We used parameters $(m, \beta, D) = (1, 1, 1)$. This continuum model was automatically discretized onto a tight-binding model using the \texttt{kwant.continuum.discretize} function.

The disorder in this effective model was implemented by adding on-site potential terms that project onto a specific ``channel'' in the pseudospin basis. The disorder Hamiltonian is $\hat{V}_{\text{dis}} = \sum_{\mathbf{R}} \xi_{\mathbf{R}} \hat{P}_{\mathbf{R}}$, where $\hat{P}_\mathbf{R}$ is a projection operator at site $\mathbf{R}$ and $\xi_\mathbf{R}$ is the random variable drawn from the same distribution as above.

\begin{itemize}
    \item \text{Symmetric (Higher-Order Node):} The potential projects onto a single basis state, corresponding to a symmetric Hybridization Orbital: $\hat{P} = \ket{\uparrow}\bra{\uparrow}$.
    \item \text{Asymmetric (First-Order Node):} The potential projects onto a state that is a superposition of the two basis states, breaking the rotation symmetry: $\hat{P} = \ket{\psi}\bra{\psi}$ where $\ket{\psi} \propto \ket{\uparrow} + 0.7\ket{\downarrow}$.
    \item \text{No Node (Localization):} The potential is a simple scalar potential, proportional to the identity matrix $\mathbf{I}_{2\times2}$.
\end{itemize}

\bibliography{ref}

@article{RVprx25,
  title = {Generalized Hydrodynamics: A Perspective},
  author = {Doyon, Benjamin and Gopalakrishnan, Sarang and M\o{}ller, Frederik and Schmiedmayer, J\"org and Vasseur, Romain},
  journal = {Phys. Rev. X},
  volume = {15},
  issue = {1},
  pages = {010501},
  numpages = {28},
  year = {2025},
  month = {Jan},
  publisher = {American Physical Society},
  doi = {10.1103/PhysRevX.15.010501},
  url = {https://link.aps.org/doi/10.1103/PhysRevX.15.010501}
}

@article{hybridization_nodes,
  title = {Heavy-fermion metals with hybridization nodes: Unconventional Fermi liquids and competing phases},
  author = {Weber, Heidrun and Vojta, Matthias},
  journal = {Phys. Rev. B},
  volume = {77},
  issue = {12},
  pages = {125118},
  numpages = {15},
  year = {2008},
  month = {Mar},
  publisher = {American Physical Society},
  doi = {10.1103/PhysRevB.77.125118},
  url = {https://link.aps.org/doi/10.1103/PhysRevB.77.125118}
}

@article{Papic23,
  title = {Superdiffusive Energy Transport in Kinetically Constrained Models},
  author = {Ljubotina, Marko and Desaules, Jean-Yves and Serbyn, Maksym and Papi\ifmmode \acute{c}\else \'{c}\fi{}, Zlatko},
  journal = {Phys. Rev. X},
  volume = {13},
  issue = {1},
  pages = {011033},
  numpages = {14},
  year = {2023},
  month = {Mar},
  publisher = {American Physical Society},
  doi = {10.1103/PhysRevX.13.011033},
  url = {https://link.aps.org/doi/10.1103/PhysRevX.13.011033}
}

@article{Romain25,
  title = {Superdiffusive Transport in Chaotic Quantum Systems with Nodal Interactions},
  author = {Wang, Yu-Peng and Ren, Jie and Gopalakrishnan, Sarang and Vasseur, Romain},
  journal = {Phys. Rev. Lett.},
  volume = {135},
  issue = {16},
  pages = {166303},
  numpages = {6},
  year = {2025},
  month = {Oct},
  publisher = {American Physical Society},
  doi = {10.1103/xx9z-4j6c},
  url = {https://link.aps.org/doi/10.1103/xx9z-4j6c}
}

@article{Chen24,
  title = {Superdiffusive transport on lattices with nodal impurities},
  author = {Wang, Yu-Peng and Ren, Jie and Fang, Chen},
  journal = {Phys. Rev. B},
  volume = {110},
  issue = {14},
  pages = {144201},
  numpages = {6},
  year = {2024},
  month = {Oct},
  publisher = {American Physical Society},
  doi = {10.1103/PhysRevB.110.144201},
  url = {https://link.aps.org/doi/10.1103/PhysRevB.110.144201}
}

@Article{Jie24,
	title={{Superdiffusive transport in quasi-particle dephasing models}},
	author={Yu-Peng Wang and Chen Fang and Jie Ren},
	journal={SciPost Phys.},
	volume={17},
	pages={150},
	year={2024},
	publisher={SciPost},
	doi={10.21468/SciPostPhys.17.6.150},
	url={https://scipost.org/10.21468/SciPostPhys.17.6.150},
}

@article{Junaid24,
  title = {Superdiffusive transport in two-dimensional fermionic wires},
  author = {Bhat, Junaid Majeed},
  journal = {Phys. Rev. B},
  volume = {110},
  issue = {11},
  pages = {115405},
  numpages = {9},
  year = {2024},
  month = {Sep},
  publisher = {American Physical Society},
  doi = {10.1103/PhysRevB.110.115405},
  url = {https://link.aps.org/doi/10.1103/PhysRevB.110.115405}
}

@article{Marko24,
  title = {Superdiffusive magnetization transport in the XX spin chain with nonlocal dephasing},
  author = {\ifmmode \check{Z}\else \v{Z}\fi{}nidari\ifmmode \check{c}\else \v{c}\fi{}, Marko},
  journal = {Phys. Rev. B},
  volume = {109},
  issue = {7},
  pages = {075105},
  numpages = {9},
  year = {2024},
  month = {Feb},
  publisher = {American Physical Society},
  doi = {10.1103/PhysRevB.109.075105},
  url = {https://link.aps.org/doi/10.1103/PhysRevB.109.075105}
}

@article{Pal25,
  title = {Measurement-Induced L\'evy Flights of Quantum Information},
  author = {Poboiko, Igor and Szyniszewski, Marcin and Turner, Christopher J. and Gornyi, Igor V. and Mirlin, Alexander D. and Pal, Arijeet},
  journal = {Phys. Rev. Lett.},
  volume = {135},
  issue = {17},
  pages = {170403},
  numpages = {7},
  year = {2025},
  month = {Oct},
  publisher = {American Physical Society},
  doi = {10.1103/tx71-1cd9},
  url = {https://link.aps.org/doi/10.1103/tx71-1cd9}
}

@article{Mantica97,
title = {Quantum intermittency in almost-periodic lattice systems derived from their spectral properties},
journal = {Physica D: Nonlinear Phenomena},
volume = {103},
number = {1},
pages = {576-589},
year = {1997},
note = {Lattice Dynamics},
issn = {0167-2789},
doi = {https://doi.org/10.1016/S0167-2789(96)00287-4},
url = {https://www.sciencedirect.com/science/article/pii/S0167278996002874},
author = {Giorgio Mantica}
}

@article{Anderson58,
  title = {Absence of Diffusion in Certain Random Lattices},
  author = {Anderson, P. W.},
  journal = {Phys. Rev.},
  volume = {109},
  issue = {5},
  pages = {1492--1505},
  numpages = {0},
  year = {1958},
  month = {Mar},
  publisher = {American Physical Society},
  doi = {10.1103/PhysRev.109.1492},
  url = {https://link.aps.org/doi/10.1103/PhysRev.109.1492}
}

@article{Thouless72,
doi = {10.1088/0022-3719/5/1/010},
url = {https://dx.doi.org/10.1088/0022-3719/5/1/010},
year = {1972},
month = {jan},
publisher = {},
volume = {5},
number = {1},
pages = {77},
author = {D J Thouless},
title = {A relation between the density of states and range of localization for one dimensional random systems},
journal = {Journal of Physics C: Solid State Physics}
}

@article{Hirota71,
  title={Exactly Soluble Models of One-Dimensional Disordered Systems},
  url = {https://doi.org/10.1143/PTP.45.1713},
  author={Hirota, T{\=o}ru and Ishii, Kazushige},
  journal={Progress of Theoretical Physics},
  volume={45},
  number={5},
  pages={1713--1715},
  year={1971},
  publisher={Oxford University Press}
}

@article{Ljubotina18,
author = {Marko Žnidarič  and Marko Ljubotina },
title = {Interaction instability of localization in quasiperiodic systems},
journal = {Proceedings of the National Academy of Sciences},
volume = {115},
number = {18},
pages = {4595-4600},
year = {2018},
doi = {10.1073/pnas.1800589115},
URL = {https://www.pnas.org/doi/abs/10.1073/pnas.1800589115},
eprint = {https://www.pnas.org/doi/pdf/10.1073/pnas.1800589115}}

@article{Reichman15,
  title = {Absence of Diffusion in an Interacting System of Spinless Fermions on a One-Dimensional Disordered Lattice},
  author = {Bar Lev, Yevgeny and Cohen, Guy and Reichman, David R.},
  journal = {Phys. Rev. Lett.},
  volume = {114},
  issue = {10},
  pages = {100601},
  numpages = {5},
  year = {2015},
  month = {Mar},
  publisher = {American Physical Society},
  doi = {10.1103/PhysRevLett.114.100601},
  url = {https://link.aps.org/doi/10.1103/PhysRevLett.114.100601}
}

@article{Demler15,
  title = {Anomalous Diffusion and Griffiths Effects Near the Many-Body Localization Transition},
  author = {Agarwal, Kartiek and Gopalakrishnan, Sarang and Knap, Michael and M\"uller, Markus and Demler, Eugene},
  journal = {Phys. Rev. Lett.},
  volume = {114},
  issue = {16},
  pages = {160401},
  numpages = {6},
  year = {2015},
  month = {Apr},
  publisher = {American Physical Society},
  doi = {10.1103/PhysRevLett.114.160401},
  url = {https://link.aps.org/doi/10.1103/PhysRevLett.114.160401}
}

@article{Altshuler06,
title = {Metal–insulator transition in a weakly interacting many-electron system with localized single-particle states},
journal = {Annals of Physics},
volume = {321},
number = {5},
pages = {1126-1205},
year = {2006},
issn = {0003-4916},
doi = {https://doi.org/10.1016/j.aop.2005.11.014},
url = {https://www.sciencedirect.com/science/article/pii/S0003491605002630},
author = {D.M. Basko and I.L. Aleiner and B.L. Altshuler}
}

@article{Karrasch17,
doi = {10.1209/0295-5075/119/37003},
url = {https://dx.doi.org/10.1209/0295-5075/119/37003},
year = {2017},
month = {oct},
publisher = {EDP Sciences, IOP Publishing and Società Italiana di Fisica},
volume = {119},
number = {3},
pages = {37003},
author = {Bar Lev, Yevgeny and Kennes, Dante M. and Klöckner, Christian and Reichman, David R. and Karrasch, Christoph},
title = {Transport in quasiperiodic interacting systems: From superdiffusion to subdiffusion},
journal = {Europhysics Letters}
}

@article{sub1,
  title = {Spectral statistics in constrained many-body quantum chaotic systems},
  author = {Moudgalya, Sanjay and Prem, Abhinav and Huse, David A. and Chan, Amos},
  journal = {Phys. Rev. Res.},
  volume = {3},
  issue = {2},
  pages = {023176},
  numpages = {27},
  year = {2021},
  month = {Jun},
  publisher = {American Physical Society},
  doi = {10.1103/PhysRevResearch.3.023176},
  url = {https://link.aps.org/doi/10.1103/PhysRevResearch.3.023176}
}

@article{sub2,
  title = {Anomalous subdiffusion from subsystem symmetries},
  author = {Iaconis, Jason and Vijay, Sagar and Nandkishore, Rahul},
  journal = {Phys. Rev. B},
  volume = {100},
  issue = {21},
  pages = {214301},
  numpages = {16},
  year = {2019},
  month = {Dec},
  publisher = {American Physical Society},
  doi = {10.1103/PhysRevB.100.214301},
  url = {https://link.aps.org/doi/10.1103/PhysRevB.100.214301}
}

@article{sub3,
  title = {Kinetically constrained freezing transition in a dipole-conserving system},
  author = {Morningstar, Alan and Khemani, Vedika and Huse, David A.},
  journal = {Phys. Rev. B},
  volume = {101},
  issue = {21},
  pages = {214205},
  numpages = {10},
  year = {2020},
  month = {Jun},
  publisher = {American Physical Society},
  doi = {10.1103/PhysRevB.101.214205},
  url = {https://link.aps.org/doi/10.1103/PhysRevB.101.214205}
}

@article{sub4,
  title = {Fracton hydrodynamics},
  author = {Gromov, Andrey and Lucas, Andrew and Nandkishore, Rahul M.},
  journal = {Phys. Rev. Res.},
  volume = {2},
  issue = {3},
  pages = {033124},
  numpages = {11},
  year = {2020},
  month = {Jul},
  publisher = {American Physical Society},
  doi = {10.1103/PhysRevResearch.2.033124},
  url = {https://link.aps.org/doi/10.1103/PhysRevResearch.2.033124}
}

@article{sub5,
  title = {Multipole conservation laws and subdiffusion in any dimension},
  author = {Iaconis, Jason and Lucas, Andrew and Nandkishore, Rahul},
  journal = {Phys. Rev. E},
  volume = {103},
  issue = {2},
  pages = {022142},
  numpages = {7},
  year = {2021},
  month = {Feb},
  publisher = {American Physical Society},
  doi = {10.1103/PhysRevE.103.022142},
  url = {https://link.aps.org/doi/10.1103/PhysRevE.103.022142}
}

@article{sub6,
  title = {Anomalous Diffusion in Dipole- and Higher-Moment-Conserving Systems},
  author = {Feldmeier, Johannes and Sala, Pablo and De Tomasi, Giuseppe and Pollmann, Frank and Knap, Michael},
  journal = {Phys. Rev. Lett.},
  volume = {125},
  issue = {24},
  pages = {245303},
  numpages = {6},
  year = {2020},
  month = {Dec},
  publisher = {American Physical Society},
  doi = {10.1103/PhysRevLett.125.245303},
  url = {https://link.aps.org/doi/10.1103/PhysRevLett.125.245303}
}

@article{sub7,
  title = {Anomalous hydrodynamics in a class of scarred frustration-free Hamiltonians},
  author = {Richter, Jonas and Pal, Arijeet},
  journal = {Phys. Rev. Res.},
  volume = {4},
  issue = {1},
  pages = {L012003},
  numpages = {8},
  year = {2022},
  month = {Jan},
  publisher = {American Physical Society},
  doi = {10.1103/PhysRevResearch.4.L012003},
  url = {https://link.aps.org/doi/10.1103/PhysRevResearch.4.L012003}
}

@article{sub8,
  title = {Subdiffusion and Many-Body Quantum Chaos with Kinetic Constraints},
  author = {Singh, Hansveer and Ware, Brayden A. and Vasseur, Romain and Friedman, Aaron J.},
  journal = {Phys. Rev. Lett.},
  volume = {127},
  issue = {23},
  pages = {230602},
  numpages = {6},
  year = {2021},
  month = {Dec},
  publisher = {American Physical Society},
  doi = {10.1103/PhysRevLett.127.230602},
  url = {https://link.aps.org/doi/10.1103/PhysRevLett.127.230602}
}

@article{sub9,
  title = {Subdiffusive transport in the Fredkin dynamical universality class},
  author = {McCarthy, Catherine and Singh, Hansveer and Gopalakrishnan, Sarang and Vasseur, Romain},
  journal = {Phys. Rev. B},
  volume = {111},
  issue = {18},
  pages = {184317},
  numpages = {13},
  year = {2025},
  month = {May},
  publisher = {American Physical Society},
  doi = {10.1103/PhysRevB.111.184317},
  url = {https://link.aps.org/doi/10.1103/PhysRevB.111.184317}
}

@article{Prosen2000,
  title = {Momentum Conservation Implies Anomalous Energy Transport in 1D Classical Lattices},
  author = {Prosen, Toma\ifmmode \check{z}\else \v{z}\fi{} and Campbell, David K.},
  journal = {Phys. Rev. Lett.},
  volume = {84},
  issue = {13},
  pages = {2857--2860},
  numpages = {0},
  year = {2000},
  month = {Mar},
  publisher = {American Physical Society},
  doi = {10.1103/PhysRevLett.84.2857},
  url = {https://link.aps.org/doi/10.1103/PhysRevLett.84.2857}
}

@article{Spohn,
	author = {Spohn, Herbert},
	date = {2014/03/01},
	doi = {10.1007/s10955-014-0933-y},
	id = {Spohn2014},
	isbn = {1572-9613},
	journal = {Journal of Statistical Physics},
	number = {5},
	pages = {1191--1227},
	title = {Nonlinear Fluctuating Hydrodynamics for Anharmonic Chains},
	url = {https://doi.org/10.1007/s10955-014-0933-y},
	volume = {154},
	year = {2014}}

@article{Narayan2002,
  title = {Anomalous Heat Conduction in One-Dimensional Momentum-Conserving Systems},
  author = {Narayan, Onuttom and Ramaswamy, Sriram},
  journal = {Phys. Rev. Lett.},
  volume = {89},
  issue = {20},
  pages = {200601},
  numpages = {4},
  year = {2002},
  month = {Oct},
  publisher = {American Physical Society},
  doi = {10.1103/PhysRevLett.89.200601},
  url = {https://link.aps.org/doi/10.1103/PhysRevLett.89.200601}
}

@article{Prosen2003,
  title = {Anomalous heat conduction in a one-dimensional ideal gas},
  author = {Casati, Giulio and Prosen, Toma\ifmmode \check{z}\else \v{z}\fi{}},
  journal = {Phys. Rev. E},
  volume = {67},
  issue = {1},
  pages = {015203},
  numpages = {4},
  year = {2003},
  month = {Jan},
  publisher = {American Physical Society},
  doi = {10.1103/PhysRevE.67.015203},
  url = {https://link.aps.org/doi/10.1103/PhysRevE.67.015203}
}

@article{Prosen2005,
    author = {Prosen, Tomaž and Campbell, David K.},
    title = {Normal and anomalous heat transport in one-dimensional classical lattices},
    journal = {Chaos: An Interdisciplinary Journal of Nonlinear Science},
    volume = {15},
    number = {1},
    pages = {015117},
    year = {2005},
    month = {03},
    issn = {1054-1500},
    doi = {10.1063/1.1868532},
    url = {https://doi.org/10.1063/1.1868532},
    eprint = {https://pubs.aip.org/aip/cha/article-pdf/doi/10.1063/1.1868532/14599171/015117\_1\_online.pdf},
}

@article{longrange2,
author = {M. K. Joshi  and F. Kranzl  and A. Schuckert  and I. Lovas  and C. Maier  and R. Blatt  and M. Knap  and C. F. Roos },
title = {Observing emergent hydrodynamics in a long-range quantum magnet},
journal = {Science},
volume = {376},
number = {6594},
pages = {720-724},
year = {2022},
doi = {10.1126/science.abk2400},
URL = {https://www.science.org/doi/abs/10.1126/science.abk2400}}

@article{longrange1,
  title = {Nonlocal emergent hydrodynamics in a long-range quantum spin system},
  author = {Schuckert, Alexander and Lovas, Izabella and Knap, Michael},
  journal = {Phys. Rev. B},
  volume = {101},
  issue = {2},
  pages = {020416},
  numpages = {6},
  year = {2020},
  month = {Jan},
  publisher = {American Physical Society},
  doi = {10.1103/PhysRevB.101.020416},
  url = {https://link.aps.org/doi/10.1103/PhysRevB.101.020416}
}

@article{longrange5,
	author = {Saha, Madhumita and Maiti, Santanu K. and Purkayastha, Archak},
	doi = {10.1103/PhysRevB.100.174201},
	issue = {17},
	journal = {Phys. Rev. B},
	month = {Nov},
	numpages = {15},
	pages = {174201},
	publisher = {American Physical Society},
	title = {Anomalous transport through algebraically localized states in one dimension},
	url = {https://link.aps.org/doi/10.1103/PhysRevB.100.174201},
	volume = {100},
	year = {2019}}

@article{longrange6,
	author = {Richter, Jonas and Lunt, Oliver and Pal, Arijeet},
	doi = {10.1103/PhysRevResearch.5.L012031},
	issue = {1},
	journal = {Phys. Rev. Res.},
	month = {Mar},
	numpages = {8},
	pages = {L012031},
	publisher = {American Physical Society},
	title = {Transport and entanglement growth in long-range random Clifford circuits},
	url = {https://link.aps.org/doi/10.1103/PhysRevResearch.5.L012031},
	volume = {5},
	year = {2023}}

@article{longrange4,
	author = {Mirlin, Alexander D. and Fyodorov, Yan V. and Dittes, Frank-Michael and Quezada, Javier and Seligman, Thomas H.},
	doi = {10.1103/PhysRevE.54.3221},
	issue = {4},
	journal = {Phys. Rev. E},
	month = {Oct},
	numpages = {0},
	pages = {3221--3230},
	publisher = {American Physical Society},
	title = {Transition from localized to extended eigenstates in the ensemble of power-law random banded matrices},
	url = {https://link.aps.org/doi/10.1103/PhysRevE.54.3221},
	volume = {54},
	year = {1996}}

@article{longrange3,
  title={Nonextensive effects in tight-binding systems with long-range hopping},
  author={Borland, Lisa and Menchero, JG},
  journal={Brazilian journal of physics},
  volume={29},
  pages={169--178},
  year={1999},
  publisher={SciELO Brasil}
}

@article{zhicheng25,
  title = {Hydrodynamic Modes and Operator Spreading in a Long-Range Center-of-Mass-Conserving Brownian Sachdev-Ye-Kitaev Model},
  author = {Cheng, Bai-Lin and Jian, Shao-Kai and Yang, Zhi-Cheng},
  journal = {Phys. Rev. Lett.},
  volume = {134},
  issue = {15},
  pages = {156301},
  numpages = {7},
  year = {2025},
  month = {Apr},
  publisher = {American Physical Society},
  doi = {10.1103/PhysRevLett.134.156301},
  url = {https://link.aps.org/doi/10.1103/PhysRevLett.134.156301}
}

@article{Titov16,
  title = {L\'evy Flights due to Anisotropic Disorder in Graphene},
  author = {Gattenl\"ohner, S. and Gornyi, I. V. and Ostrovsky, P. M. and Trauzettel, B. and Mirlin, A. D. and Titov, M.},
  journal = {Phys. Rev. Lett.},
  volume = {117},
  issue = {4},
  pages = {046603},
  numpages = {5},
  year = {2016},
  month = {Jul},
  publisher = {American Physical Society},
  doi = {10.1103/PhysRevLett.117.046603},
  url = {https://link.aps.org/doi/10.1103/PhysRevLett.117.046603}
}

@article{Fibonacci1,
  title = {One-Dimensional Schr\"odinger Equation with an Almost Periodic Potential},
  author = {Ostlund, Stellan and Pandit, Rahul and Rand, David and Schellnhuber, Hans Joachim and Siggia, Eric D.},
  journal = {Phys. Rev. Lett.},
  volume = {50},
  issue = {23},
  pages = {1873--1876},
  numpages = {0},
  year = {1983},
  month = {Jun},
  publisher = {American Physical Society},
  doi = {10.1103/PhysRevLett.50.1873},
  url = {https://link.aps.org/doi/10.1103/PhysRevLett.50.1873}
}

@article{Fibonacci2,
  title = {Localization Problem in One Dimension: Mapping and Escape},
  author = {Kohmoto, Mahito and Kadanoff, Leo P. and Tang, Chao},
  journal = {Phys. Rev. Lett.},
  volume = {50},
  issue = {23},
  pages = {1870--1872},
  numpages = {0},
  year = {1983},
  month = {Jun},
  publisher = {American Physical Society},
  doi = {10.1103/PhysRevLett.50.1870},
  url = {https://link.aps.org/doi/10.1103/PhysRevLett.50.1870}
}

@article{Fibonacci3,
  title={Dynamics of an Electron in Quasiperiodic Systems. I. Fibonacci Model},
  author={Hisashi Hiramoto and Shuji Abe},
  journal={Journal of the Physical Society of Japan},
  volume={57},
  number={1},
  pages={230-240},
  year={1988},
  doi={10.1143/JPSJ.57.230}
}

@article{Phillips1990,
  title = {Absence of localization in a random-dimer model},
  author = {Dunlap, David H. and Wu, H-L. and Phillips, Philip W.},
  journal = {Phys. Rev. Lett.},
  volume = {65},
  issue = {1},
  pages = {88--91},
  numpages = {0},
  year = {1990},
  month = {Jul},
  publisher = {American Physical Society},
  doi = {10.1103/PhysRevLett.65.88},
  url = {https://link.aps.org/doi/10.1103/PhysRevLett.65.88}
}

@article{Harmonic1971,
    author = {Rubin, Robert J. and Greer, William L.},
    title = "{Abnormal Lattice Thermal Conductivity of a One‐Dimensional, Harmonic, Isotopically Disordered Crystal}",
    journal = {Journal of Mathematical Physics},
    volume = {12},
    number = {8},
    pages = {1686-1701},
    year = {1971},
    month = {08},
    issn = {0022-2488},
    doi = {10.1063/1.1665793},
    url = {https://doi.org/10.1063/1.1665793},
    eprint = {https://pubs.aip.org/aip/jmp/article-pdf/12/8/1686/19114109/1686\_1\_online.pdf},
}

@article{Dhar2001,
  title = {Heat Conduction in the Disordered Harmonic Chain Revisited},
  author = {Dhar, Abhishek},
  journal = {Phys. Rev. Lett.},
  volume = {86},
  issue = {26},
  pages = {5882--5885},
  numpages = {0},
  year = {2001},
  month = {Jun},
  publisher = {American Physical Society},
  doi = {10.1103/PhysRevLett.86.5882},
  url = {https://link.aps.org/doi/10.1103/PhysRevLett.86.5882}
}

@article{Cane2021,
doi = {10.1088/1742-5468/ac32b8},
url = {https://dx.doi.org/10.1088/1742-5468/ac32b8},
year = {2021},
month = {nov},
publisher = {IOP Publishing and SISSA},
volume = {2021},
number = {11},
pages = {113204},
author = {Gaëtan Cane and Junaid Majeed Bhat and Abhishek Dhar and Cédric Bernardin},
title = {Localization effects due to a random magnetic field on heat transport in a harmonic chain},
journal = {Journal of Statistical Mechanics: Theory and Experiment}
}

@article{Marko11,
  title = {Spin Transport in a One-Dimensional Anisotropic Heisenberg Model},
  author = {\ifmmode \check{Z}\else \v{Z}\fi{}nidari\ifmmode \check{c}\else \v{c}\fi{}, Marko},
  journal = {Phys. Rev. Lett.},
  volume = {106},
  issue = {22},
  pages = {220601},
  numpages = {4},
  year = {2011},
  month = {May},
  publisher = {American Physical Society},
  doi = {10.1103/PhysRevLett.106.220601},
  url = {https://link.aps.org/doi/10.1103/PhysRevLett.106.220601}
}

@article{Prosen17,
  title={Spin diffusion from an inhomogeneous quench in an integrable system},
  author={Ljubotina, Marko and {\v{Z}}nidari{\v{c}}, Marko and Prosen, Toma{\v{z}}},
  journal={Nature communications},
  volume={8},
  number={1},
  pages={16117},
  year={2017},
  publisher={Nature Publishing Group UK London},
  url = {https://doi.org/10.1038/ncomms16117}
}

@article{Prosen19,
  title = {Kardar-Parisi-Zhang Physics in the Quantum Heisenberg Magnet},
  author = {Ljubotina, Marko and \ifmmode \check{Z}\else \v{Z}\fi{}nidari\ifmmode \check{c}\else \v{c}\fi{}, Marko and Prosen, Toma\ifmmode \check{z}\else \v{z}\fi{}},
  journal = {Phys. Rev. Lett.},
  volume = {122},
  issue = {21},
  pages = {210602},
  numpages = {6},
  year = {2019},
  month = {May},
  publisher = {American Physical Society},
  doi = {10.1103/PhysRevLett.122.210602},
  url = {https://link.aps.org/doi/10.1103/PhysRevLett.122.210602}
}

@article{Romain19,
  title = {Kinetic Theory of Spin Diffusion and Superdiffusion in $XXZ$ Spin Chains},
  author = {Gopalakrishnan, Sarang and Vasseur, Romain},
  journal = {Phys. Rev. Lett.},
  volume = {122},
  issue = {12},
  pages = {127202},
  numpages = {6},
  year = {2019},
  month = {Mar},
  publisher = {American Physical Society},
  doi = {10.1103/PhysRevLett.122.127202},
  url = {https://link.aps.org/doi/10.1103/PhysRevLett.122.127202}
}

@article{Brayden21,
  title = {Superuniversality of Superdiffusion},
  author = {Ilievski, Enej and De Nardis, Jacopo and Gopalakrishnan, Sarang and Vasseur, Romain and Ware, Brayden},
  journal = {Phys. Rev. X},
  volume = {11},
  issue = {3},
  pages = {031023},
  numpages = {31},
  year = {2021},
  month = {Jul},
  publisher = {American Physical Society},
  doi = {10.1103/PhysRevX.11.031023},
  url = {https://link.aps.org/doi/10.1103/PhysRevX.11.031023}
}

@article{Claeys22,
  title = {Absence of Superdiffusion in Certain Random Spin Models},
  author = {Claeys, Pieter W. and Lamacraft, Austen and Herzog-Arbeitman, Jonah},
  journal = {Phys. Rev. Lett.},
  volume = {128},
  issue = {24},
  pages = {246603},
  numpages = {5},
  year = {2022},
  month = {Jun},
  publisher = {American Physical Society},
  doi = {10.1103/PhysRevLett.128.246603},
  url = {https://link.aps.org/doi/10.1103/PhysRevLett.128.246603}
}

@Article{Lucas21,
	title={{Hydrodynamics in lattice models with continuous non-Abelian symmetries}},
	author={Paolo Glorioso and Luca V. Delacrétaz and Xiao Chen and Rahul M. Nandkishore and Andrew Lucas},
	journal={SciPost Phys.},
	volume={10},
	pages={015},
	year={2021},
	publisher={SciPost},
	doi={10.21468/SciPostPhys.10.1.015},
	url={https://scipost.org/10.21468/SciPostPhys.10.1.015},
}

@article{AnnualReviews24,
   author = "Gopalakrishnan, Sarang and Vasseur, Romain",
   title = "Superdiffusion from Nonabelian Symmetries in Nearly Integrable Systems", 
   journal= "Annual Review of Condensed Matter Physics",
   year = "2024",
   volume = "15",
   number = "Volume 15, 2024",
   pages = "159-176",
   doi = "https://doi.org/10.1146/annurev-conmatphys-032922-110710",
   url = "https://www.annualreviews.org/content/journals/10.1146/annurev-conmatphys-032922-110710",
   publisher = "Annual Reviews",
   issn = "1947-5462",
   type = "Journal Article",
  }

@article{RMP23,
  title = {Finite-temperature transport in one-dimensional quantum lattice models},
  author = {Bertini, B. and Heidrich-Meisner, F. and Karrasch, C. and Prosen, T. and Steinigeweg, R. and \ifmmode \check{Z}\else \v{Z}\fi{}nidari\ifmmode \check{c}\else \v{c}\fi{}, M.},
  journal = {Rev. Mod. Phys.},
  volume = {93},
  issue = {2},
  pages = {025003},
  numpages = {71},
  year = {2021},
  month = {May},
  publisher = {American Physical Society},
  doi = {10.1103/RevModPhys.93.025003},
  url = {https://link.aps.org/doi/10.1103/RevModPhys.93.025003}
}

@article{Norman22,
  title = {Universal Kardar-Parisi-Zhang Dynamics in Integrable Quantum Systems},
  author = {Ye, Bingtian and Machado, Francisco and Kemp, Jack and Hutson, Ross B. and Yao, Norman Y.},
  journal = {Phys. Rev. Lett.},
  volume = {129},
  issue = {23},
  pages = {230602},
  numpages = {7},
  year = {2022},
  month = {Nov},
  publisher = {American Physical Society},
  doi = {10.1103/PhysRevLett.129.230602},
  url = {https://link.aps.org/doi/10.1103/PhysRevLett.129.230602}
}

@article{Romain24,
  title = {Slow crossover from superdiffusion to diffusion in isotropic spin chains},
  author = {McCarthy, Catherine and Gopalakrishnan, Sarang and Vasseur, Romain},
  journal = {Phys. Rev. B},
  volume = {110},
  issue = {18},
  pages = {L180301},
  numpages = {6},
  year = {2024},
  month = {Nov},
  publisher = {American Physical Society},
  doi = {10.1103/PhysRevB.110.L180301},
  url = {https://link.aps.org/doi/10.1103/PhysRevB.110.L180301}
}

@article{Moessner24,
  title = {Parametrically Long Lifetime of Superdiffusion in Nonintegrable Spin Chains},
  author = {McRoberts, Adam J. and Moessner, Roderich},
  journal = {Phys. Rev. Lett.},
  volume = {133},
  issue = {25},
  pages = {256301},
  numpages = {7},
  year = {2024},
  month = {Dec},
  publisher = {American Physical Society},
  doi = {10.1103/PhysRevLett.133.256301},
  url = {https://link.aps.org/doi/10.1103/PhysRevLett.133.256301}
}

@article{boundarydrivenRMP,
  title = {Nonequilibrium boundary-driven quantum systems: Models, methods, and properties},
  author = {Landi, Gabriel T. and Poletti, Dario and Schaller, Gernot},
  journal = {Rev. Mod. Phys.},
  volume = {94},
  issue = {4},
  pages = {045006},
  numpages = {58},
  year = {2022},
  month = {Dec},
  publisher = {American Physical Society},
  doi = {10.1103/RevModPhys.94.045006},
  url = {https://link.aps.org/doi/10.1103/RevModPhys.94.045006}
}

@article{KwantRef,
doi = {10.1088/1367-2630/16/6/063065},
url = {https://doi.org/10.1088/1367-2630/16/6/063065},
year = {2014},
month = {jun},
publisher = {IOP Publishing},
volume = {16},
number = {6},
pages = {063065},
author = {Groth, Christoph W and Wimmer, Michael and Akhmerov, Anton R and Waintal, Xavier},
title = {Kwant: a software package for quantum transport},
journal = {New Journal of Physics}
}

@article{Anderson1961,
  title = {Localized Magnetic States in Metals},
  author = {Anderson, P. W.},
  journal = {Phys. Rev.},
  volume = {124},
  issue = {1},
  pages = {41--53},
  numpages = {0},
  year = {1961},
  month = {Oct},
  publisher = {American Physical Society},
  doi = {10.1103/PhysRev.124.41},
  url = {https://link.aps.org/doi/10.1103/PhysRev.124.41}
}

@book{hewson1997kondo, place={Cambridge}, series={Cambridge Studies in Magnetism}, title={The Kondo Problem to Heavy Fermions}, publisher={Cambridge University Press}, author={Hewson, Alexander Cyril}, year={1993}, collection={Cambridge Studies in Magnetism}}

@book{altland2010condensed, place={Cambridge}, edition={2}, title={Condensed Matter Field Theory}, publisher={Cambridge University Press}, author={Altland, Alexander and Simons, Ben D.}, year={2010}}

@book{akkermans2007mesoscopic, place={Cambridge}, title={Mesoscopic Physics of Electrons and Photons}, publisher={Cambridge University Press}, author={Akkermans, Eric and Montambaux, Gilles}, year={2007}}

@incollection{bernevig2013topological,
  title={Topological insulators and topological superconductors},
  author={Bernevig, B Andrei},
  booktitle={Topological Insulators and Topological Superconductors},
  year={2013},
  publisher={Princeton university press}
}

@book{bruus2004many,
    author = {Bruus, Henrik and Flensberg, Karsten},
    title = {Many–Body Quantum Theory in Condensed Matter Physics: An Introduction},
    publisher = {Oxford University Press},
    year = {2004},
    month = {09},
    isbn = {9780198566335},
    doi = {10.1093/oso/9780198566335.001.0001},
    url = {https://doi.org/10.1093/oso/9780198566335.001.0001},
}

@article{gattenlohner2016levy,
  title = {L\'evy Flights due to Anisotropic Disorder in Graphene},
  author = {Gattenl\"ohner, S. and Gornyi, I. V. and Ostrovsky, P. M. and Trauzettel, B. and Mirlin, A. D. and Titov, M.},
  journal = {Phys. Rev. Lett.},
  volume = {117},
  issue = {4},
  pages = {046603},
  numpages = {5},
  year = {2016},
  month = {Jul},
  publisher = {American Physical Society},
  doi = {10.1103/PhysRevLett.117.046603},
  url = {https://link.aps.org/doi/10.1103/PhysRevLett.117.046603}
}

@article{chiu2016classification,
  title = {Classification of topological quantum matter with symmetries},
  author = {Chiu, Ching-Kai and Teo, Jeffrey C. Y. and Schnyder, Andreas P. and Ryu, Shinsei},
  journal = {Rev. Mod. Phys.},
  volume = {88},
  issue = {3},
  pages = {035005},
  numpages = {63},
  year = {2016},
  month = {Aug},
  publisher = {American Physical Society},
  doi = {10.1103/RevModPhys.88.035005},
  url = {https://link.aps.org/doi/10.1103/RevModPhys.88.035005}
}

@article{po2020symmetry,
   title={Symmetry indicators of band topology},
   volume={32},
   ISSN={1361-648X},
   url={http://dx.doi.org/10.1088/1361-648X/ab7adb},
   DOI={10.1088/1361-648x/ab7adb},
   number={26},
   journal={Journal of Physics: Condensed Matter},
   publisher={IOP Publishing},
   author={Po, Hoi Chun},
   year={2020},
   month=apr, pages={263001} }

@book{dresselhaus2007group,
  title={Group theory: application to the physics of condensed matter},
  author={Dresselhaus, Mildred S and Dresselhaus, Gene and Jorio, Ado},
  year={2007},
  publisher={Springer Science \& Business Media}
}

@article{po2017symmetry,
   title={Symmetry-based indicators of band topology in the 230 space groups},
   volume={8},
   ISSN={2041-1723},
   url={http://dx.doi.org/10.1038/s41467-017-00133-2},
   DOI={10.1038/s41467-017-00133-2},
   number={1},
   journal={Nature Communications},
   publisher={Springer Science and Business Media LLC},
   author={Po, Hoi Chun and Vishwanath, Ashvin and Watanabe, Haruki},
   year={2017},
   month=jun }

@article{bradlyn2017topological,
   title={Topological quantum chemistry},
   volume={547},
   ISSN={1476-4687},
   url={http://dx.doi.org/10.1038/nature23268},
   DOI={10.1038/nature23268},
   number={7663},
   journal={Nature},
   publisher={Springer Science and Business Media LLC},
   author={Bradlyn, Barry and Elcoro, L. and Cano, Jennifer and Vergniory, M. G. and Wang, Zhijun and Felser, C. and Aroyo, M. I. and Bernevig, B. Andrei},
   year={2017},
   month=jul, pages={298–305} }

@article{jin2022chern,
  title = {Chern numbers of topological phonon band crossing determined with inelastic neutron scattering},
  author = {Jin, Zhendong and Hu, Biaoyan and Liu, Yiran and Li, Yangmu and Zhang, Tiantian and Iida, Kazuki and Kamazawa, Kazuya and Kolesnikov, A. I. and Stone, M. B. and Zhang, Xiangyu and Chen, Haiyang and Wang, Yandong and Zaliznyak, I. A. and Tranquada, J. M. and Fang, Chen and Li, Yuan},
  journal = {Phys. Rev. B},
  volume = {106},
  issue = {22},
  pages = {224304},
  numpages = {9},
  year = {2022},
  month = {Dec},
  publisher = {American Physical Society},
  doi = {10.1103/PhysRevB.106.224304},
  url = {https://link.aps.org/doi/10.1103/PhysRevB.106.224304}
}

@article{fang2012multi,
  title = {Multi-Weyl Topological Semimetals Stabilized by Point Group Symmetry},
  author = {Fang, Chen and Gilbert, Matthew J. and Dai, Xi and Bernevig, B. Andrei},
  journal = {Phys. Rev. Lett.},
  volume = {108},
  issue = {26},
  pages = {266802},
  numpages = {5},
  year = {2012},
  month = {Jun},
  publisher = {American Physical Society},
  doi = {10.1103/PhysRevLett.108.266802},
  url = {https://link.aps.org/doi/10.1103/PhysRevLett.108.266802}
}

@article{sobota2021angle,
  title = {Angle-resolved photoemission studies of quantum materials},
  author = {Sobota, Jonathan A. and He, Yu and Shen, Zhi-Xun},
  journal = {Rev. Mod. Phys.},
  volume = {93},
  issue = {2},
  pages = {025006},
  numpages = {72},
  year = {2021},
  month = {May},
  publisher = {American Physical Society},
  doi = {10.1103/RevModPhys.93.025006},
  url = {https://link.aps.org/doi/10.1103/RevModPhys.93.025006}
}

@article{luttingerpr,
  title = {Analytic Properties of Single-Particle Propagators for Many-Fermion Systems},
  author = {Luttinger, J. M.},
  journal = {Phys. Rev.},
  volume = {121},
  issue = {4},
  pages = {942--949},
  numpages = {0},
  year = {1961},
  month = {Feb},
  publisher = {American Physical Society},
  doi = {10.1103/PhysRev.121.942},
  url = {https://link.aps.org/doi/10.1103/PhysRev.121.942}
}

@book{coleman2015introduction,
  title={Introduction to many-body physics},
  author={Coleman, Piers},
  year={2015},
  publisher={Cambridge University Press}
}

@article{anderson1999site,
title = {Site-distributions of Fe alloying additions to B2-ordered NiAl},
journal = {Intermetallics},
volume = {7},
number = {9},
pages = {1017-1024},
year = {1999},
issn = {0966-9795},
doi = {https://doi.org/10.1016/S0966-9795(99)00013-8},
url = {https://www.sciencedirect.com/science/article/pii/S0966979599000138},
author = {I.M Anderson and A.J Duncan and J Bentley}
}

@article{soluyanov2011wannier,
  title = {Wannier representation of ${\mathbb{Z}}_{2}$ topological insulators},
  author = {Soluyanov, Alexey A. and Vanderbilt, David},
  journal = {Phys. Rev. B},
  volume = {83},
  issue = {3},
  pages = {035108},
  numpages = {11},
  year = {2011},
  month = {Jan},
  publisher = {American Physical Society},
  doi = {10.1103/PhysRevB.83.035108},
  url = {https://link.aps.org/doi/10.1103/PhysRevB.83.035108}
}

@article{thonhauser2006insulator,
  title = {Insulator/Chern-insulator transition in the Haldane model},
  author = {Thonhauser, T. and Vanderbilt, David},
  journal = {Phys. Rev. B},
  volume = {74},
  issue = {23},
  pages = {235111},
  numpages = {8},
  year = {2006},
  month = {Dec},
  publisher = {American Physical Society},
  doi = {10.1103/PhysRevB.74.235111},
  url = {https://link.aps.org/doi/10.1103/PhysRevB.74.235111}
}

@book{ashcroft2011solid,
  title={Solid State Physics},
  author={Ashcroft, N.W. and Mermin, N.D.},
  isbn={9788131500521},
  lccn={74009772},
  url={https://books.google.com/books?id=x_s_YAAACAAJ},
  year={2011},
  publisher={Cengage Learning}
}

@misc{muller2010disorder,
      title={Disorder and interference: localization phenomena}, 
      author={Cord A. Müller and Dominique Delande},
      year={2016},
      eprint={1005.0915},
      archivePrefix={arXiv},
      primaryClass={cond-mat.dis-nn},
      url={https://arxiv.org/abs/1005.0915}, 
}

@article{Wölfle1984interference,
  title = {Electron localization in anisotropic systems},
  author = {W\"olfle, P. and Bhatt, R. N.},
  journal = {Phys. Rev. B},
  volume = {30},
  issue = {6},
  pages = {3542--3544},
  numpages = {0},
  year = {1984},
  month = {Sep},
  publisher = {American Physical Society},
  doi = {10.1103/PhysRevB.30.3542},
  url = {https://link.aps.org/doi/10.1103/PhysRevB.30.3542}
}

@article{Abrahams1979Scaling,
  title = {Scaling Theory of Localization: Absence of Quantum Diffusion in Two Dimensions},
  author = {Abrahams, E. and Anderson, P. W. and Licciardello, D. C. and Ramakrishnan, T. V.},
  journal = {Phys. Rev. Lett.},
  volume = {42},
  issue = {10},
  pages = {673--676},
  numpages = {0},
  year = {1979},
  month = {Mar},
  publisher = {American Physical Society},
  doi = {10.1103/PhysRevLett.42.673},
  url = {https://link.aps.org/doi/10.1103/PhysRevLett.42.673}
}

@article{PhysRevB.64.165303,
  title = {Deconstructing Kubo formula usage: Exact conductance of a mesoscopic system from weak to strong disorder limit},
  author = {Nikoli\ifmmode \acute{c}\else \'{c}\fi{}, Branislav K.},
  journal = {Phys. Rev. B},
  volume = {64},
  issue = {16},
  pages = {165303},
  numpages = {7},
  year = {2001},
  month = {Oct},
  publisher = {American Physical Society},
  doi = {10.1103/PhysRevB.64.165303},
  url = {https://link.aps.org/doi/10.1103/PhysRevB.64.165303}
}

@article{xu2011chern,
  title = {Chern Semimetal and the Quantized Anomalous Hall Effect in ${\mathrm{HgCr}}_{2}{\mathrm{Se}}_{4}$},
  author = {Xu, Gang and Weng, Hongming and Wang, Zhijun and Dai, Xi and Fang, Zhong},
  journal = {Phys. Rev. Lett.},
  volume = {107},
  issue = {18},
  pages = {186806},
  numpages = {5},
  year = {2011},
  month = {Oct},
  publisher = {American Physical Society},
  doi = {10.1103/PhysRevLett.107.186806},
  url = {https://link.aps.org/doi/10.1103/PhysRevLett.107.186806}
}

@article{liu2017predicted,
  title = {Predicted Realization of Cubic Dirac Fermion in Quasi-One-Dimensional Transition-Metal Monochalcogenides},
  author = {Liu, Qihang and Zunger, Alex},
  journal = {Phys. Rev. X},
  volume = {7},
  issue = {2},
  pages = {021019},
  numpages = {14},
  year = {2017},
  month = {May},
  publisher = {American Physical Society},
  doi = {10.1103/PhysRevX.7.021019},
  url = {https://link.aps.org/doi/10.1103/PhysRevX.7.021019}
}

@article{RevModPhys.57.287,
  title = {Disordered electronic systems},
  author = {Lee, Patrick A. and Ramakrishnan, T. V.},
  journal = {Rev. Mod. Phys.},
  volume = {57},
  issue = {2},
  pages = {287--337},
  numpages = {0},
  year = {1985},
  month = {Apr},
  publisher = {American Physical Society},
  doi = {10.1103/RevModPhys.57.287},
  url = {https://link.aps.org/doi/10.1103/RevModPhys.57.287}
}
\end{document}